\documentclass[
  aps,
  prd,
  twocolumn,
  nofootinbib,
  floatfix,
  superscriptaddress
]{revtex4-2}

\usepackage{amsmath,amssymb,bm}
\usepackage{graphicx}
\usepackage[dvipsnames]{xcolor}
\usepackage{tikz}
\usetikzlibrary{arrows.meta,calc,positioning,shapes.geometric}
\usepackage[
  colorlinks=true,
  linkcolor=MidnightBlue,
  citecolor=ForestGreen,
  urlcolor=MidnightBlue
]{hyperref}
\usepackage{xcolor}

\hypersetup{
  pdftitle={Coherent detector responses to quantum-controlled spherical-shell geometries},
  pdfauthor={Shoshauna Gauvin and Robert B. Mann},
  pdfsubject={Detector responses to quantum-controlled spherical-shell geometries},
  pdfkeywords={Unruh-DeWitt detector, quantum-controlled spacetime, spherical shell, relational position}
}

\graphicspath{{figures/}}

\newcommand{\dd}{\mathrm{d}}
\newcommand{\ii}{\mathrm{i}}
\newcommand{\F}{\mathcal{F}}
\newcommand{\Hilbert}{\mathcal{H}}
\newcommand{\order}{\mathcal{O}}
\newcommand{\sinc}{\operatorname{sinc}}
\newcommand{\ketbra}[2]{|#1\rangle\!\langle#2|}

\definecolor{BranchBlue}{RGB}{0,114,178}
\definecolor{BranchVermillion}{RGB}{213,94,0}
\tikzset{
  shellone/.style={draw=BranchBlue,line width=1.15pt},
  shelltwo/.style={draw=BranchVermillion,line width=1.15pt,densely dashed},
  qguide/.style={draw=black!55,line width=0.55pt,densely dotted},
  qdim/.style={<->,>=Latex,line width=0.65pt},
  detector/.style={diamond,draw=black,fill=black,
    minimum size=5pt,inner sep=0pt},
  centerone/.style={circle,draw=BranchBlue,fill=white,
    minimum size=5pt,inner sep=0pt,line width=1pt},
  centertwo/.style={rectangle,draw=BranchVermillion,fill=white,
    minimum size=5pt,inner sep=0pt,line width=1pt}
}

\begin{document}

\title{Coherent detector responses to quantum-controlled spherical-shell geometries}

\author{Shoshauna Gauvin}
\email{s3gauvin@uwaterloo.ca}
\affiliation{Department of Physics and Astronomy, University of Waterloo,
Waterloo, Ontario, Canada N2L 3G1}
\affiliation{Perimeter Institute for Theoretical Physics, 31 Caroline Street
North, Waterloo, Ontario, Canada N2L 2Y5}

\author{Robert B.\ Mann}
\email{rbmann@uwaterloo.ca}
\affiliation{Department of Physics and Astronomy, University of Waterloo,
Waterloo, Ontario, Canada N2L 3G1}
\affiliation{Perimeter Institute for Theoretical Physics, 31 Caroline Street
North, Waterloo, Ontario, Canada N2L 2Y5}

\begin{abstract}
A detector inside a static spherical shell samples globally matched field
modes even though its local geometry is flat. We construct the response of a
compactly switched Unruh-DeWitt detector on a quantum-controlled pair of
transparent spherical-shell spacetimes with common Arnowitt-Deser-Misner
mass. A stipulated common-incoming scattering identification places the
branch excitations in one Hilbert space. Their positive Gram matrix determines joint, conditional, and dephased
detector responses at a finite common readout, retaining the physical
detector-clock phase in the uncompensated protocol. An additional coherent
phase-gate protocol compensates free detector-clock propagation while
preserving the shell interaction response. For
concentric shells the centered response is exactly $s$ wave. A common incoming
direction basis extends the construction to a lab-fixed detector and displaced
shell centers, preserving common-translation covariance. The equal-radius
benchmark exhibits conditional changes of order $25\%$ without dark-port
amplification. Combining radius and position produces an assignment-odd response
that reverses under reflection and vanishes at the relational midpoint. The
relative normalized-state contrast reaches $34.15\%$ among the computed samples; the
relative clock-resolved probe-influence contrast reaches $8.09\%$ for the declared
readout, while larger values belong to its separate shared-phase envelope.
These effects already occur in the dominant-energy-condition-compatible core.
They are relational, model-dependent detector responses, not a
background-independent witness of quantum geometry. First-order coefficients
have finite pointlike limits; finite-coupling statements are made only at a
fixed shared regulator.
\end{abstract}
\maketitle

\section{Introduction}
\label{sec:introduction}

Unruh-DeWitt detectors probe quantum fields through a local two-level
coupling over a finite switching interval \cite{Unruh76,DeWitt:1980hx}.
Inside a static spherical shell the metric is flat up to a constant time
rescaling. Nevertheless, field modes are normalized by matching the regular
interior solution through the shell to the Schwarzschild exterior
\cite{Ng2016HollowShell,Cong2020LocalVsGlobal}. In particular, a compactly switched detector can distinguish
the shell vacuum from the Minkowski vacuum even when the interaction
ends before a light signal could reach the shell and return
\cite{Cong2020LocalVsGlobal}.

Quantum-controlled backgrounds and trajectories allow interference between
branch-dependent field amplitudes \cite{FooOnoeMannZych2021,RMann2022,RMann2023}.
A related shell construction by M{\'o}ller, Sahdo, and Yokomizo uses controlled
shell arrangements to implement indefinite causal order
\cite{MollerSahdoYokomizo2024}. Here the question is instead how coherent shell
geometry changes a local detector response. Static-trajectory superpositions
in a single Schwarzschild background connect detector coherence to
field-output distinguishability \cite{PaczosBarbado2023}; position-superposed
Ba{\~n}ados-Teitelboim-Zanelli (BTZ) backgrounds provide a complementary setting for relational detector
readout \cite{WalleghemCepollaro2026}.

We consider alternative transparent shells of common
Arnowitt-Deser-Misner (ADM) mass and different
areal radii, and then allow their centers to differ relative to one fixed
detector. Unequal-radius branches are globally nonisometric, with different
interior lapses and globally matched amplitudes and phases. Equal-radius
translated configurations require an external relational reference:
translation of the complete system alone is not an observable displacement
\cite{FooSuryaatmadjaMannZych2026}. Our displaced-center construction therefore
uses detector-shell relative positions in one declared asymptotic frame,
consistent with quantum-reference-frame descriptions
\cite{GiacominiCastroRuizBrukner2019,deLaHametteKabelCastroRuizBrukner2023,FooOnoeZych2020}.

A cross-branch field response also requires a choice of how field algebras
are identified. Locally covariant quantum field theory does not itself select
an isometry between distinct global backgrounds
\cite{BrunettiFredenhagenVerch2003}. We stipulate a common incoming scattering
basis at the shared asymptotic mass and Killing time. Displaced shells are
expressed in its common direction basis; partial-wave labels around different
origins are not equated. This identification is a physical assumption of the
controlled model and is held fixed throughout.

The resulting positive excitation Gram matrix organizes the joint detector
and control probabilities, conditional detector response, and dephased
mixture. Keeping these quantities distinct avoids attributing a normalization
factor to geometric enhancement. We propagate the detector to one finite
post-switch event, so its branch-dependent proper-time phase remains in the
coherent cross term. A full phase scan distinguishes fringe translation from
visibility; the magnitude of a normalized overlap alone does not determine
the clock-resolved port response.

Our contribution is the common-direction construction for controlled shell
scattering and its nonzero finite radius-position assignment response at a
specified clock readout. The centered and lab-fixed equal-radius cases provide
benchmarks. Combined paths exchange which center carries the smaller shell:
reflection fixes the midpoint null and sign reversal as consistency checks.
An independent control holds $b/R$ fixed, where $b$ is the separation
of the shell centers and $R$ is their mean radius, to test whether
changing relative separation alone explains the contrast. Normalized-state and leading
probe-influence contrasts distinguish conditioning from yield imbalance and
detector clock history. Compactness parametrizes these finite paths; every
sample remains exterior, and the continuation beyond the dominant energy
condition is separately marked.

In a companion manuscript~\cite{GauvinMannCompactnessCompanion2026} we employ the same
shell model and the mode, switching, Gram-matrix, and regulator methods to investigate
 conditioned-state geometry under compactness changes, receiver
information, and the strict cavity limit; here we evaluate finite
radius-center paths.

Sections~\ref{sec:geometry}-\ref{sec:center} construct the modes, protocol, and
centered response; Sec.~\ref{sec:relational-center} extends the Gram matrix to
displaced centers. Sections~\ref{sec:numerics} and \ref{sec:results} specify and
evaluate the benchmarks. The appendices supply the spectral, perturbative,
and numerical foundations.

\section{Static thin-shell branch geometries}
\label{sec:geometry}

We use geometrized units $G=c=\hbar=1$ and metric signature $(-,+,+,+)$.
We model a shell in a coherent superposition of two static configurations,
whose branch geometries are labeled by $i\in\{1,2\}$.  They have the same
ADM mass $M>0$ and shell radii
\begin{equation}
  R_i>2M, \qquad R_1\neq R_2
  \quad\text{for the centered benchmark}.
  \label{eq:radius-domain}
\end{equation}
The centered baseline has a common center.  The mode construction itself
requires only $R_i>2M$; the profile benchmark in
Sec.~\ref{sec:relational-numerics} sets $R_1=R_2$ to isolate position, whereas
the joint benchmark combines the nearby radii in Eq.~\eqref{eq:nearby-radii}
with two displaced centers.  In every case the branches are alternative
geometries, not simultaneous material layers.
The asymptotic Schwarzschild time $t$ and the areal radius $r$ are used in
both branches.  Defining
\begin{equation}
  f(r)=1-\frac{2M}{r}, \qquad f_i=f(R_i)=1-\frac{2M}{R_i},
  \label{eq:lapses}
\end{equation}
we use $\mathcal C_i:=2M/R_i=1-f_i$ for the dimensionless compactness
of branch $i$.  The branch metric is
\begin{equation}
  \dd s_i^2=-N_i(r)^2\dd t^2+B_i(r)^2\dd r^2+r^2\dd\Omega^2,
  \label{eq:piecewise-metric}
\end{equation}
where
\begin{equation}
  (N_i^2,B_i^2)=
  \begin{cases}
    (f_i,1), & 0\leq r<R_i,\\
    (f(r),f(r)^{-1}), & r>R_i.
  \end{cases}
  \label{eq:metric-functions}
\end{equation}
The induced metric is continuous at $r=R_i$.  The interior is Minkowski
space in the time coordinate $T_i=\sqrt{f_i}\,t$, whereas $t$ is normalized
to proper time at infinity.  A static detector at the center has branchwise
proper time
\begin{equation}
  \dd\tau=\sqrt{f_i}\,\dd t.
  \label{eq:proper-time}
\end{equation}
The switching profile and detector gap are the same functions of proper time
on both branches.  Equation~\eqref{eq:proper-time} relates this operational
clock to the shared asymptotic time.  Fixing an interval of $t$ instead would
define a different detector protocol. The calculations here keep the same
proper-time pulse and detector gap on both branches.
For a common asymptotic readout event $t=t_{\rm r}$, the accumulated detector
times are
\begin{equation}
  \tau_i^{\rm r}:=\tau_i(t_{\rm r})=\sqrt{f_i}\,t_{\rm r}.
  \label{eq:readout-proper-times}
\end{equation}
Both clocks have their origin at the shared pulse center $t=0$.
The difference $\tau_1^{\rm r}-\tau_2^{\rm r}$ is retained below as a physical
clock record.

The distributional stress at the shell follows from the Israel junction
conditions \cite{Israel1966}. The classical matter source is confined
to the shell surface; the interior contains no shell matter.
For a Minkowski interior and a Schwarzschild
exterior, the surface stress tensor in an orthonormal shell frame is
$S^{a}{}_{b}=\operatorname{diag}(-\Sigma_i,p_i,p_i)$, with
\begin{align}
  \Sigma_i&=\frac{1-\sqrt{f_i}}{4\pi R_i},
  \label{eq:surface-density-shell}\\
  p_i&=\frac{1-M/R_i-\sqrt{f_i}}
  {8\pi R_i\sqrt{f_i}}
  =\frac{(1-\sqrt{f_i})^2}{16\pi R_i\sqrt{f_i}}\geq0.
  \label{eq:pressure-shell}
\end{align}
Thus $M$ is the ADM mass, not the proper rest mass $4\pi R_i^2\Sigma_i$.
The tangential pressure needed to keep the shell static diverges
as its radius approaches $2M$ from above.  The weak and null energy conditions hold for
every $R_i>2M$.  The dominant energy condition (DEC) holds if and only if
$R_i\geq25M/12$, equivalently $\mathcal C_i\leq24/25$.  The centered and
relational-center benchmarks obey this bound.  The combined radius-translation
scan extends beyond it as a formal sequence of prescribed static
thin-shell backgrounds with DEC-violating surface stress; every sampled branch
nevertheless remains strictly outside $R_i=2M$.

We treat each shell and the control preparation defined in
Sec.~\ref{sec:detector} as prescribed static inputs.
The shell is transparent to the scalar field: it carries no localized scalar
charge and induces no additional surface interaction.  These assumptions
separate the field matching problem from a material model of a semitransparent
or fluctuating boundary.

\begin{figure*}[t]
\centering
\begin{tikzpicture}[x=0.92cm,y=0.92cm,font=\small]
\begin{scope}
  \node[anchor=west,font=\bfseries] at (0,5.05)
    {(a) Centered radius branches};
  \node at (3.15,4.60)
    {$|C\rangle=\alpha_C|1\rangle_C+\beta_C e^{\ii\phi}|2\rangle_C$};

  \coordinate (ca) at (3.15,2.20);
  \draw[shellone] (ca) circle[radius=1.28];
  \draw[shelltwo] (ca) circle[radius=2.02];

  \draw[shellone,-{Latex[length=2mm]}]
    (ca) -- ++(150:1.28)
    node[midway,above right,text=BranchBlue] {$R_1$};
  \draw[shelltwo,-{Latex[length=2mm]}]
    (ca) -- ++(28:2.02)
    node[midway,below right,text=BranchVermillion] {$R_2$};

  \node[detector] (xa) at (ca) {};
  \node[below=4pt of xa,align=center]
    {$\bm x=\bm c_1=\bm c_2$\\[-1pt] detector};

  \node[text=BranchBlue,anchor=east] at (1.72,3.32)
    {solid: $|1\rangle_C$};
  \node[text=BranchVermillion,anchor=west] at (4.55,3.90)
    {dashed: $|2\rangle_C$};

  \node at (3.15,-0.05)
    {$\rho_1=\rho_2=0,\qquad \ell=0$};
\end{scope}

\draw[black!25] (7.20,-0.25) -- (7.20,5.10);

\begin{scope}[xshift=8.0cm]
  \node[anchor=west,font=\bfseries] at (0,5.05)
    {(b) Lab-fixed translated centers};

  \coordinate (c1) at (2.00,2.20);
  \coordinate (c2) at (4.50,2.20);
  \coordinate (c0) at ($(c1)!0.5!(c2)$);
  \coordinate (xb) at (3.60,2.20);

  \draw[->,black!65] (0.10,2.20) -- (6.85,2.20)
    node[right] {scan axis};
  \draw[shellone] (c1) circle[radius=1.85];
  \draw[shelltwo] (c2) circle[radius=1.85];

  \node[centerone,label={[text=BranchBlue]below:$\bm c_1$}] at (c1) {};
  \node[centertwo,label={[text=BranchVermillion]below:$\bm c_2$}] at (c2) {};
  \node[detector,label=below:$\bm x$] at (xb) {};

  \draw[black!55] ($(c0)+(-0.08,-0.08)$) -- ($(c0)+(0.08,0.08)$);
  \draw[black!55] ($(c0)+(-0.08,0.08)$) -- ($(c0)+(0.08,-0.08)$);
  \node[above=3pt of c0,text=black!65] {$\bm c_0$};

  \draw[qguide] (c1) -- ++(0,2.22);
  \draw[qguide] (c2) -- ++(0,2.22);
  \draw[qdim] ($(c1)+(0,2.15)$) -- ($(c2)+(0,2.15)$)
    node[midway,fill=white,inner sep=1pt] {$|\bm b|$};

  \draw[qguide] (c1) -- ++(0,-1.38);
  \draw[qguide] (xb) -- ++(0,-2.36);
  \draw[qdim,draw=BranchBlue]
    ($(c1)+(0,-1.12)$) -- ($(xb)+(0,-1.12)$)
    node[midway,fill=white,inner sep=1pt,text=BranchBlue] {$\rho_1$};

  \draw[qguide] (c2) -- ++(0,-2.36);
  \draw[qdim,draw=BranchVermillion]
    ($(xb)+(0,-2.25)$) -- ($(c2)+(0,-2.25)$)
    node[midway,fill=white,inner sep=1pt,text=BranchVermillion] {$\rho_2$};

  \draw[shelltwo,-{Latex[length=2mm]}]
    (c2) -- ++(58:1.85)
    node[midway,right,text=BranchVermillion] {$R$};

  \node[text=BranchBlue,anchor=east] at (1.48,4.18) {$|1\rangle_C$};
  \node[text=BranchVermillion,anchor=west] at (5.02,4.18) {$|2\rangle_C$};
  \node[align=center] at (3.35,-0.55)
    {$\rho_i=|\bm x-\bm c_i|<R,\quad
      \widehat D_{\bm x}\mapsto\{\rho_1,\rho_2\}$};
\end{scope}
\end{tikzpicture}
\caption{Shell-geometry protocols.  Solid blue and dashed vermillion denote
alternative control branches, not simultaneous material shells.  (a)
Concentric shells have $R_1\ne R_2$ and a common center; only $\ell=0$
survives at the centered detector.  (b) Intrinsically identical shells have
centers $\bm c_1\ne\bm c_2$ and one detector at $\bm x$.  The midpoint
$\bm c_0$ is a reference, not an effective center.  A distance measurement
returns $\rho_i=|\bm x-\bm c_i|$ with branch probabilities, whereas coherence
enters the clock-resolved cross term $\F_{12}^{\rm r}(t_{\rm r})$.}
\label{fig:shell-branch-geometry}
\end{figure*}
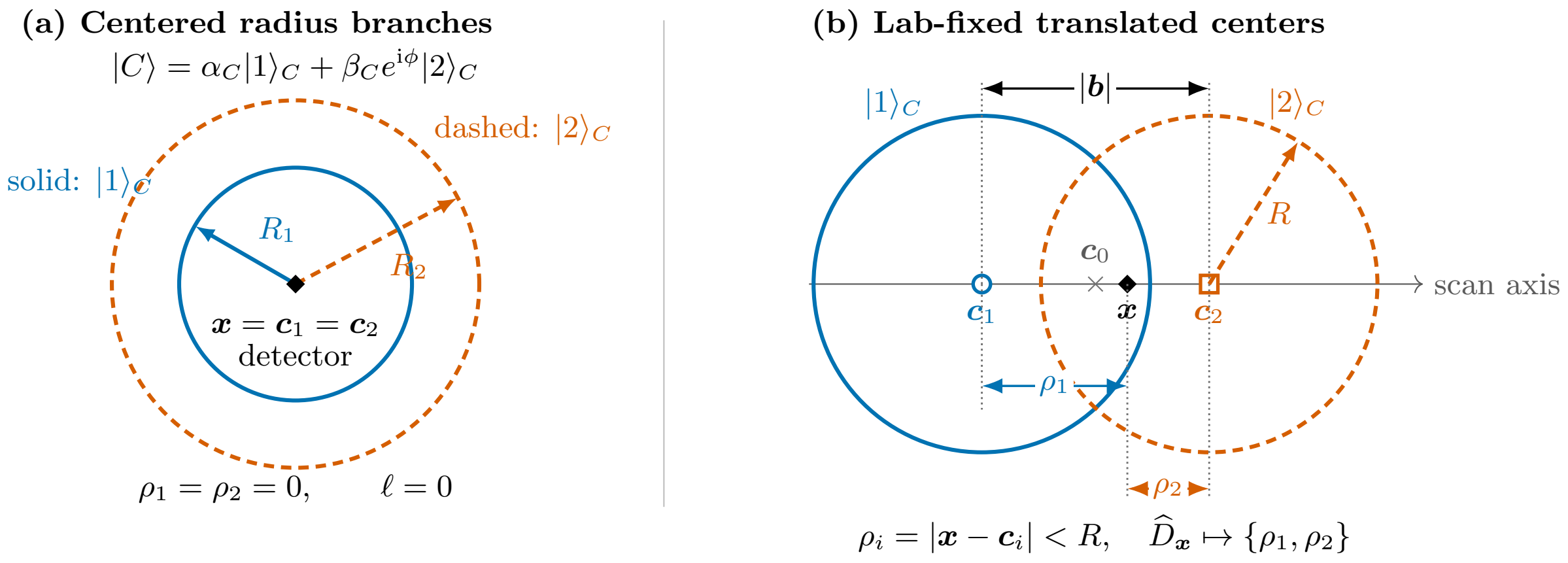

Figure~\ref{fig:shell-branch-geometry} summarizes the centered-radius and
lab-fixed translated-center protocols; its overlaid branches are alternatives,
not simultaneous material shells.

\section{Common scattering modes and field state}
\label{sec:field}

We begin by constructing the global scalar modes sampled by the detector.  The
central point is that a flat shell interior does not determine their
normalization: the regular interior solution must be joined to the common
Schwarzschild scattering problem at infinity.

\subsection{Branchwise Klein-Gordon modes}

Consider a real, massless, minimally coupled scalar field,
\begin{equation}
  \Box_i\Phi_i=0.
  \label{eq:kg-equation}
\end{equation}
Positive-frequency modes are labeled using the shared asymptotic Killing
frequency $\omega>0$:
\begin{equation}
  v^{(i)}_{\omega\ell m}(x)=
  \frac{e^{-\ii\omega t}}{\sqrt{4\pi\omega}}
  Y_{\ell m}(\vartheta,\varphi)\frac{\widetilde u^{(i)}_{\omega\ell}(r)}{r}.
  \label{eq:global-mode}
\end{equation}
Inside branch $i$, regularity at the origin gives
\begin{equation}
  \frac{u^{(i)}_{\omega\ell}(r)}{r}
  =A^{(i)}_{\omega\ell}
  j_\ell(k_i r),
  \qquad
  k_i=\frac{\omega}{\sqrt{f_i}},
  \label{eq:interior-mode}
\end{equation}
where $u^{(i)}$ is initially chosen real.  The coefficient
$A^{(i)}_{\omega\ell}$ cannot be determined from the interior solution alone.
It contains the effect of propagation through the shell and the entire
exterior scattering region.

For $r>R_i$, introduce the standard Schwarzschild tortoise coordinate
\begin{equation}
  r_*=r+2M\ln\left(\frac{r}{2M}-1\right),
  \qquad \frac{\dd r_*}{\dd r}=f(r)^{-1}.
  \label{eq:tortoise}
\end{equation}
The same additive convention is used in both branches.
Although the shell branches have $M>0$, we also use the continuous
$M\to0$ limit as a normalization check; in this limit $f\to1$ and
$r_*\to r$.
The exterior radial function obeys the Regge-Wheeler equation
\begin{equation}
  \left[\frac{\dd^2}{\dd r_*^2}+\omega^2-V_\ell(r)\right]
  u^{(i)}_{\omega\ell}(r)=0,
  \label{eq:regge-wheeler}
\end{equation}
with
\begin{equation}
  V_\ell(r)=f(r)\left[
  \frac{\ell(\ell+1)}{r^2}+\frac{2M}{r^3}
  \right].
  \label{eq:scalar-potential}
\end{equation}
Although the detector remains in a flat interior, this exterior equation is
essential because it fixes both the magnitude and phase of the mode sampled
by the detector.

\subsection{Transparent-shell junction conditions}

Integrating Eq.~\eqref{eq:kg-equation} through an infinitesimal
Gaussian-normal pillbox around the shell gives continuity of the field and of
its normal derivative.  For the
radial field $\psi=u/r$, these conditions are
\begin{align}
  \psi^{(i)}_{\rm out}(R_i)&=\psi^{(i)}_{\rm in}(R_i),
  \label{eq:field-junction}\\
  \sqrt{f_i}\,
  \partial_r\psi^{(i)}_{\rm out}(R_i)&=
  \partial_r\psi^{(i)}_{\rm in}(R_i).
  \label{eq:derivative-junction}
\end{align}
The factor $\sqrt{f_i}$ is the exterior component of the unit normal.  These
are scalar-field matching conditions; the gravitational Israel junction conditions
instead determine Eqs.~\eqref{eq:surface-density-shell} and
\eqref{eq:pressure-shell}.

Starting with unit interior coefficient, Eqs.~\eqref{eq:field-junction} and
\eqref{eq:derivative-junction} supply the exterior initial data at $R_i$.
The real solution is propagated to large radius and fitted to
\begin{equation}
  u^{(i)}_{\omega\ell}(r)\longrightarrow
  2\sin\!\left(\omega r_*+\delta^{(i)}_{\omega\ell}\right).
  \label{eq:standing-normalization}
\end{equation}
The factor of two fixes $A^{(i)}_{\omega\ell}$ rather than leaving it as an
unspecified matching constant.  Appendix~\ref{app:junction} proves that this
choice gives unit Klein-Gordon normalization and the correct Minkowski
limit.  We now convert the standing solution to the common-incoming
convention needed for cross-branch interference.

The standing solution has an incoming coefficient with a branch-dependent
phase.  Multiplication by $-\ii e^{\ii\delta^{(i)}_{\omega\ell}}$ makes that
coefficient unity.  We therefore define
\begin{equation}
  \widetilde u^{(i)}_{\omega\ell}
  =-\ii e^{\ii\delta^{(i)}_{\omega\ell}}
  u^{(i)}_{\omega\ell},
  \qquad
  \mathcal A^{(i)}_{\omega\ell}
  :=A^{(i)}_{\omega\ell}e^{\ii\delta^{(i)}_{\omega\ell}}.
  \label{eq:incoming-mode}
\end{equation}
The common factor $-\ii$ cancels from all response-matrix elements.  The
relative phase in $\mathcal A^{(1)}\mathcal A^{(2)*}$ does not cancel and must
be retained in the mixed response.

\begin{figure*}[t]
\centering
\begin{tikzpicture}[x=0.90cm,y=0.86cm,font=\footnotesize]
\node[font=\small\bfseries,align=center] at (7.45,5.82)
  {Flat interiors $T_i=\sqrt{f_i}\,t$ matched to branchwise Schwarzschild
   exteriors with common $M_{\rm ADM}=M$};

\begin{scope}
  \fill[black!4] (0.45,0.35) rectangle (2.75,4.75);
  \fill[black!1] (2.75,0.35) rectangle (6.55,4.75);
  \draw[->] (0.35,0.30) -- (0.35,5.10) node[above] {$t$};
  \draw[->] (0.35,0.30) -- (6.90,0.30) node[right] {$r$};
  \node[font=\bfseries] at (3.55,5.12)
    {(a) branch $|1\rangle_C$: shell $R_1$};
  \node[text=black!60] at (1.55,4.40) {interior};
  \node[text=black!60] at (4.62,4.40) {exterior};

  \draw[line width=1.15pt] (0.72,0.52) -- (0.72,4.20);
  \node[anchor=west] at (0.80,3.55) {$D:\ r=0$};
  \draw[shellone] (2.75,0.35) -- (2.75,4.75);
  \node[text=BranchBlue,anchor=west] at (2.82,0.60) {$r=R_1$};
  \draw[qguide] (6.45,0.35) -- (6.45,4.75);
  \node[anchor=east] at (6.32,0.62) {$r_*\to+\infty$};

  \draw[line width=0.75pt]
    (6.15,1.38) -- (2.75,2.08) -- (0.75,2.80);
  \draw[line width=0.75pt]
    (0.75,2.80) -- (2.75,3.20) -- (6.15,3.72);
  \draw[-{Latex[length=2mm]},line width=0.75pt]
    ($(6.15,1.38)!0.30!(2.75,2.08)$) --
    ($(6.15,1.38)!0.57!(2.75,2.08)$);
  \draw[-{Latex[length=2mm]},line width=0.75pt]
    ($(2.75,3.20)!0.30!(6.15,3.72)$) --
    ($(2.75,3.20)!0.57!(6.15,3.72)$);
  \node[anchor=north,fill=white,inner sep=0.8pt] at (5.00,1.22)
    {incoming};
  \node[anchor=south,fill=white,inner sep=0.8pt] at (5.10,3.86)
    {$S_\ell^{(1)}$ outgoing};
  \fill[BranchBlue] (2.75,2.08) circle[radius=1.7pt];
  \fill[BranchBlue] (2.75,3.20) circle[radius=1.7pt];
\end{scope}

\begin{scope}[xshift=8.0cm]
  \fill[black!4] (0.45,0.35) rectangle (3.35,4.75);
  \fill[black!1] (3.35,0.35) rectangle (6.55,4.75);
  \draw[->] (0.35,0.30) -- (0.35,5.10) node[above] {$t$};
  \draw[->] (0.35,0.30) -- (6.90,0.30) node[right] {$r$};
  \node[font=\bfseries] at (3.55,5.12)
    {(b) branch $|2\rangle_C$: shell $R_2$};
  \node[text=black!60] at (1.80,4.40) {interior};
  \node[text=black!60] at (4.88,4.40) {exterior};

  \draw[line width=1.15pt] (0.72,0.52) -- (0.72,4.20);
  \node[anchor=west] at (0.80,3.55) {$D:\ r=0$};
  \draw[shelltwo] (3.35,0.35) -- (3.35,4.75);
  \node[text=BranchVermillion,anchor=west] at (3.42,0.60) {$r=R_2$};
  \draw[qguide] (6.45,0.35) -- (6.45,4.75);
  \node[anchor=east] at (6.32,0.62) {$r_*\to+\infty$};

  \draw[line width=0.75pt]
    (6.15,1.38) -- (3.35,2.08) -- (0.75,2.80);
  \draw[line width=0.75pt]
    (0.75,2.80) -- (3.35,3.20) -- (6.15,3.72);
  \draw[-{Latex[length=2mm]},line width=0.75pt]
    ($(6.15,1.38)!0.30!(3.35,2.08)$) --
    ($(6.15,1.38)!0.57!(3.35,2.08)$);
  \draw[-{Latex[length=2mm]},line width=0.75pt]
    ($(3.35,3.20)!0.30!(6.15,3.72)$) --
    ($(3.35,3.20)!0.57!(6.15,3.72)$);
  \node[anchor=north,fill=white,inner sep=0.8pt] at (5.00,1.22)
    {incoming};
  \node[anchor=south,fill=white,inner sep=0.8pt] at (5.10,3.86)
    {$S_\ell^{(2)}$ outgoing};
  \fill[BranchVermillion] (3.35,2.08) circle[radius=1.7pt];
  \fill[BranchVermillion] (3.35,3.20) circle[radius=1.7pt];
\end{scope}

\node[draw=black!20,rounded corners=1.5pt,fill=black!1,
      align=center,font=\scriptsize,inner xsep=5pt,inner ysep=2.5pt]
  at (7.45,-0.55)
  {shared incoming data at infinity:
   $t,\ \omega,\ (\ell,m),\ a_{\omega\ell m}$\\[-1pt]
   transparent matching at $R_i$:
   $\psi_{\rm out}=\psi_{\rm in}$,\quad
   $\sqrt{f_i}\,\partial_r\psi_{\rm out}=\partial_r\psi_{\rm in}$};
\end{tikzpicture}
\caption{Branchwise radial matching.  Each panel is one controlled background
with flat interior $T_i=\sqrt{f_i}\,t$, shell radius $R_i>2M$, and a
Schwarzschild exterior of common ADM mass $M$.  The branches share asymptotic
time, incoming operators, and unit-incoming normalization, while
$S_\ell^{(i)}=-e^{2\ii\delta^{(i)}_{\omega\ell}}$ remains branch dependent.
Arrows indicate the ingoing and outgoing parts of the schematic radial
scattering construction. Blue and vermillion dots mark crossings of the
respective shell surfaces; they do not represent additional detectors or
interactions. The connection at $r=0$ expresses regularity of the radial
solution. The detector does not act as a reflecting boundary, and the drawn
slopes are schematic.}
\label{fig:shell-matching-spacetime}
\end{figure*}
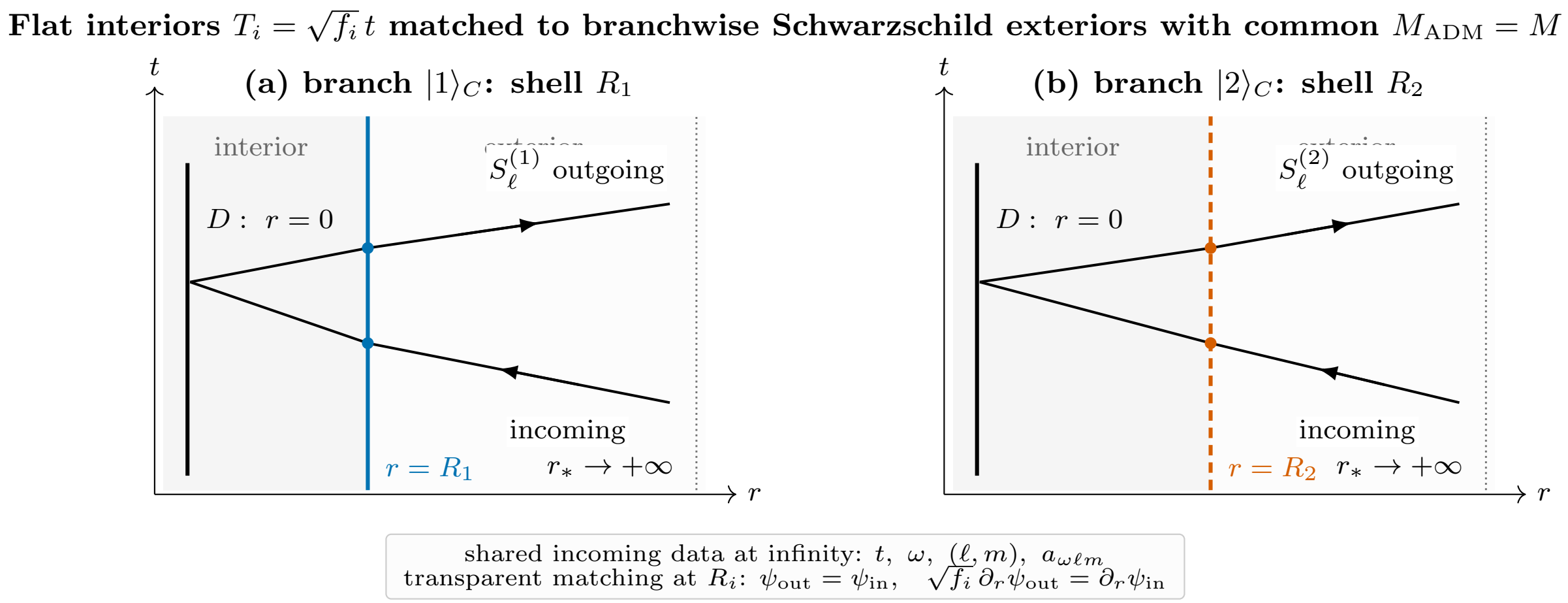

The branch clock, transparent-shell junction, and common-incoming
normalization conventions are collected in
Fig.~\ref{fig:shell-matching-spacetime}.  Each panel is a separate controlled
background.  This is important: the common labels at infinity identify the
mode algebra, whereas the branch-dependent clock and matching data determine
the interior field sampled by the detector.

\subsection{Field-algebra identification}
\label{sec:field-algebra-identification}

We now state the cross-branch model explicitly.  The common incoming modes
define one set of operators satisfying
\begin{equation}
  [a_{\omega\ell m},a^\dagger_{\omega'\ell'm'}]
  =\delta(\omega-\omega')\delta_{\ell\ell'}\delta_{mm'}.
  \label{eq:common-commutator}
\end{equation}
For either branch,
\begin{equation}
  \Phi_i(x)=\sum_{\ell,m}\int_0^\infty\dd\omega
  \left[a_{\omega\ell m}v^{(i)}_{\omega\ell m}(x)
  +a^\dagger_{\omega\ell m}v^{(i)*}_{\omega\ell m}(x)\right].
  \label{eq:field-expansion}
\end{equation}
The static incoming vacuum is defined by
$a_{\omega\ell m}|0\rangle=0$.  It is the positive-frequency ground state
associated with the shared asymptotic Killing time $t$; ``incoming'' fixes the
scattering normalization and does not denote an incoming particle flux.
Branchwise and mixed two-point functions then have the unambiguous mode
representation
\begin{align}
  W_{ij}(x,x')
  &=\langle0|\Phi_i(x)\Phi_j(x')|0\rangle\notag\\
  &=\sum_{\ell,m}\int_0^\infty\dd\omega\,
  v^{(i)}_{\omega\ell m}(x)v^{(j)*}_{\omega\ell m}(x').
  \label{eq:mixed-wightman}
\end{align}
For each branch and angular channel, the transparent junction selects a
self-adjoint, nonnegative Friedrichs radial operator.  The spectral theorem
proved in Appendix~\ref{app:radial-completeness} shows that its spectrum is
purely absolutely continuous on $[0,\infty)$, with multiplicity one and no
bound-state, normalizable zero-mode, or singular-continuous sector.  The
normalized scattering modes therefore give the full radial resolution of the
identity, with no omitted discrete sum.  This theorem assumes minimal coupling,
the regular center, and the transparent interface.

This identification is natural here because the asymptotic mass, time, mode
labels, and incoming normalization agree.  Operationally, it is singled out by
an input preparation in which the same asymptotic incident wave packet is
prepared and the control selects which static shell scatters it.  The
calculation then uses the vacuum of those common input operators.  The
common asymptotic phase reference fixes the field preparation and complementary
control basis.  It does not identify the branch detector clocks or remove their
relative free evolution to the readout event.  A simultaneous unitary change of
incoming basis in both branches
leaves $W_{ij}$ and every Gram entry invariant.  A branch-dependent isometry,
by contrast, changes the off-diagonal kernel and defines a different
controlled-field model even though the diagonal responses may be unchanged.

The same distinction applies to phase conventions. For the field state 
 $|\eta_i\rangle$ in the interacting picture (defined below), a change of phase
$|\eta_i\rangle\mapsto e^{\ii\gamma_i}|\eta_i\rangle$, 
accompanied by an inverse rephasing of
the corresponding control amplitude leaves the joint state and the physical
fringe combination later appearing as
$\kappa_{\rm r}\F_{12}^{\rm r}$ in
Eq.~\eqref{eq:joint-response} invariant.  Holding the control phase fixed while
rephasing only one branch instead changes the preparation.  The unit-incoming
condition in Eq.~\eqref{eq:incoming-mode} and the asymptotic field reference
fix the convention used below.  The detector-clock phase is instead fixed by
Eq.~\eqref{eq:readout-proper-times} and the declared readout schedule.  This
remains a phenomenological identification
of the branch field algebras; we do not assume that an arbitrary pair of
spacetimes admits it.
For the displaced-center extension, Eq.~\eqref{eq:field-expansion} is first
unitarily rewritten in the common incoming direction basis; branch-centered
$(\ell,m)$ labels are never identified directly.

\section{Controlled detector and postselection}
\label{sec:detector}

We now connect the branchwise field construction to the measurement protocol.
The kinematic space is
$\Hilbert_C\otimes\Hilbert_\Phi\otimes\Hilbert_D$, ordered as control,
common incoming field algebra, and detector.

The geometry control has orthonormal branch states $|1\rangle_C$ and
$|2\rangle_C$.  Its initial state is
\begin{equation}
  |C\rangle=\alpha_C|1\rangle_C+
  \beta_C e^{\ii\phi}|2\rangle_C,
  \qquad |\alpha_C|^2+|\beta_C|^2=1.
  \label{eq:control-state}
\end{equation}
The phase $\phi$ is the preparation phase.  We take
$H_D=\Omega|e\rangle\langle e|$, so that the ground state energy defines the zero of
the detector model and $\Omega>0$ is its free gap referred to branchwise proper
time.  The detector begins in its ground state $|g\rangle$, and the field
in the common incoming vacuum. In the concentric centered benchmark the detector
is held at the common shell center; in the displaced-center protocols
of Sec.~\ref{sec:relational-center}, it is held at one fixed laboratory
position inside both alternative shells.  The complete initial state is therefore
\begin{equation}
  |\Psi_0\rangle=|C\rangle\otimes|0\rangle\otimes|g\rangle.
  \label{eq:initial-joint-state}
\end{equation}
Absorbing the monopole transition matrix element into a dimensionless
coupling $\lambda$, the branch-$i$ proper-time interaction generator is
\begin{equation}
  H_{I,i}^{(\tau_i)}(\tau_i)
  =\lambda\chi(\tau_i)\mu(\tau_i)\Phi_i[x_i(\tau_i)],
  \label{eq:interaction}
\end{equation}
where $\chi(\tau_i)$ is the switching function along each branch and
\begin{equation}
\mu(\tau)=\sigma^+e^{\ii\Omega\tau}
  +\sigma^-e^{-\ii\Omega\tau}
  \label{eq:monopole}
\end{equation}
is    the detector monopole operator; 
$\sigma^+=|e\rangle\langle g|$ and
$\sigma^-=|g\rangle\langle e|$ are its raising and lowering operators,
and $\tau$ is detector proper time.
Point evaluation in Eq.~\eqref{eq:interaction} does not by itself define
higher-order time-ordered products.  For full detector outputs we use the
common incoming-mode regulator $\mathfrak r=(\kappa,\Lambda,J)$ specified and
controlled in Appendix~\ref{app:uv-prescription}: a finite positive-frequency
band and angular cutoff in one shared incoming frame.  All statements about
full $U_i$, $|E_i\rangle$, probabilities, and higher-order output expansions
are at fixed $\mathfrak r$.  In such statements a suppressed regulator index
applies to all associated states and Gram data as well; we write
$\order_{\mathfrak r}$ to expose the remainder's dependence.  Unadorned
first-order vectors and coefficients denote their finite pointlike limits.
These conventions never identify a finite-cutoff coefficient with its
pointlike value inside a remainder estimate.  The weak-coupling expansion
is taken first and the regulator is removed only from the proven finite
coefficients, as in Eq.~\eqref{eq:coefficient-first-point-limit}; no full
point-detector propagator or cutoff-uniform error is assumed.

Equal branch proper times are not identified, so the controlled interaction is
ordered instead in the shared asymptotic time.  With
$\tau_i=\sqrt{f_i}\,t$, $x_i(t):=x_i(\sqrt{f_i}t)$, and
$\mu_i^{(t)}(t):=\mu(\sqrt{f_i}t)$, the standard quantum-reference-frame
transformation of detector-field couplings
\cite{GiacominiKempf2022,WalleghemCepollaro2026} gives
\begin{multline}
  H_I^{(t)}(t)=\lambda\sum_{i=1}^2\ketbra{i}{i}_C\,
  \sqrt{f_i}\,\chi(\sqrt{f_i}t)\\
  {}\times\mu_i^{(t)}(t)\Phi_i[x_i(t)].
  \label{eq:asymptotic-time-interaction}
\end{multline}
The factor $\sqrt{f_i}=\dd\tau_i/\dd t$ is the lapse Jacobian.  Projector
orthogonality gives, at fixed $\mathfrak r$, the block unitary
$U_I=\mathcal T_t\exp[-\ii\int\dd t\,H_I^{(t)}]=
\sum_i\ketbra{i}{i}_C U_i$; monotonicity of $t\mapsto\tau_i$ makes its ordering
equivalent to proper-time ordering within each block.  We do not remove the
detector's branch-dependent free phase.  Instead, we propagate both branches to
the common event $t=t_{\rm r}$ in Eq.~\eqref{eq:readout-proper-times}.  Any
separately specified control Hamiltonian is retained in the readout amplitudes
$c_i(t_{\rm r})$ below.
We use an exactly compact switching with a nonnegative Fourier transform
and $A_\chi>0$. Let $b_a$ be the normalized rectangular pulse of half-width
$a=\sigma/2$. We construct $\chi_c$ by convolving twelve copies of this pulse:
\begin{equation}
 \begin{aligned}
 b_a(\tau)&=\frac{\mathbf 1_{[-a,a]}(\tau)}{2a},
 &a&=\frac{\sigma}{2},\\
 \chi_c(\tau)&=A_\chi\sqrt{2\pi}\sigma\,b_a^{*12}(\tau).
 \end{aligned}
 \label{eq:compact-switching}
\end{equation}
The indicator $\mathbf 1_{[-a,a]}(\tau)$ equals one for $|\tau|\leq a$
and zero otherwise. The notation $b_a^{*12}:=b_a*\cdots*b_a$ denotes
twelvefold convolution, where
\[
 (f*g)(\tau):=\int_{-\infty}^{\infty}f(u)g(\tau-u)\,\dd u.
\]
This order-12 centered B-spline is real, even, nonnegative, globally
$C^{10}$, and supported on $[-T_\chi,T_\chi]$ with
$T_\chi=12a=6\sigma$.  Its normalization matches both the pulse area and
temporal variance of $A_\chi e^{-\tau^2/(2\sigma^2)}$:
\begin{equation}
 \int\chi_c\,\dd\tau=A_\chi\sqrt{2\pi}\sigma,\qquad
 \frac{\int\tau^2\chi_c(\tau)\dd\tau}{\int\chi_c(\tau)\dd\tau}
 =\sigma^2.
 \label{eq:compact-switching-moments}
\end{equation}
With $\sinc x:=\sin x/x$, its Fourier transform is the closed expression
\begin{equation}
 \widehat\chi_c(\nu)
 =A_\chi\sqrt{2\pi}\sigma
 \sinc^{12}\!\left(\frac{\sigma\nu}{2}\right)\geq0 .
 \label{eq:compact-switching-transform}
\end{equation}
The even exponent twelve makes $\widehat\chi_c$ nonnegative:
its isolated zeros do not introduce $\pi$ phase jumps, so the switching
contributes no spectral phase.

Writing $\chi_c=A_\chi\chi_0$, where $\chi_0$ is the pulse with unit amplitude, the interaction in Eq.~\eqref{eq:interaction} then becomes
\[
 H_{I,i}^{(\tau_i)}(\tau_i)
 =(\lambda A_\chi)\chi_0(\tau_i)\mu(\tau_i)\Phi_i[x_i(\tau_i)].
\]
Consequently the scalar parameters $\lambda$ and $A_\chi$ enter only
through their product $\lambda_{\rm eff}:=\lambda A_\chi$.
We retain $A_\chi$ in analytic formulas to display the pulse normalization,
and use $A_\chi=1$ for all numerical configurations, figures, and quoted
numerical coefficients. Restoring a common $A_\chi$ at fixed
bare $\lambda$ multiplies the leading response and attenuation coefficients
by $A_\chi^2$; normalized branch overlaps are unchanged.

At first order in the coupling, the raising part of $\mu(\tau)$ creates the completed
interaction-picture field vector
\begin{equation}
  |\eta_i(\Omega)\rangle=
  \int_{-T_\chi}^{T_\chi}\dd\tau\,
  \chi_c(\tau)e^{\ii\Omega\tau}
  \Phi_i[x_i(\tau)]|0\rangle.
  \label{eq:excitation-vector}
\end{equation}
Every physical readout is required to satisfy
$\tau_i^{\rm r}\geq T_\chi$ on every branch.
Let $H_\Phi^{\rm free}$ be the common asymptotic free-field Hamiltonian, acting
as multiplication by $\omega$ on one-particle spectral amplitudes.  The full
Schr\"odinger-picture readout vector and its pullback to a common fixed field
slice are
\begin{equation}
  \begin{aligned}
  |\zeta_i^{S}(t_{\rm r})\rangle
  &:=e^{-\ii H_\Phi^{\rm free}t_{\rm r}}
     e^{-\ii\Omega\tau_i^{\rm r}}|\eta_i\rangle,\\
  |\zeta_i^{\rm r}(t_{\rm r})\rangle
  &:=e^{+\ii H_\Phi^{\rm free}t_{\rm r}}
     |\zeta_i^{S}(t_{\rm r})\rangle
   =e^{-\ii\Omega\tau_i^{\rm r}}|\eta_i\rangle.
  \end{aligned}
  \label{eq:readout-excitation-vector}
\end{equation}
The pullback is harmless for a Gram matrix whose branches are compared at one
common event, because it is then the same unitary on every branch.  It cannot
be discarded when differentiating a curve whose readout event itself varies
with the curve parameter.
The detector-excited component at the common readout is therefore
\begin{equation}
  |\Psi_e^{(1)}(t_{\rm r})\rangle=-\ii\lambda
  \sum_{i=1}^2c_i(t_{\rm r})|i\rangle_C
  |\zeta_i^{S}(t_{\rm r})\rangle|e\rangle,
  \label{eq:first-order-state}
\end{equation}
where $c_i(t_{\rm r})$ includes any declared free control evolution and reduces
to $(\alpha_C,\beta_Ce^{\ii\phi})$ when that evolution is absent.  Both branch
excitations lie in the stipulated common field Hilbert space, and complementary
control readout probes their physical readout overlap.

In fixed-event inner products below, we abbreviate
$|\zeta_i(t_{\rm r})\rangle:=|\zeta_i^{\rm r}(t_{\rm r})\rangle$; the
corresponding inner products of the full Schr\"odinger-picture vectors are
identical because the suppressed field evolution is common.
Define the interaction and readout Gram matrices by
\begin{equation}
  \F_{ij}(\Omega)
  :=\langle\eta_i(\Omega)|\eta_j(\Omega)\rangle,
  \label{eq:gram-definition}
\end{equation}
and
\begin{align}
  \F_{ij}^{\rm r}(t_{\rm r};\Omega)
  &:=\langle\zeta_i(t_{\rm r};\Omega)|
             \zeta_j(t_{\rm r};\Omega)\rangle
  \nonumber\\
  &=e^{\ii\Omega(\tau_i^{\rm r}-\tau_j^{\rm r})}
    \F_{ij}(\Omega).
  \label{eq:readout-gram-definition}
\end{align}
Thus $\bm{\F}^{\rm r}=D^\dagger\bm{\F}D$, with
$D_{ii}=e^{-\ii\Omega\tau_i^{\rm r}}$, and the readout Gram matrix is positive.
In terms of Eq.~\eqref{eq:mixed-wightman}, this convention is explicitly
\begin{multline}
  \F_{ij}(\Omega)
  =\int_{-T_\chi}^{T_\chi}\dd\tau
   \int_{-T_\chi}^{T_\chi}\dd\tau'\,
  \chi_c(\tau)\chi_c(\tau')e^{-\ii\Omega(\tau-\tau')}\\
  {}\times W_{ij}\bigl(x_i(\tau),x_j(\tau')\bigr).
  \label{eq:wightman-response}
\end{multline}
Consequently, with
\begin{equation}
  \Delta_D(t_{\rm r})
  :=\Omega(\tau_1^{\rm r}-\tau_2^{\rm r}),
  \qquad
  \F_{12}^{\rm r}(t_{\rm r})
  =e^{\ii\Delta_D(t_{\rm r})}\F_{12},
  \label{eq:detector-clock-phase}
\end{equation}
the relation is exact and contains no phase lock or tail approximation.
For the static interior protocols considered here,
$\Delta_D(t_{\rm r})
=\Omega(\sqrt{f_1}-\sqrt{f_2})t_{\rm r}$.
It follows without further assumptions that
\begin{equation}
  \F_{ji}^{\rm r}=\F_{ij}^{\rm r*},\qquad
  \F_{ii}^{\rm r}\geq0,\qquad
  |\F_{12}^{\rm r}|^2\leq\F_{11}^{\rm r}\F_{22}^{\rm r}.
  \label{eq:gram-properties}
\end{equation}
These are physical consistency conditions and numerical regression tests.

We read out the control in the complementary basis
\begin{equation}
  |\pm_\theta\rangle=
  \frac{|1\rangle_C\pm e^{\ii\theta}|2\rangle_C}{\sqrt2},
  \qquad \Pi_\pm=\ketbra{\pm_\theta}{\pm_\theta}.
  \label{eq:readout-basis}
\end{equation}
For compactness let
\begin{align}
  L_{\rm r}&=|c_1(t_{\rm r})|^2\F_{11}^{\rm r}
  +|c_2(t_{\rm r})|^2\F_{22}^{\rm r},
  \label{eq:L-mixture}\\
  \kappa_{\rm r}&=c_1(t_{\rm r})^*c_2(t_{\rm r})e^{-\ii\theta}.
  \label{eq:kappa}
\end{align}
The order-$\lambda^2$ coefficient of the joint probability for a
detector excitation and control result $\pm$ is
\begin{align}
  Q_\pm(t_{\rm r})&=\frac12\left[
  L_{\rm r}\pm2\operatorname{Re}
  \!\left(\kappa_{\rm r}\F_{12}^{\rm r}(t_{\rm r})\right)
  \right],
  \nonumber\\
  P_{\mathfrak r}(e,\pm;t_{\rm r})
  &=\lambda^2Q_{\pm,\mathfrak r}(t_{\rm r})
    +\order_{\mathfrak r}(\lambda^4).
  \label{eq:joint-response}
\end{align}
Here $Q_{\pm,\mathfrak r}$ uses the regulated Gram entries in the first
line; $Q_\pm=\lim_{\mathfrak r\to\mathrm{pt}}Q_{\pm,\mathfrak r}$.
The corresponding zeroth-order postselection probability is
\begin{equation}
  P_\pm^{(0)}(t_{\rm r})=\frac12\left[
  1\pm2\operatorname{Re}\kappa_{\rm r}
  \right].
  \label{eq:postselection-probability}
\end{equation}
When $P_\pm^{(0)}>0$, the normalized conditional detector response is
\begin{align}
  \F_\pm^{\mathrm{cond}}(t_{\rm r})&=
  \frac{Q_\pm(t_{\rm r})}{P_\pm^{(0)}(t_{\rm r})},
  \nonumber\\
  P_{\mathfrak r}(e|\pm;t_{\rm r})
  &=\lambda^2\F_{\pm,\mathfrak r}^{\mathrm{cond}}(t_{\rm r})
  +\order_{\mathfrak r}(\lambda^4).
  \label{eq:conditional-response}
\end{align}
Near a dark port, where $P_\pm^{(0)}$ is comparable to neglected
perturbative corrections or the statistical resolution, we report $Q_\pm$
rather than dividing by a small number.

If the control is ignored, or is dephased in the branch basis before the
interaction, the response is the mixture
\begin{equation}
  \F_{\mathrm{mix}}(t_{\rm r})=L_{\rm r}.
  \label{eq:mixture-response}
\end{equation}
Completeness of the two readout projectors gives the exact leading-order
identities
\begin{align}
  Q_+(t_{\rm r})+Q_-(t_{\rm r})
  &=\F_{\mathrm{mix}}(t_{\rm r}),
  \nonumber\\
  \sum_{s=\pm}P_s^{(0)}(t_{\rm r})\F_s^{\mathrm{cond}}(t_{\rm r})
  &=\F_{\mathrm{mix}}(t_{\rm r}).
  \label{eq:control-trace}
\end{align}
The second identity is written only where the conditional responses are
defined; the joint identity remains valid even at a dark port.  In
particular, tracing the control recovers the mixture, while identical
branches give the one-branch conditional response rather than twice that
response.  Appendix~\ref{app:positivity} derives these consistency limits
and the positivity bounds used below.

Gram positivity bounds every phase scan.  Appendix~\ref{app:positivity} gives
the visibility and geometry-excess limits, including why unit port visibility
need not imply branch distinguishability.

\section{Centered-detector response}
\label{sec:center}

We first evaluate the response at the common center.  This case is both the
cleanest physical baseline and the strongest normalization check, because
regularity removes every multipole except the $s$ wave.

On branch $i$ the centered worldline is
\begin{equation}
  x_i(\tau)=\bigl(t=\tau/\sqrt{f_i},r=0\bigr).
  \label{eq:center-worldline}
\end{equation}
Because $j_0(0)=1$ and $j_\ell(0)=0$ for every $\ell>0$, only the
$(\ell,m)=(0,0)$ mode contributes.  The response is therefore not an
approximate angular-momentum truncation.  The interaction-picture, full
Schr\"odinger-picture, and common-slice-pulled-back post-switch amplitudes are
\begin{align}
  q_i^I(\omega;\Omega)
  &=\frac{\mathcal A^{(i)*}_{\omega0}}{4\pi\sqrt{\omega}}
  \widehat\chi_c
  \!\left(\Omega+\frac{\omega}{\sqrt{f_i}}\right),
  \nonumber\\
  q_i^{S}(\omega;t_{\rm r},\Omega)
  &=e^{-\ii\omega t_{\rm r}}
    e^{-\ii\Omega\tau_i^{\rm r}}q_i^I(\omega;\Omega),
  \nonumber\\
  q_i^{\rm r}(\omega;t_{\rm r},\Omega)
  &:=e^{+\ii\omega t_{\rm r}}q_i^S(\omega;t_{\rm r},\Omega)
   =e^{-\ii\Omega\tau_i^{\rm r}}q_i^I(\omega;\Omega).
  \label{eq:finite-readout-amplitude}
\end{align}
At one common $t_{\rm r}$ the field factor cancels from every Gram entry, so
the equal-event response calculation may use $q_i^{\rm r}$.  Across unequal
readout events it is part of the physical state and must be retained.  We write
$q_i:=q_i^I$.  The interaction and equal-event readout Gram entries are therefore
\begin{align}
  \F_{ij}(\Omega)&=\int_0^\infty\dd\omega\,q_i^*q_j.
  \nonumber\\
  \F_{ij}^{\rm r}(t_{\rm r};\Omega)
  &=e^{\ii\Omega(\tau_i^{\rm r}-\tau_j^{\rm r})}\F_{ij}(\Omega).
  \label{eq:center-gram-integral}
\end{align}
The compact Fourier factors are real and nonnegative, whereas the cross
response can remain complex because the two shell radii produce different
scattering phases.

In the flat limit, $f_i\to1$ and $\mathcal A_{\omega0}\to2\omega$,
so Eq.~\eqref{eq:center-gram-integral} gives the norm density
$\omega|\widehat\chi_c(\Omega+\omega)|^2/(4\pi^2)$.
Equation~\eqref{eq:compact-switching-transform} and
$|\sinc x|\leq\min(1,|x|^{-1})$ imply
$\widehat\chi_c(\nu)=O(|\nu|^{-12})$.
At fixed $\Omega$ and $\sigma$, the ultraviolet envelope is therefore
\[
 \omega|\widehat\chi_c(\Omega+\omega)|^2
 =\omega\,O(\omega^{-24})=O(\omega^{-23}).
\]
The flat-space first-order response vector is consequently ultraviolet
normalizable.  This statement does not extend to higher-order point-detector
Dyson vectors; Appendix~\ref{app:uv-prescription} gives the distinct
regulated construction and an explicit second-order UV obstruction.
Appendix~\ref{app:uv-normalizability} proves first-order normalizability, including the curved
matched modes and translated response, and explains why a Gaussian truncated
at a finite nonzero tail would instead give a logarithmically divergent norm.

\subsection{Branch overlap and probe-induced control attenuation}
\label{sec:control-attenuation}

For nonzero diagonal response norms, define
\begin{equation}
 N_i:=\F_{ii},\qquad
 \mu_{12}:=\frac{\F_{12}}{\sqrt{\F_{11}\F_{22}}},\qquad
 D_{12}:=-\ln|\mu_{12}|^2 .
 \label{eq:renyi-half-quantum}
\end{equation}
Writing $|\widehat\eta_i\rangle:=|\eta_i\rangle/\sqrt{N_i}$ and
$\rho_i:=|\widehat\eta_i\rangle\langle\widehat\eta_i|$, we have
$\mu_{12}=\langle\widehat\eta_1|\widehat\eta_2\rangle$.
Gram positivity gives $0\leq|\mu_{12}|\leq1$.
The squared pure-state trace distance is $1-|\mu_{12}|^2$, while their
sandwiched R\'enyi-$1/2$ divergence is
\[
 D_{12}=\widetilde D_{1/2}(\rho_1\Vert\rho_2)=-\ln|\mu_{12}|^2.
\]
This divergence vanishes for identical rays, is finite for nonorthogonal
states, and diverges for orthogonal states.
The overlap magnitude and the resulting distances are unchanged by the scalar
detector phase at a common readout. A coherent control observable retains that phase.

At fixed $\mathfrak r$, tracing the detector and field gives
$[\rho_{C,\mathfrak r}^{\rm out}(t_{\rm r})]_{12}
=[\rho_C^{\rm in}]_{12}\mathcal I_{12,\mathfrak r}(t_{\rm r})$, where $|E_{i,\mathfrak r}\rangle$ is the normalized branch detector-field
output at the common readout and
$\mathcal I_{12,\mathfrak r}=\langle E_{2,\mathfrak r}|E_{1,\mathfrak r}\rangle$.
Branchwise unitarity gives the expansion below; convergence of the
first-order Gram matrix then gives its unadorned coefficient limit:
\begin{align}
  |\mathcal I_{12,\mathfrak r}(t_{\rm r})|
  &=1-\lambda^2\Gamma_{12,\mathfrak r}^{(2)}(t_{\rm r})
    +\order_{\mathfrak r}(\lambda^4),
  \nonumber\\
  \Gamma_{12}^{(2)}(t_{\rm r})
  &=\frac12\left(
    \F_{11}^{\rm r}+\F_{22}^{\rm r}
    -2\operatorname{Re}\F_{12}^{\rm r}\right)
  \nonumber\\
  &=\frac12\|\zeta_1(t_{\rm r})
  -\zeta_2(t_{\rm r})\|^2\geq0 .
  \label{eq:leading-control-attenuation}
\end{align}
For every allowed post-switch readout this is exactly
\begin{equation}
  \Gamma_{12}^{(2)}(t_{\rm r})
  =\frac12(\F_{11}+\F_{22})
  -\operatorname{Re}\!\left[
  e^{\ii\Delta_D(t_{\rm r})}\F_{12}\right].
  \label{eq:post-switch-clock-attenuation}
\end{equation}
For completeness, take $|G\rangle:=|g\rangle|0\rangle$ and apply the
norm-convergent regulated Dyson series, followed by the known free
propagation to the common readout:
\begin{align}
  |E_{i,\mathfrak r}\rangle
  &=|G\rangle+\lambda|A_{i,\mathfrak r}\rangle
     +\lambda^2|C_{i,\mathfrak r}\rangle
     +\order_{\mathfrak r}(\lambda^3),
  \nonumber\\
  |A_{i,\mathfrak r}\rangle
  &=-\ii|e\rangle|\zeta_{i,\mathfrak r}(t_{\rm r})\rangle,
  \qquad
  |C_{i,\mathfrak r}\rangle:=U_{i,\mathfrak r}^{(2)}|G\rangle .
  \label{eq:dyson-environment-expansion}
\end{align}
The coefficients include that fixed readout propagation.  Detector parity
gives $\langle G|A_{i,\mathfrak r}\rangle=0$, while unitarity gives
\begin{equation}
  2\operatorname{Re}\langle G|C_{i,\mathfrak r}\rangle
  =-\langle A_{i,\mathfrak r}|A_{i,\mathfrak r}\rangle=-N_{i,\mathfrak r}.
  \label{eq:dyson-unitarity-identity}
\end{equation}
Consequently,
\begin{align}
  \mathcal I_{12,\mathfrak r}
  &=1+\lambda^2\left(
    \langle C_{2,\mathfrak r}|G\rangle
    +\langle G|C_{1,\mathfrak r}\rangle\right.
    \nonumber\\
  &\hspace{27mm}\left.
    +\langle A_{2,\mathfrak r}|A_{1,\mathfrak r}\rangle\right)
    +\order_{\mathfrak r}(\lambda^4),
  \nonumber\\
  |\mathcal I_{12,\mathfrak r}|
  &=1-\frac{\lambda^2}{2}\left(
    N_{1,\mathfrak r}+N_{2,\mathfrak r}
    -2\operatorname{Re}\F_{12,\mathfrak r}^{\rm r}\right)
    \nonumber\\
  &\quad+\order_{\mathfrak r}(\lambda^4).
  \label{eq:dyson-overlap-derivation}
\end{align}
The real part of $\langle A_{2,\mathfrak r}|A_{1,\mathfrak r}\rangle$
is $\operatorname{Re}\F_{12,\mathfrak r}^{\rm r}$.
Even/odd detector parity of the exact regulated branch states excludes
odd overlap powers.  The imaginary vacuum coefficients
$\operatorname{Im}\langle G|C_{i,\mathfrak r}\rangle$ can depend on both
branch and cutoff.  They affect $\arg\mathcal I_{12,\mathfrak r}$ but not
the modulus coefficient through order $\lambda^2$; they are not discarded
or absorbed into the prepared control phase.  Equation~\eqref{eq:first-order-regulator-limit}
then proves $\Gamma_{12,\mathfrak r}^{(2)}\to\Gamma_{12}^{(2)}\geq0$,
without asserting a limit for the full higher-order state or its remainder.

For the centered response, $q_i^r$ is the common-readout spectral amplitude defined in Eq. (47): $q_i^r(\omega;t_r,\Omega) = e^{-i\Omega\tau_i^r}q_i^I(\omega;\Omega)$. Here $i = 1,2$ labels the shell branches and the superscript r denotes readout, it is distinct from the regulator $\mathcal{r}$. Suppressing the fixed gap $\Omega$ in its arguments, we evaluate the stable positive integral
\begin{equation}
  \Gamma_{12}^{(2)}(t_{\rm r})
  =\frac12\int_0^\infty\dd\omega\,
    |q_1^{\rm r}(\omega;t_{\rm r})
     -q_2^{\rm r}(\omega;t_{\rm r})|^2 ,
  \label{eq:attenuation-spectral-integral}
\end{equation}
which is checked against Eq.~\eqref{eq:leading-control-attenuation} without
subtracting nearly equal Gram entries.  After compact-pulse switch-off its
exact detector-clock-phase envelope is
\begin{align}
  \Gamma_{\min}^{(2)}
  &=\frac12(\F_{11}+\F_{22})-|\F_{12}|,
  \nonumber\\
  \Gamma_{\max}^{(2)}
  &=\frac12(\F_{11}+\F_{22})+|\F_{12}|,
  \nonumber\\
  \overline{\Gamma}_{\rm det\,scan}^{(2)}
  &=\frac12(\F_{11}+\F_{22}).
  \label{eq:attenuation-phase-envelope}
\end{align}
These are the lower and upper values over a deterministic readout-phase
scan and its uniform arithmetic mean. They retain one declared clock protocol;
no phase correction is applied to the physical observable. Each coefficient
is accumulated over one switching window and is not a temporal decay rate.

\subsection{Radius-pair convention}

At fixed ADM mass, the combined paths use a mean radius $R$, nominal
compactness $\mathcal C:=2M/R$, and a symmetric separation $\delta R$:
\begin{equation}
 R_\pm=R\pm\frac{\delta R}{2},\qquad R_\pm>2M.
 \label{eq:nearby-radii}
\end{equation}
We assign $R_1=R_-$ and $R_2=R_+$, so that
$R=(R_1+R_2)/2$ and $\delta R=R_2-R_1>0$.
The comparisons below use finite $\delta R$; no coincident-radius
limit is taken.
The separation schedule and common post-switch readout are specified in
Sec.~\ref{sec:compactness-numerics}. No local radius derivative or horizon
state is required for these finite-pair comparisons.

\section{Lab-fixed detector and relational shell centers}
\label{sec:relational-center}

For  branches in a displaced superposition, we fix one asymptotic Cartesian frame, shell centers
$\bm c_i$, and a detector at one laboratory position $\bm x$, so
$\bm\rho_i=\bm x-\bm c_i$.  Partial waves about different origins are not
identified.  Instead, the common incoming data are translated into a direction
basis with amplitudes $q_i$ defined in Appendix~\ref{app:direction-basis}; the
response is the positive Gram integral
\begin{equation}
  \F_{ij}(\bm x;\Omega)=\int_0^\infty\dd\omega
  \int\dd\Omega_{\hat{\bm n}}\,
  q_i(\omega,\hat{\bm n};\bm x)^*
  q_j(\omega,\hat{\bm n};\bm x).
  \label{eq:relational-gram}
\end{equation}
At a finite common post-switch readout, its amplitudes acquire the same
detector phase factors as Eq.~\eqref{eq:finite-readout-amplitude}; hence
$\F_{12}^{\rm r}=e^{\ii\Delta_D}\F_{12}$ exactly after switch-off, while the
diagonal entries and $|\mu_{12}|$ are unchanged.
It is invariant under a common translation and reduces continuously to the
centered and Minkowski limits.  Diagonal entries encode branch-relative
distances, whereas coherence enters $\F_{12}$.  Appendix~\ref{app:direction-basis}
gives the distance observable, convention-dependent translation phase,
near-center expansion, collinear quadrature, and benchmark.  As proved there,
each diagonal response is $C^4$ through its branch center and admits the even
finite-order expansion in Eq.~\eqref{eq:near-center-local-response}; no
all-orders spatial analyticity is asserted.  Its finite width is relational
rather than a regularization of a central singularity.

\subsection{Combined radius-translation paths}
\label{sec:combined-radius-translation}

To vary radius and center together, take the nearby radii in
Eq.~\eqref{eq:nearby-radii} and, along a common Cartesian axis, define
\begin{equation}
  c_\pm=c_0\pm\frac b2,
  \qquad x=c_0+s .
  \label{eq:combined-center-parameterization}
\end{equation}
The setup is depicted in the right panel of Fig.~\ref{fig:shell-branch-geometry}.
The two assignments are
\begin{align}
  \mathrm{SL}:&(R_1,c_1;R_2,c_2)
    =(R_-,c_-;R_+,c_+),
  \nonumber\\
  \mathrm{LS}:&(R_1,c_1;R_2,c_2)
    =(R_-,c_+;R_+,c_-).
  \label{eq:radius-center-assignments}
\end{align}
Thus SL places the smaller-radius alternative to the left and LS places it to
the right.  These remain pairs of alternative single-shell geometries; they
are not nested shells.  For $o\in\{\mathrm{SL},\mathrm{LS}\}$, let
$\F_{ij}^{(o)}(s)$ be Eq.~\eqref{eq:relational-gram} evaluated with the
corresponding assignment and define
\begin{align}
  \mu_o(s)&:=\frac{\F_{12}^{(o)}(s)}
  {\sqrt{\F_{11}^{(o)}(s)\F_{22}^{(o)}(s)}},
  \nonumber\\
  D_o(s)&:=-\ln|\mu_o(s)|^2,
  \label{eq:assignment-renyi}\\
  D_{\rm odd}(s)&:=\frac{D_{\rm SL}(s)-D_{\rm LS}(s)}2,
  \nonumber\\
  \mathcal A_D(s)&:=
  \frac{D_{\rm SL}(s)-D_{\rm LS}(s)}
       {D_{\rm SL}(s)+D_{\rm LS}(s)} .
  \label{eq:assignment-renyi-contrast}
\end{align}
Here $D_o$ is the finite sandwiched R\'enyi-$1/2$ divergence in
Eq.~\eqref{eq:renyi-half-quantum}; $\mathcal A_D$ is reported only when its
denominator and numerical error are resolved.

The corresponding clock-resolved single-switching probe-influence coefficient
is evaluated without Gram-entry subtraction.  At fixed $(s,t_{\rm r})$,
abbreviate $\Gamma_o^{(2)}:=\Gamma_o^{(2)}(s;t_{\rm r})$ after its first line:
\begin{align}
  \Gamma_o^{(2)}(s;t_{\rm r})
  &:=\frac12\int_0^\infty\dd\omega
  \int\dd\Omega_{\hat{\bm n}}\,
  |q_{1,o}^{\rm r}(t_{\rm r})-q_{2,o}^{\rm r}(t_{\rm r})|^2,
  \label{eq:assignment-influence-direct}\\
  \Gamma_{\rm odd}^{(2)}(t_{\rm r})
  &:=\frac{\Gamma_{\rm SL}^{(2)}-\Gamma_{\rm LS}^{(2)}}2,
  \nonumber\\
  \mathcal A_\Gamma(t_{\rm r})
  &:=\frac{\Gamma_{\rm SL}^{(2)}-\Gamma_{\rm LS}^{(2)}}
       {\Gamma_{\rm SL}^{(2)}+\Gamma_{\rm LS}^{(2)}} .
  \label{eq:assignment-influence-contrast}
\end{align}
After compact-pulse switch-off it equals exactly
\begin{align}
  \Gamma_o^{(2)}(s;t_{\rm r})
  &=\frac12\left(\F_{11}^{(o)}+\F_{22}^{(o)}\right)
  \nonumber\\
  &\quad-\operatorname{Re}\!\left(
  e^{\ii\Delta_D(t_{\rm r})}\F_{12}^{(o)}\right)
  \label{eq:assignment-clock-influence}
\end{align}
and has the positive split
\begin{align}
  \Gamma_o^{(2)}(t_{\rm r})
  &=
  \frac12\left(\sqrt{\F_{11}^{(o)}}-\sqrt{\F_{22}^{(o)}}\right)^2
  \nonumber\\
  &\quad+\sqrt{\F_{11}^{(o)}\F_{22}^{(o)}}
  \left[1-\operatorname{Re}\!\left(
  e^{\ii\Delta_D(t_{\rm r})}\mu_o\right)\right].
  \label{eq:assignment-influence-split}
\end{align}
The first term is yield-amplitude imbalance and the second is the
clock-resolved normalized-vector mismatch.  Accordingly,
$\mathcal A_\Gamma(t_{\rm r})$ and $\mathcal A_D$ answer different questions;
attenuation at fixed $\mathfrak r$ is
$\lambda^2\Gamma_{o,\mathfrak r}^{(2)}(t_{\rm r})
+\order_{\mathfrak r}(\lambda^4)$ for one switching window and a specified
readout.  The plotted $\Gamma_o^{(2)}$ is its coefficient limit.  A phase envelope varies the one common
$\Delta_D$ shared by SL and LS; minimizing the two assignments independently
would not describe one clock protocol.

For completeness, the shared-phase envelope of the contrast itself can
also be obtained without a phase scan.  At the declared post-switch readout put
\begin{align}
  a_o^{\rm r}&:=\frac12\left(
    \F_{11}^{\rm r,(o)}+\F_{22}^{\rm r,(o)}\right),
  \nonumber\\
  \mathcal Z_o^I&:=e^{-\ii\Delta_D(t_{\rm r})}
                    \F_{12}^{\rm r,(o)}(t_{\rm r}),
  \nonumber\\
  \bm h_o&:=\left(-\operatorname{Re}\mathcal Z_o^I,
                    \operatorname{Im}\mathcal Z_o^I\right),
  \nonumber\\
  \bm u(\varphi)&:=(\cos\varphi,\sin\varphi),
  \nonumber\\
  \Gamma_o^{(2)}(\varphi;t_{\rm r})
    &:=a_o^{\rm r}+\bm h_o\mathbin{\cdot}\bm u(\varphi),
  \label{eq:assignment-common-phase-form}
\end{align}
where $\varphi=\Delta_D(t_{\rm r})$ gives the physical readout and a scan holds
the compact-pulse interaction Gram entries fixed.  The demodulation defining
$\mathcal Z_o^I$ exposes, but does not alter, the interaction entry; it is not
applied to the physical observable.  With
\begin{align}
  n_0&:=a_{\rm SL}^{\rm r}-a_{\rm LS}^{\rm r},
  &\bm n&:=\bm h_{\rm SL}-\bm h_{\rm LS},
  \nonumber\\
  d_0&:=a_{\rm SL}^{\rm r}+a_{\rm LS}^{\rm r},
  &\bm d&:=\bm h_{\rm SL}+\bm h_{\rm LS},
  \label{eq:assignment-common-phase-coefficients}
\end{align}
one has
$\mathcal A_\Gamma(\varphi)=(n_0+\bm n\cdot\bm u)/
(d_0+\bm d\cdot\bm u)$.  Provided the physical denominator is strictly
positive for every common phase, its two extrema are the real roots $r_\pm$
of
\begin{align}
  0={}&\left(d_0^2-\lVert\bm d\rVert^2\right)r^2
  \nonumber\\
  &+2\left(\bm n\mathbin{\cdot}\bm d-n_0d_0\right)r
  +n_0^2-\lVert\bm n\rVert^2,
  \label{eq:assignment-common-phase-extrema}
  \\
  \mathcal A_{\Gamma,\min}&=\min(r_-,r_+),
  \nonumber\\
  \mathcal A_{\Gamma,\max}&=\max(r_-,r_+).
  \nonumber
\end{align}
This follows by requiring
$(n_0-rd_0)+(\bm n-r\bm d)\cdot\bm u=0$ to be tangent to the unit circle.
It is the exact common-phase band used below and is distinct from the pair of
generally unattainable bounds obtained by extremizing SL and LS separately.

Reflection about $c_0$ gives
\begin{align*}
  D_{\rm SL}(s)&=D_{\rm LS}(-s),
  \\
  \Gamma_{\rm SL}^{(2)}(s;t_{\rm r})
  &=\Gamma_{\rm LS}^{(2)}(-s;t_{\rm r})
  \quad\text{(common readout)}.
\end{align*}
Both assignment-odd terms therefore vanish exactly at $s=0$.  The finite scan
below uses $s=b/4$ and finite $b/M$, so its reported contrast is a finite-path
observable.

\section{Numerical implementation and validation}
\label{sec:numerics}

Regular interior modes are matched to the common-incoming exterior convention
and integrated with adaptive error control.  Appendix~\ref{app:numerics} gives
the benchmark parameters, grids, tolerances, fitting domains, validation comparisons, and explicitly declared acceptance gates.  The response plots report leading coefficients or their ratios, not finite-coupling detector evolution.  A small
$\lambda^2\F_{ii,\mathfrak r}$ is a weak-click condition, but does not alone
bound the higher-order remainder.  Finite-coupling probability claims
additionally require the fixed-regulator remainder to be negligible at the
chosen $\lambda$; no such numerical remainder bound is supplied by these
first-order quadrature checks.

\subsection{Readout and radius-translation scan}
\label{sec:compactness-numerics}

The pulse ends at $T_\chi=6\sigma$. For each pair the physical common readout is
\begin{equation}
 t_{\rm pair}(R_-,R_+;n_{\rm r})
 :=\frac{n_{\rm r}\sigma}{\min_{i\in\{-,+\}}\sqrt{f_i}},
 \qquad n_{\rm r}=6.
 \label{eq:common-readout-schedule}
\end{equation}
The same formula applies to the centered $(R_1,R_2)$ benchmark. Both branches
have completed their switching at this event. Because the compact pulse and
its first ten derivatives vanish at the support endpoint, equality at
switch-off needs no tail extrapolation. Checks at $n_{\rm r}=7,8$ compare
\begin{equation}
 \F_{11}^{\rm r},\qquad \F_{22}^{\rm r},\qquad
 e^{-\ii\Delta_D(t_{\rm r})}\F_{12}^{\rm r}.
 \label{eq:post-switch-readout-diagnostics}
\end{equation}
These phase-demodulated entries are invariant after switch-off. The physical
clock-resolved $\Gamma_{12}^{(2)}(t_{\rm r})$ need not be invariant under a
change of readout event.

The combined benchmark uses $M=0.20$, $\sigma=\Omega=1$, and $A_\chi=1$.
Its 20 compactness rows use the bounded lapse-stretched plotting coordinate
\begin{equation}
 x=1-\sqrt{1-\mathcal C}=1-\sqrt{f(R)}.
 \label{eq:stretched-compactness-coordinate}
\end{equation}
Here $1-x$ is the interior-to-asymptotic clock factor. Ticks are labeled by
$\mathcal C$, and $x=1$ is not a data point.

A fixed radius separation cannot remain exterior arbitrarily close to $2M$.
Let $\rho=R-2M > 0$. We use
\begin{align}
    \frac{\delta R(\mathcal{C})}{M} := min[0.015, q_{pair}(\frac{2}{\mathcal{C}}-2)], 
\end{align}
\begin{align}
    R = \frac{2M}{\mathcal{C}}, \qquad
    a(\mathcal{C};q_{pair}) := \frac{\delta R(\mathcal{C})}{0.015M},
    \qquad q_{pair} = 0.75
    \\
    \delta R(\mathcal{C}) = \min{(0.015M, q_{pair}(R-2M))}
\label{eq:adaptive-compactness-pair}
\end{align}
with the dimensionless parameter  $q_{pair}$ fixed;  we shall set    $q_{pair} =0.75$  
throughout this scan.
Note that
$a$ is the radius-separation reduction factor, not an independently varied parameter. For $\mathcal{C} \leq \mathcal{C_*} = 100/101 \approx 0.9901, a =1$ and $\partial R = 0.015M$. Above this onset the separation decreases as $\partial R = q_{pair}(R - 2M)$, so $\partial R/(R-2M)$ remains equal to the same fixed value 0.75. It is $a$, not $q_{pair}$ that changes. Thus the sampled separation remains 0.015M through $\mathcal{C} = 0.990$. The inward branch remains strictly exterior because $R_- - 2M = \rho - \frac{\partial R}{2} \geq (1 - \frac{q_{pair}}{2})\rho = \frac{5\rho}{8} > 0$.

Thus $q_{\rm pair}=\delta R/(R-2M)$ after contraction. The separation is
$0.015M$ through $\mathcal C=0.990$ and contracts beyond
$\mathcal C_*=0.9901$. The inward branch remains strictly exterior. It first
violates the dominant energy condition at nominal $\mathcal C=0.956556$;
the sampled all-branch-compatible core ends at $\mathcal C=0.950$.
Rows above the analytic boundary are a formal continuation on prescribed
static backgrounds with DEC-violating surface stress. These rows lie outside the
DEC-compatible thin-shell regime. The centered and equal-radius relational
benchmarks are DEC compatible.

\section{Matched-mode results}
\label{sec:results}

All numerical figures are evaluated from globally matched modes using the
response and Gram-matrix formulas above. Appendix~\ref{app:numerics}
describes the numerical methods and summarizes the validation checks.

\subsection{Mode normalization and scattering phase}

\begin{figure*}[t]
  \centering
  \includegraphics[width=\textwidth]{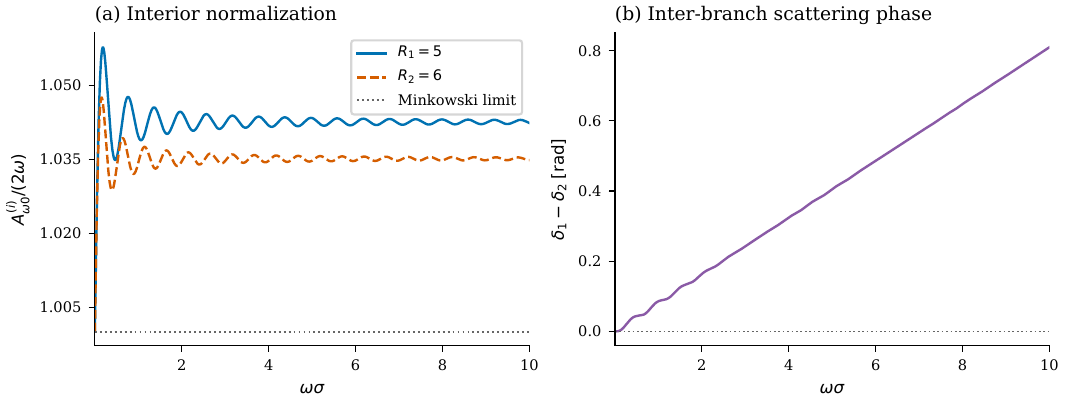}
  \caption{Matched $\ell=0$ modes for $M=0.20$, $R_1=5$, $R_2=6$, and
  $\sigma=1$.  (a) Interior normalization relative to $2\omega$.  (b)
  Unwrapped phase $\delta_1-\delta_2$.  Shared asymptotic frequency, tortoise
  convention, and unit-incoming normalization make both differences
  shell-matching effects.}
  \label{fig:publication-modes}
\end{figure*}

Figure~\ref{fig:publication-modes} shows why local flatness is not enough:
global shell matching changes the interior amplitude by up to about $6\%$ and
the relative scattering phase by $0.81$ rad.  Coherent detector predictions
must therefore retain the exterior matching rather than substitute a locally
flat mode in Eq.~\eqref{eq:center-gram-integral}.

\subsection{Detector and control responses}

\begin{figure*}[t]
  \centering
  \includegraphics[width=\textwidth]{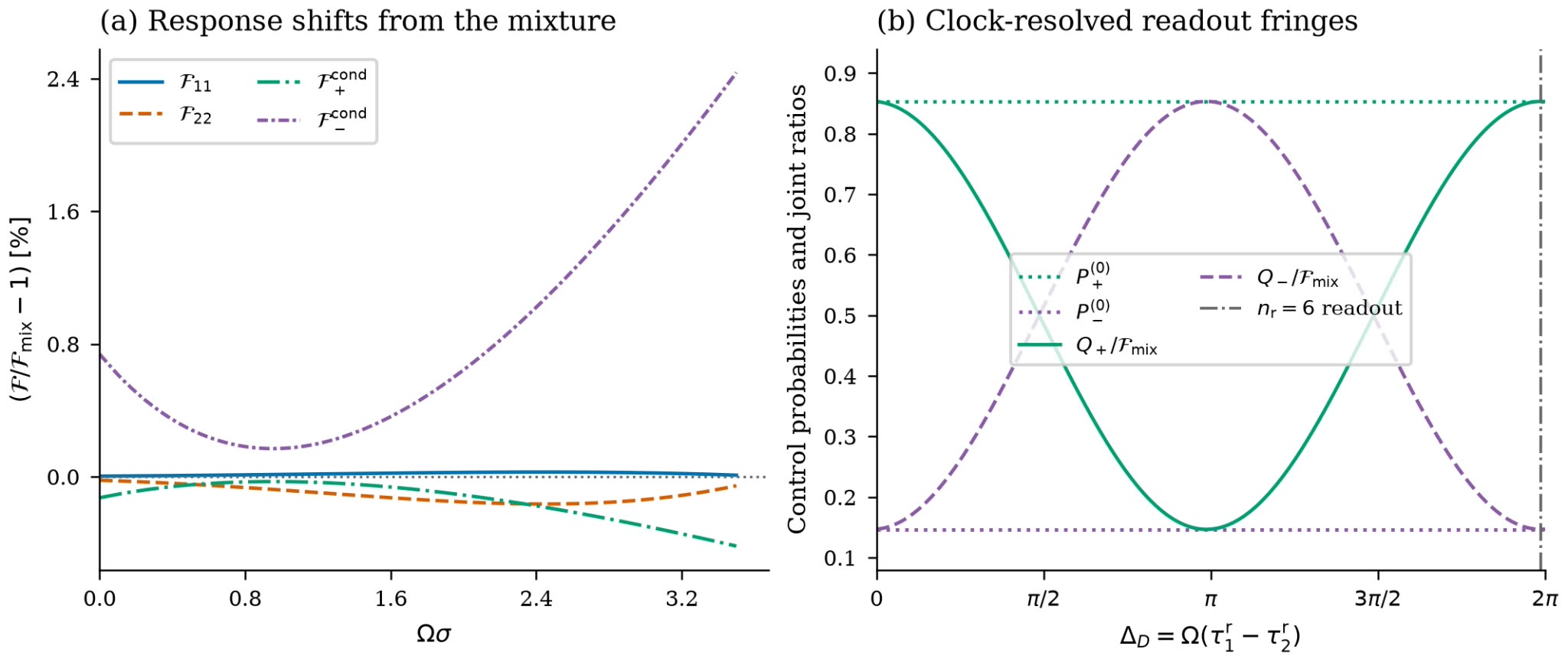}
  \caption{Responses for the concentric shell superposition, with $M=0.20$, $R_1=5$, $R_2=6$, and
  $\sigma=1$.  (a) Local and conditional deviations from the dephased mixture
  for the $n_{\rm r}=6$ common-readout schedule and fixed analyzer phase.
  (b) $P_\pm^{(0)}$ and $Q_\pm/\F_{\rm mix}$ under a detector-clock-phase scan at
  $\Omega\sigma=1$, $\theta=0$; $\Delta_D(t_{\rm r})$ shifts the click fringe
  and is not removed.  The chosen amplitudes avoid dark ports.  The plotted
  $Q_\pm$ are leading coefficients; finite-coupling accuracy requires the
  fixed-regulator conditions in Sec.~\ref{sec:numerics}.}
  \label{fig:publication-response}
\end{figure*}

Figure~\ref{fig:publication-response} shows that coherent postselection
redistributes excitation probability between control ports according to the
proper-time phase accumulated by the declared common readout.  The full
  detector-clock-phase scan separates this horizontal fringe shift from its
phase-independent visibility.  The absence of dark-port normalization and the
control-trace identity rule out the apparent factor-of-two interpretation that
can result from comparing joint and conditional responses.

\subsection{Clock-compensated excitation and de-excitation}
\label{sec:detector-deexcitation}

To include decay, in this subsection we use $\Omega=E_f-E_i$
as a signed transition energy and write the positive physical
level spacing as $E=|\Omega|>0$, with
$H_D=E|e\rangle\langle e|$.
Thus $\Omega=-E<0$ describes $|e\rangle\to|g\rangle$,
whereas $\Omega=+E>0$ describes $|g\rangle\to|e\rangle$.
All field-mode frequencies $\omega$ remain positive.
For the centered detector, the interaction-reference
amplitudes are
\begin{align}
 q_i^I(\omega;\Omega)
 &=\frac{\mathcal A_{\omega0}^{(i)*}}{4\pi\sqrt{\omega}}\,
 \widehat\chi_c\!\left(
 \frac{\omega}{\sqrt{f_i}}+\Omega\right),
 \nonumber\\
 F_{ij}^I(\Omega)
 &=\int_0^\infty d\omega\,
 q_i^I(\omega;\Omega)^*q_j^I(\omega;\Omega).
 \label{eq:signed-transition-gram}
\end{align}

\begin{figure*}[t]
 \centering
 \includegraphics[width=\textwidth]
 {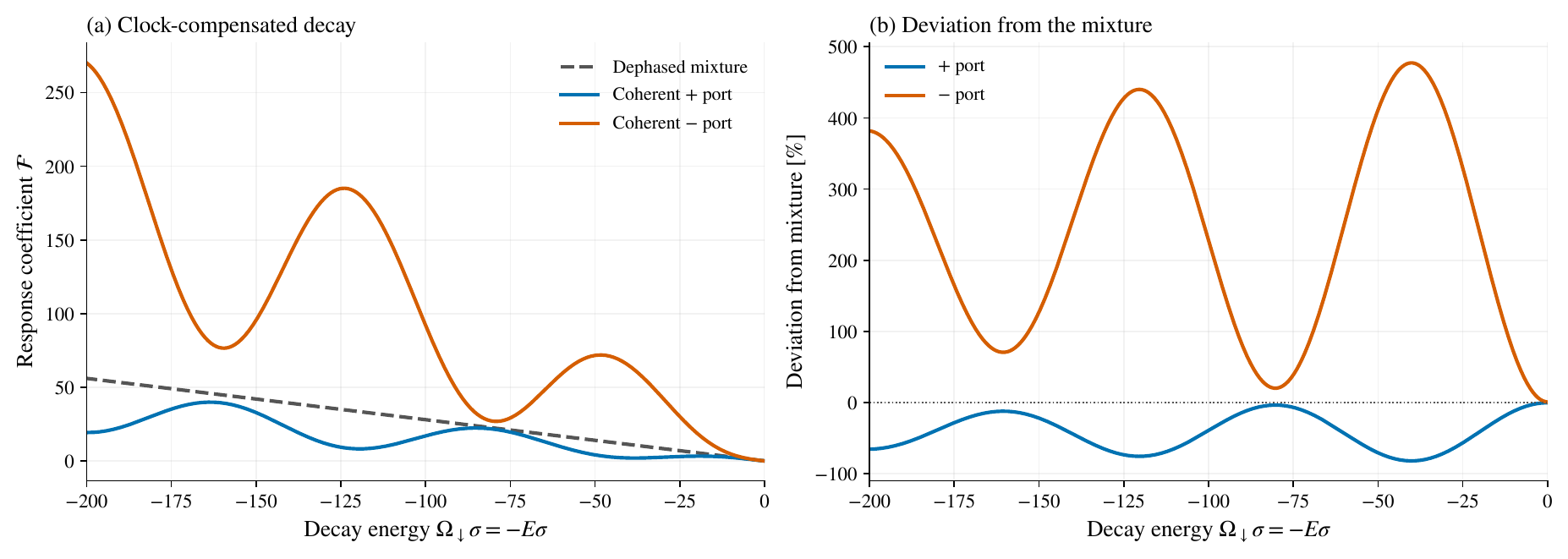}
 \caption{Clock-compensated de-excitation at the common center of a
 coherent superposition of alternative single-shell geometries, compared
 with the branch-dephased mixture. Parameters are $M=0.20$, $R_1=5$,
 $R_2=6$, $\sigma=A_\chi=1$, order-12 compact switching,
 $\alpha_C=\cos(\pi/8)$, and $\beta_C=\sin(\pi/8)$.
 Preparation and readout use Eq.~\eqref{eq:compensated-readout-schedule},
 the coherent phase gates in Eq.~\eqref{eq:clock-compensation-gates}, and
 initial control phase $\phi=0$ and analyzer phase $\theta=0$.
 (a) Leading decay-response coefficients.
 (b) Percentage deviations from the mixed response.
 The signed transition energy is $\Omega=-E$ with
 $-200\leq\Omega\sigma\leq0$; zero denotes the continuous limit.
 Conditional responses use the constant physical port probabilities in
 Eq.~\eqref{eq:compensated-conditional-response}.
 Free detector-clock phases are compensated; shell mode phases and
 lapse-dependent switching arguments are retained.}
 \label{fig:deexcitation-response}
\end{figure*}

\begin{figure*}[t]
 \centering
 \includegraphics[width=\textwidth]
 {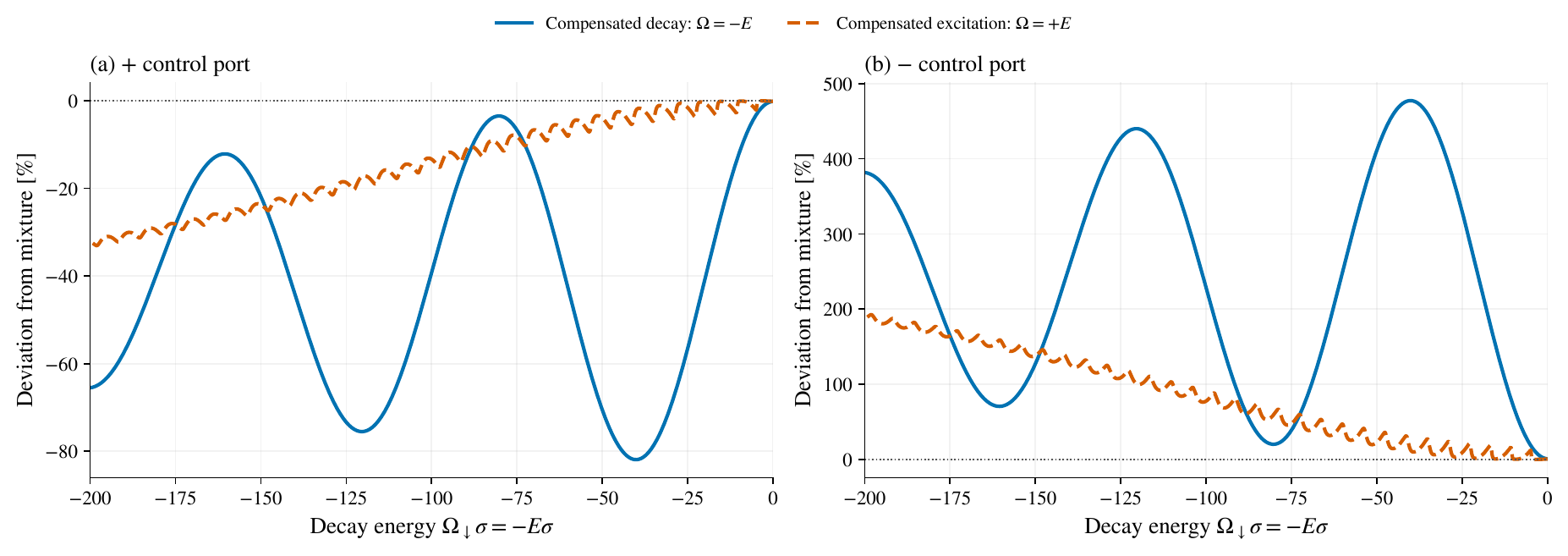}
 \caption{Clock-compensated coherent-versus-mixed conditional-response
 contrast for (a) the $+$ and (b) the $-$ control port, using the parameters
 and physical compensation protocol of
 Fig.~\ref{fig:deexcitation-response}.
 Solid curves show decay at $\Omega=-E$ and dashed curves excitation at
 $\Omega=+E$, compared at the same positive level spacing $E$.
 The horizontal coordinate is the negative decay transition energy
 $\Omega_\downarrow\sigma=-E\sigma\in[-200,0]$.
 Each direction is normalized to its own mixed response and uses
 $P_+^{(0)}=0.85355339$ and $P_-^{(0)}=0.14644661$.
 Zero is the dephased reference. Compensation removes free detector-clock
 propagation while retaining shell transfer functions and finite-window
 spectral weighting; large excitation percentages need not imply a large
 absolute excitation yield.}
 \label{fig:excitation-decay-contrast}
\end{figure*}

We retain $M=0.20$, $R_1=5$, $R_2=6$, $\sigma=A_\chi=1$,
the order-12 compact switching, and a detector at the common shell center.
The control amplitudes are $\alpha_C=\cos(\pi/8)$ and
$\beta_C=\sin(\pi/8)$. Unlike the uncompensated readout of
Fig.~\ref{fig:publication-response}, Figs.~\ref{fig:deexcitation-response}
and~\ref{fig:excitation-decay-contrast} use an explicitly compensated
preparation and measurement protocol. By clock compensation we mean the calibrated pre- and post-interaction phase gates that cancel the detector's branch-dependent free-evolution phases in both the transition and no-transition sectors. This choice separates the shell-dependent field response from controllable
detector-clock phases through physical preparation and readout operations,
while retaining the shell scattering amplitudes and phases.
Compensating both transition and no-transition sectors ensures that the
transition probability and its postselection normalization describe the
same experimental protocol. Without compensation, we would measure the combined effect of the shell on the field and of the different ticking rates of the detector’s clock. Time passes at slightly different rates in the two shell configurations, so the detector accumulates different quantum phases in the two branches. When we recombine those branches, these phase differences affect whether their contributions reinforce or cancel. They also affect how often the selected measurement outcome occurs. Consequently, a peak could arise partly because the detector-clock phases line up favourably, rather than because the shell scatters the field especially strongly. This is a valid measurement, but it mixes those two effects. Compensation removes the controllable clock contribution, making the remaining dependence on the shell geometry easier to interpret.
The shell mode amplitudes and phases, and the lapse-dependent switching
arguments, are retained. Put $n_i=\sqrt{f_i}$ and
$P_e=|e\rangle\langle e|$, and choose
\begin{align}
 t_{\rm p}&=-T,\qquad t_{\rm r}=T,\nonumber\\
 T&=\frac{7\sigma}{\min_i n_i},\qquad
 \tau_i^{\rm p,r}=n_i t_{\rm p,r}.
 \label{eq:compensated-readout-schedule}
\end{align}
Each branch then has at least $\sigma$ of proper-time padding outside the
switching support $|\tau_i|\leq6\sigma$.

Prepare the control with $\phi=0$ and apply coherent phase gates before
and after the interaction,
\begin{align}
 V_{\rm pre}&=\sum_i |i\rangle\langle i|_C\otimes
      e^{-iE\tau_i^{\rm p}P_e},\nonumber\\
 V_{\rm post}&=\sum_i |i\rangle\langle i|_C\otimes
      e^{+iE\tau_i^{\rm r}P_e}.
 \label{eq:clock-compensation-gates}
\end{align}
These gates act without measuring the shell branch. Their calibrated
phases depend on the gap, lapses, and endpoint times, but not on the
field frequency or shell scattering phase. Finite diagonal control pulses
can be placed in the padding, with their integrated phase areas calibrated
including the commuting free detector evolution.

To see why both gates are needed, write
\begin{equation}
 \mathsf D(t)=\sum_i |i\rangle\langle i|_C\otimes
      e^{-iE\tau_i(t)P_e}.
\end{equation}
Suppressing the common free-field evolution, the physical propagator and
the compensated sequence satisfy
\begin{align}
 U_S&=\mathsf D(t_{\rm r})U_I\mathsf D^\dagger(t_{\rm p}),
 \nonumber\\
 V_{\rm post}U_SV_{\rm pre}&=U_I.
 \label{eq:compensated-propagator}
\end{align}
Thus the transition amplitudes and the no-transition control state are
compensated together. The interaction-reference amplitudes in
Eq.~\eqref{eq:signed-transition-gram} are unchanged. This is a modified
physical protocol; a change of analyzer phase alone does not give the same
cancellation. The other readouts in this paper retain their stated protocols.

Use the original complementary analyzer with $\theta=0$. For
$s\in\{\downarrow,\uparrow\}$ define
\begin{align}
 F_{ij}^{\downarrow}(E)&:=F_{ij}^{I}(-E),\nonumber\\
 F_{ij}^{\uparrow}(E)&:=F_{ij}^{I}(+E),\nonumber\\
 L^s&:=\alpha_C^2F_{11}^s+\beta_C^2F_{22}^s.
\end{align}
Both transition directions now have the physical port probabilities and
leading response coefficients
\begin{align}
 P_\pm^{(0),s}&=\frac{1\pm2\alpha_C\beta_C}{2},\nonumber\\
 Q_\pm^s&=\frac{L^s\pm2\alpha_C\beta_C
                         \operatorname{Re}F_{12}^s}{2},\nonumber\\
 \mathcal F_\pm^{{\rm cond},s}
 &=\frac{L^s\pm2\alpha_C\beta_C\operatorname{Re}F_{12}^s}
          {1\pm2\alpha_C\beta_C}.
 \label{eq:compensated-conditional-response}
\end{align}
Numerically, $P_+^{(0),s}=0.85355339$ and
$P_-^{(0),s}=0.14644661$ throughout the scan. Branch dephasing gives
$Q_\pm^s=L^s/2$ and $P_\pm^{(0),s}=1/2$, so both mixed conditional
responses equal $\mathcal F_{\rm mix}^s=L^s$.

Figure~\ref{fig:deexcitation-response} shows the compensated decay response
on $-200\leq\Omega\sigma\leq0$. The largest sampled minus-port enhancement
is approximately $477.36\%$ near $\Omega\sigma=-40.15$, where the plus-port
response is approximately $81.90\%$ below the mixture. A second minus-port
maximum of approximately $440.05\%$ occurs near $\Omega\sigma=-120.40$.
These broad oscillations follow the relative shell mode propagation phase;
the approximately $80.6$ spacing in $E\sigma$ agrees with its smooth
high-frequency phase slope. The port probabilities are constant, so the
oscillations are not produced by a varying postselection denominator.
Increasing separation of the branch emission spectra reduces their overlap
and the successive relative maxima.

Figure~\ref{fig:excitation-decay-contrast} compares compensated decay and
excitation at matched $E\sigma=|\Omega|\sigma$. Its horizontal coordinate
is $\Omega_\downarrow\sigma=-E\sigma$: each dashed excitation point is
evaluated at $\Omega=+E$. The finer excitation oscillations arise as the
switching sidelobes weight different positive field frequencies carrying
different relative mode phases. They persist for a smooth propagation-phase
approximation and do not require narrow scattering resonances. The gap-dependent oscillations are qualitatively similar to those reported for superposed periodically identified Minkowski
spacetimes~\cite{RMann2023} and their accelerated extension~\cite{GoelEtAl2025}.
Here they survive clock compensation and reflect the shell mode phases and finite switching. Percentage
contrast should be distinguished from absolute yield: at $E\sigma=200$,
$L^\downarrow\simeq56.0652$, whereas
$L^\uparrow\simeq2.04\times10^{-48}$.

Compensation removes the additional free detector-clock propagation while
retaining the exact shell transfer amplitudes and phases. Unequal-lapse
spectral weighting and finite switching also remain. The compensated
curves therefore describe the clock-independent interaction response;
they should not be identified with a scattering-only quantity independent
of the detector protocol.

Tracing out the control gives $L^\downarrow$ for
both coherent and dephased preparation.
At fixed common regulator $\mathfrak r$,
\[
 P_{\mathfrak r}(g|\pm)
 =\lambda^2
 \mathcal F_{\pm,\mathfrak r}^{\rm cond,\downarrow}
 +O_{\mathfrak r}(\lambda^4).
\]
The plotted quantities are coefficient limits for one
switching window, rather than stationary transition rates.

\subsection{Cross-branch overlap}

With $\mu_{12}^{I}(\Omega):=F_{12}^{I}(\Omega)/
\sqrt{F_{11}^{I}(\Omega)F_{22}^{I}(\Omega)}$, the branch excitation states
are nearly parallel over the original positive-gap scan of
Fig.~\ref{fig:publication-response}. This does not extend to the full
range of Fig.~\ref{fig:excitation-decay-contrast}: at $E\sigma=200$,
$|\mu_{12}^{I}|\simeq0.5858$ for decay and $0.6364$ for excitation.
For uncompensated excitation, port redistribution depends on both the
interaction overlap phase and the detector-clock phase. The compensated
readout removes the latter, retaining the interaction overlap and its phase.
The overlap characterizes the stipulated controlled-field model but is not,
by itself, a quantum-geometry witness.
The independent convergence audit, and the archived
observable-by-observable validation comparisons are collected in
Appendix~\ref{app:numerics}.

\subsection{Relational-center profile benchmark}

\begin{figure*}[t]
  \centering
  \includegraphics[width=\textwidth]{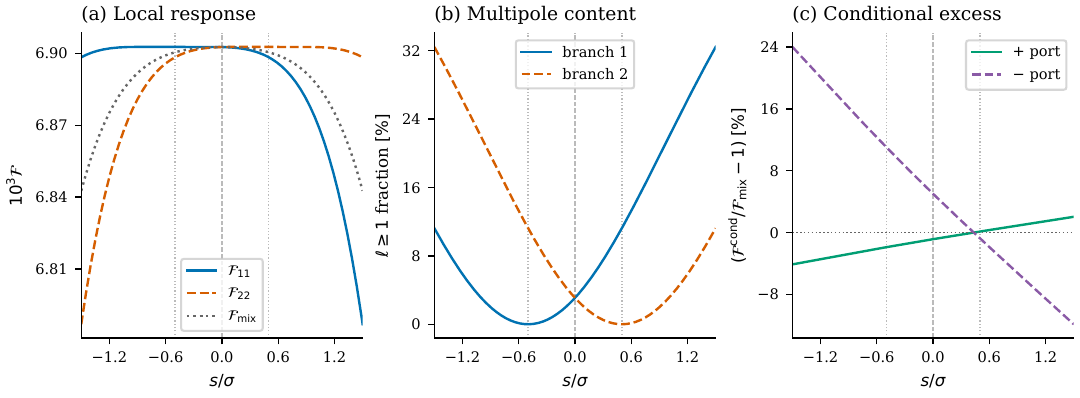}
  \caption{Relational-center benchmark with $M=1$, $R=3$,
  $c_{1,2}=\mp0.5$, $\sigma=\Omega=1$, equal weights, and
  $\phi-\theta=\pi/4$.  Lines mark $c_1,c_0,c_2$.  (a) Local responses and
  mixture.  (b) Higher-multipole fractions.  (c) Conditional excesses, whose
  cross terms vanish under control dephasing.  Here the radii and lapses are
  equal, so $\Delta_D(t_{\rm r})=0$ at every common readout.}
  \label{fig:publication-relational-center}
\end{figure*}

Figure~\ref{fig:publication-relational-center} depicts the results for  the lab-fixed
scenario (right panel of figure~\ref{fig:shell-branch-geometry}): its smooth diagonal profiles encode branch-relative distances,
whereas the cross term contains coherent cross-branch response structure absent
from either diagonal profile.  Conditional changes of order $25\%$ occur
without dark-port amplification, showing a coherent positional effect rather
than normalization by a vanishing postselection probability.  The setup,
multipole content, and center-regularity checks are given in
Appendices~\ref{app:direction-basis} and \ref{app:numerics}.

\subsection{Explicit interference and geometry-dependent fringe changes}
\label{sec:interference-diagnostics}

The relational-center Gram entries also determine an explicit interference
scan without solving new scattering problems. We keep one common post-switch
readout fixed and vary the relative analyzer phase
$\psi:=\phi-\theta$ for balanced control amplitudes. Define
\begin{equation}
 L:=\F_{\mathrm{mix}}=\frac{\F_{11}+\F_{22}}2,
 \qquad z:=\frac{\F_{12}^{\rm r}}{L}.
 \label{eq:interference-normalized-cross}
\end{equation}
For this equal-radius benchmark the lapses agree, so
$\F_{12}^{\rm r}=\F_{12}$ at every common post-switch readout.
Equations~\eqref{eq:joint-response} and
\eqref{eq:postselection-probability} give
\begin{align}
 \frac{Q_\pm(\psi)}{L}
 &=\frac12\left[1\pm\operatorname{Re}(e^{\ii\psi}z)\right],
 \nonumber\\
 P_\pm^{(0)}(\psi)&=\frac12(1\pm\cos\psi).
 \label{eq:interference-port-fringes}
\end{align}
The first line is the leading probability for a control port conditioned on
a detector click; it is distinct from the detector response conditioned on
a control port, $\F_\pm^{\mathrm{cond}}=Q_\pm/P_\pm^{(0)}$.
Branch dephasing sets the cross contribution to zero, giving
$Q_\pm/L=1/2$. Identical branches instead have $z=1$ and retain the ordinary
control fringe. Consequently, high fringe visibility alone does not isolate
the effect of the branch geometry.

To remove that ordinary fringe, define the connected coefficient
\begin{align}
 K(\psi)&:=(Q_+-Q_-)
       -L(P_+^{(0)}-P_-^{(0)}),
 \nonumber\\
 W(\psi)&:=\frac{K(\psi)}L
       =\operatorname{Re}\!\left[e^{\ii\psi}(z-1)\right]
       \quad\text{(coherent preparation)}.
 \label{eq:interference-connected}
\end{align}
For the excitation indicator $n_e$ and binary control-port outcome
$S=\pm1$, the fixed-regulator covariance is
$\operatorname{Cov}_{\mathfrak r}(n_e,S)
=\lambda^2K_{\mathfrak r}+\order_{\mathfrak r}(\lambda^4)$;
$K$ is its first-order-Gram coefficient limit. The definition in the first
line of Eq.~\eqref{eq:interference-connected} vanishes for dephased preparation
as well as for identical branches. In the dephased case both the click-port
asymmetry and the reference control asymmetry vanish; the coherent formula
in the second line is not applied with an unchanged reference fringe.
No small control-port post\-selection probability divides $K$ or $W$.

\begin{figure*}[t]
 \centering
 \includegraphics[width=\textwidth]{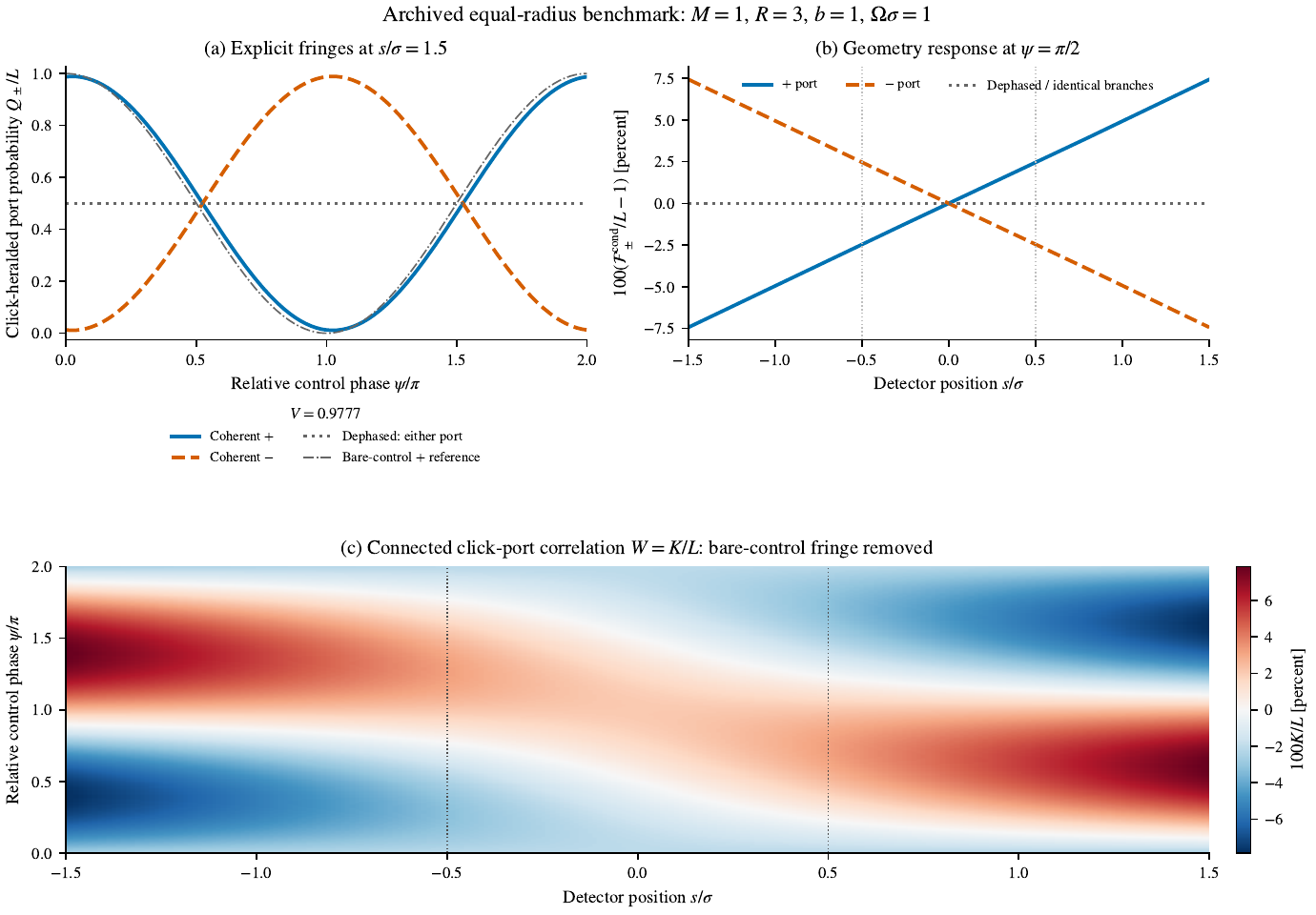}
 \caption{Explicit interference among the computed equal-radius relational-center
 benchmark: $M=1$, $R=3$, $b=1$, $\sigma=\Omega=1$, and balanced control
 amplitudes. The common readout is held fixed while
 $\psi=\phi-\theta$ varies; equal lapses give $\Delta_D=0$.
 (a) Leading control-port probabilities conditioned on a detector click,
 $Q_\pm/L$, at $s/\sigma=1.5$. Branch dephasing gives $1/2$ at either port;
 the bare-control $+$ reference is $P_+^{(0)}$ and is also the identical-branch
 limit. (b) Detector responses conditioned on a control port, shown as
 $100(\F_\pm^{\mathrm{cond}}/L-1)$ at $\psi=\pi/2$.
 Here both zeroth-order port probabilities are $1/2$.
 (c) The connected coefficient $100W=100K/L$ from
 Eq.~\eqref{eq:interference-connected}, with the ordinary control fringe
 subtracted. The connected coefficient vanishes for identical branches and
 for branch-dephased preparation. Vertical guides mark the two shell centers.
 The curves and map are algebraic reconstructions from the same Gram
 entries as Fig.~\ref{fig:publication-relational-center}, with no new radial
 integrations. These normalized quantities are not joint click probabilities;
 those are $\lambda^2Q_{\pm,\mathfrak r}+\order_{\mathfrak r}(\lambda^4)$
 at fixed regulator.
 }
 \label{fig:interference-diagnostics}
\end{figure*}

Figure~\ref{fig:interference-diagnostics} displays the resulting fringes,
a quadrature-phase detector response, and the connected coefficient.
At $\psi=\pi/2$, $P_+^{(0)}=P_-^{(0)}=1/2$ and
$\F_\pm^{\mathrm{cond}}/L-1=\mp\operatorname{Im}z$.
The computed spatial scan gives conditional shifts of up to $7.43\%$ at
this phase, while the maximum phase-dependent connected magnitude is about
$7.85\%$ of $L$. These are coherence-dependent detector/control responses
within the stated field-algebra identification; tracing the control still
gives $L$.

\subsection{Interference for unequal shell radii}
\label{sec:radius-interference}

\begin{figure*}[t]
 \centering
 \includegraphics[width=\textwidth]{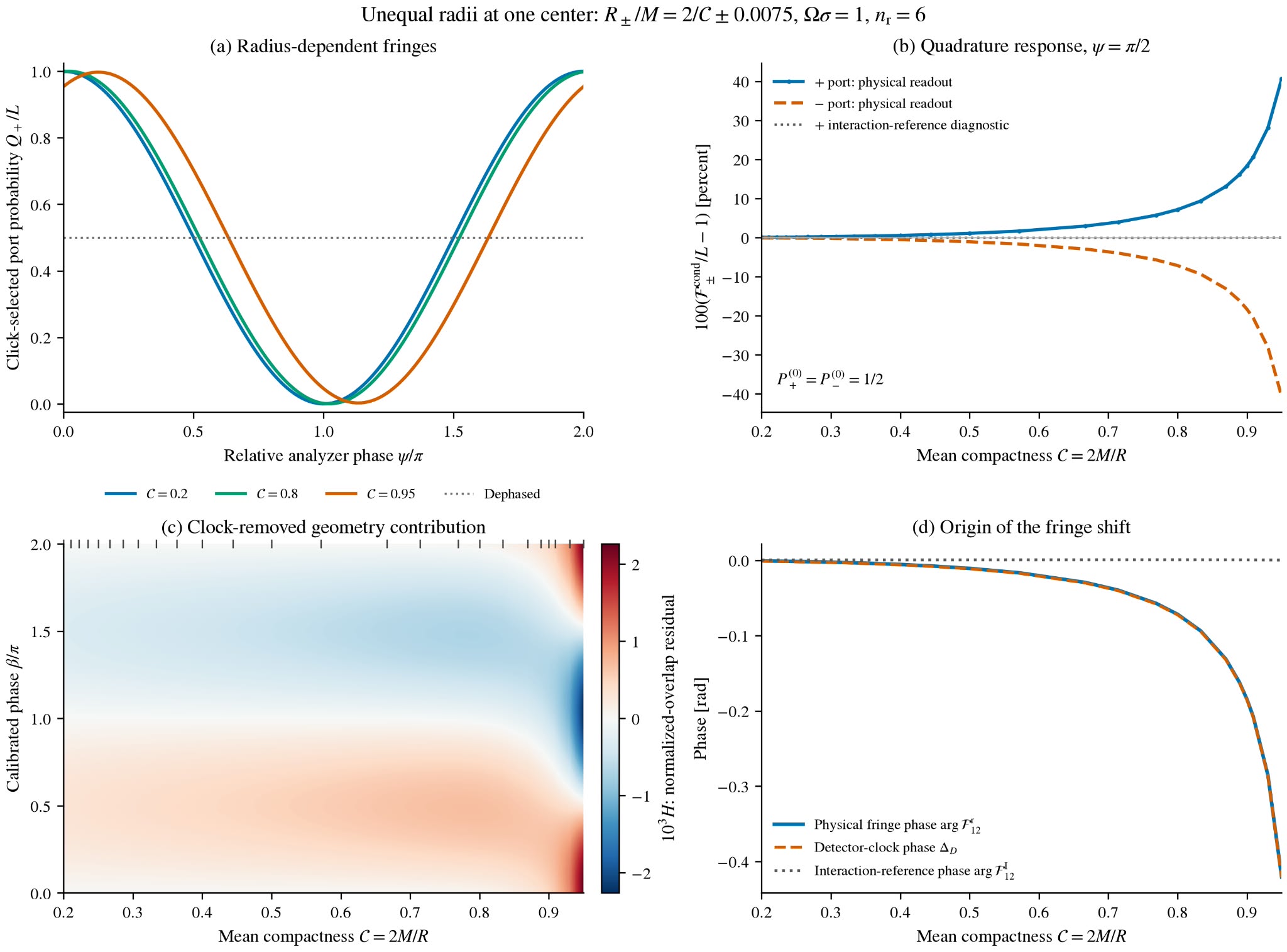}
 \caption{Interference for a detector at the common center of two alternative concentric shells. This extends the centered benchmark of Fig. 4 to a scan of mean compactness, using
 $M=0.20$, $\sigma=\Omega=1$, $R_\pm/M=2/\mathcal C\pm0.0075$, and the
 $n_{\rm r}=6$ common-readout schedule within each pair.
 (a) Click-selected $+$-port fringes for three mean compactnesses ; branch
 dephasing gives $1/2$.
 (b) Conditional detector excesses at $\psi=\pi/2$, where
 $P_\pm^{(0)}=1/2$. The dotted interaction-reference curve is a diagnostic;
 no clock-phase correction is applied to the physical curves. (c) Clock-calibrated normalized-overlap residual, and interaction-reference phase show why the large quadrature response is
 predominantly a clock-induced fringe shift.
 Only the computed DEC-compatible samples $0.20\leq\mathcal C\leq0.95$
are used; upper-edge ticks mark all 25 computed compactness values.
The map fill linearly interpolates the real and imaginary parts of
$1-\mu^{\rm I}$ between those samples. The phase dependence is evaluated analytically. Underlying compactness resolution
is set by the computed compactness grid, rather than the pulse width $\sigma$.
No extra physical samples or resolved intermediate features are implied.
Lines connect computed samples. The displayed quantities are evaluated
algebraically from the centered finite-pair Gram entries using
Eqs.~\eqref{eq:interference-normalized-cross}--\eqref{eq:interference-connected}
and the overlap residual defined below, with no additional radial integrations.
(d) The physical fringe phase, detector-clock phase, $10^3H$, with
 $\beta=\psi+\Delta_D$, the deterministic free clock phase removed, and
 branch-yield imbalance corrected as in
 Eqs.~\eqref{eq:geometry-calibrated-overlap} and
 \eqref{eq:geometry-fringe-residual}.
 }
 \label{fig:unequal-radius-interference}
\end{figure*}

The same phase scan can be made for a centered detector in a superposition
of shells of unequal radii. Figure~\ref{fig:unequal-radius-interference} uses the companion centered finite-pair scan, restricted to its 25 DEC-compatible samples, with $M=0.20$, $\sigma=\Omega=1$, and
$R_\pm/M=2/\mathcal C\pm0.0075$. Within each pair, the common readout is
chosen so that the slower detector clock reaches $6\sigma$; the readout
event therefore depends on $\mathcal C$. The analyzer phase varies at that
fixed event, rather than changing the detector-clock phase. Equations~
\eqref{eq:interference-normalized-cross}--\eqref{eq:interference-connected}
apply with the physical cross response
$\F_{12}^{\rm r}=e^{\ii\Delta_D}\F_{12}^{\rm I}$, where
$\F_{12}^{\rm I}$ denotes the interaction-reference cross response.

The quadrature conditional shift reaches $40.62\%$ at
$\mathcal C=0.95$, with both zeroth-order control-port probabilities equal
to $1/2$. This large redistribution is primarily a detector-clock effect:
the physical fringe phase is $-0.420716$ rad, while
$\Delta_D=-0.420962$ rad and $\arg\F_{12}^{\rm I}=0.00024547$ rad.
It is not a $40\%$ change of visibility or a pure shell-scattering effect.
At this point $|z|=0.994700$ and $L=4.50663\times10^{-4}$; absolute joint
click probabilities still carry the factor $\lambda^2$. The figure extends
the centered response benchmark to a radius scan while retaining the clock
phase required by the physical readout protocol.
To expose the smaller geometry-dependent part, panel (c) removes the known
free detector phase and the branch-yield balance, using the normalized-overlap
residual $H$ defined in Eq.~\eqref{eq:geometry-fringe-residual} below.
Its maximum magnitude at $\mathcal C=0.95$ is $0.226224\%$ of unit overlap,
compared with the clock-dominated $40.62\%$ physical quadrature shift.

\subsection{Joint radius-translation response}
\label{sec:combined-results}

We next use $M=0.20$, $\sigma=\Omega=1$, the adaptive $R_\pm$ pair in
Eq.~\eqref{eq:adaptive-compactness-pair}, and the assignments in
Eq.~\eqref{eq:radius-center-assignments}.  The compactness panels use the
bounded coordinate $x$ in Eq.~\eqref{eq:stretched-compactness-coordinate}, with
$b/M\in\{0.5,1,2,2.5\}$ at $s=b/4$, hence $2s/b=0.5$.  The $b/M=2$ position
profile spans the natural center-to-center parity interval
$2s/b\in[-1,1]$, which lies inside the full containment domain.  A separate
$b/R=1/3$ control co-scales the separation with the radius.

\begin{figure*}[t]
  \centering
  \includegraphics[width=\textwidth]{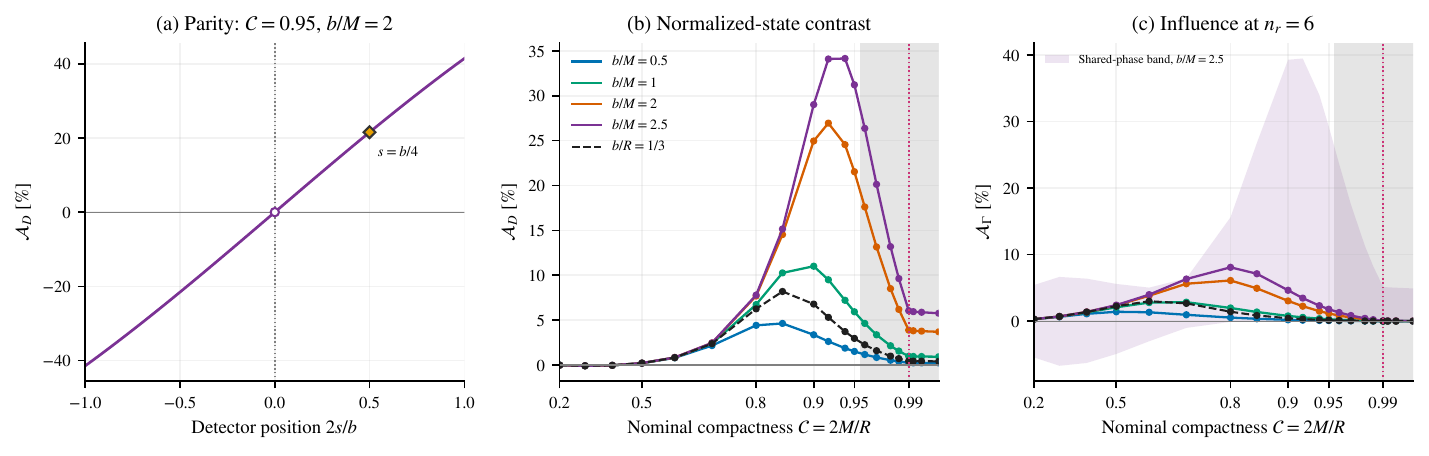}
  \caption{Combined radius-translation response for alternative shell
  geometries with $M=0.20$ and $\sigma=\Omega=1$.  (a) Assignment contrast
  across detector position at $\mathcal C=0.95$ and $b/M=2$; reflection
  exchanges SL and LS and enforces the midpoint null.  The production point
  $2s/b=0.5$ is marked; the profile spans the natural center-to-center parity
  interval inside the full containment domain.  (b) Normalized-state contrast
  $\mathcal A_D$ for four fixed-$b/M$ protocols
  and the independent fixed-$b/R=1/3$ control.  (c) Relative direct-integral influence contrast
  $\mathcal A_\Gamma(t_{\rm r})$ for the $n_{\rm r}=6$ common readout, together
  with the envelope generated by one shared $\Delta_D$ for SL and LS; no phase
  correction is applied.  The physical curve peaks at $8.09\%$ in the displayed
  scan; the roughly $39\%$ upper value belongs to the phase envelope, not to the
  declared readout. The coefficients accumulate over one switching window;
  the contrast is not a decay rate.  Circular markers identify rows resolved under the
  frequency-density, angular-quadrature, and $\ell=2\to3$ spot-check envelope,
  while lines connect all computed rows as diagnostics.  Gray shading
  marks the formal continuation with DEC-violating surface stress, and the
  magenta line marks adaptive pair-contraction onset $\mathcal C_*=0.9901$.
  Panels (b),(c) use the compactness coordinate $x$ in
  Eq.~\eqref{eq:stretched-compactness-coordinate}.  The production curves use
  $2s/b=0.5$; the validated $b/M=2.5$ curve is an
  extended finite-offset result, not an optimum.}
  \label{fig:publication-offset-compactness}
\end{figure*}

Figure~\ref{fig:publication-offset-compactness} shows a nonzero finite response
to exchanging the radius-center assignments. Reflection provides the midpoint
null and sign-reversal checks; the fixed-$b/R$ control excludes increasing
relative separation as the sole explanation, without eliminating
assignment-dependent diagonal yields. Both contrasts peak in the
DEC-compatible core before adaptive pair contraction. At
$\mathcal C=0.94$, $b/M=2.5$, the phase-invariant $\mathcal A_D=34.15\%$
compares $D_{\rm SL}=0.00565326$ and $D_{\rm LS}=0.00277508$.
For the declared unlocked $n_{\rm r}=6$ readout,
$\max|\mathcal A_\Gamma|=0.0809009$ at $\mathcal C=0.80$, $b/M=2.5$,
with $\Gamma_{\rm SL}^{(2)}=1.13555\times10^{-5}$ and
$\Gamma_{\rm LS}^{(2)}=9.65565\times10^{-6}$.
These percentages compare logarithmic divergences and influence coefficients,
respectively. At the latter peak, the SL-minus-LS coefficient difference is
$\Delta\Gamma^{(2)}=1.69982\times10^{-6}$: the SL-minus-LS diagonal mean in
Eq.~\eqref{eq:assignment-clock-influence} contributes $6.79977\times10^{-7}$
(about $40\%$), and the clock-dressed cross term contributes
$1.01984\times10^{-6}$. This is an algebraic decomposition of the contrast
numerator. The leading control-coherence attenuation difference is
$\lambda^2$ times the corresponding regulated coefficient difference, not an
$8.09$ percentage-point coherence loss.

Over the compactness rows,
$-2.8961\leq\Delta_D\leq-1.13\times10^{-3}$ rad. The exact
shared-phase envelope is much broader,
$-0.06710\leq\mathcal A_\Gamma\leq0.39468$.  In particular, the zero-detector-clock-phase
value $0.3946451$ at $\mathcal C=0.92$, $b/M=2.5$, is a phase-sensitive envelope
point, not the deterministic $n_{\rm r}=6$ prediction.  For $b/M=2.5$ the upper envelope
retains this large peak, whereas the physical curve falls to
$|\mathcal A_\Gamma|=3.13\times10^{-4}$ by $\mathcal C=0.999$.  The band varies
one common detector-clock phase for the two assignments rather than aligning
them independently.
As in the position-superposition response in BTZ
\cite{WalleghemCepollaro2026}, the odd signal is tied to a relational
assignment; the $b/M=2.5$ curve is a finite-offset diagnostic, not an
optimization over offset.

\subsection{Interference readout of the assignment contrasts}
\label{sec:assignment-interference}

The combined radius-position result has a direct interference readout.
For each assignment $o\in\{\mathrm{SL},\mathrm{LS}\}$, write
$N_{i,o}=\F_{ii,o}^{\rm r}$, $L_o=(N_{1,o}+N_{2,o})/2$, and
$C_o=\F_{12,o}^{\rm r}$. With balanced coherent preparation, the measured
fringe visibility and a separately calibrated branch-yield balance give
\begin{align}
 V_o&=\frac{|C_o|}{L_o},\qquad
 \mathcal B_o^{\rm y}=\frac{\sqrt{N_{1,o}N_{2,o}}}{L_o},
 \nonumber\\
 D_o&=-2\ln\!\left(\frac{V_o}{\mathcal B_o^{\rm y}}\right).
 \label{eq:assignment-fringe-distance}
\end{align}
Here $\mathcal B_o^{\rm y}$ is a yield-balance factor, distinct from the frequency-space Bhattacharyya coefficient
\cite{Bhattacharyya1943}. Thus fringe amplitudes together
with branch-resolved yields reconstruct the phase-invariant
$\mathcal A_D=34.15\%$ result; the percentage compares small logarithmic
distances rather than percentage-point changes of fringe visibility.

A complementary readout uses the initially dark minus port at $\psi=0$:
\begin{equation}
 2Q_{-,o}(0)=L_o-\operatorname{Re}C_o=\Gamma_o^{(2)}(t_{\rm r}).
 \label{eq:assignment-dark-port-influence}
\end{equation}
Consequently the relative SL/LS contrast of these joint click coefficients
is exactly $\mathcal A_\Gamma$, including its $8.09\%$ peak. We plot
$Q_-$ itself because $P_-^{(0)}(0)=0$; no conditional detector response is
formed by division by that probability. The analyzer is aligned with the
prepared control, without additionally compensating $\Delta_D$.
The detector-clock phase and the same analyzer phase are retained for both
assignments. An analyzer scan reconstructs the physical cross response;
it does not vary the physical attenuation coefficient or the detector-clock
phase envelope in Fig.~\ref{fig:publication-offset-compactness}.

To display the off-diagonal part separately, define
\begin{align}
 J_-(\psi)&:=Q_{-,\mathrm{SL}}-Q_{-,\mathrm{LS}}
 \nonumber\\
 &\hspace{1em}-\frac{L_{\mathrm{SL}}-L_{\mathrm{LS}}}{2}
 \nonumber\\
 &=-\frac12\operatorname{Re}\!\left[
 e^{\ii\psi}(C_{\mathrm{SL}}-C_{\mathrm{LS}})\right].
 \label{eq:assignment-fringe-cross-residual}
\end{align}
We normalize it by the common positive yield
$L_{\rm ref}:=(L_{\mathrm{SL}}+L_{\mathrm{LS}})/2$.
This residual vanishes under branch dephasing and removes the diagonal-yield
contribution. The full $8.09\%$ influence contrast contains both contributions,
as quantified above. The arbitrary-phase residual is an off-diagonal
assignment diagnostic; by itself it does not subtract the radius-only and
position-only responses needed to establish nonadditivity.

\begin{figure*}[t]
 \centering
 \includegraphics[width=\textwidth]{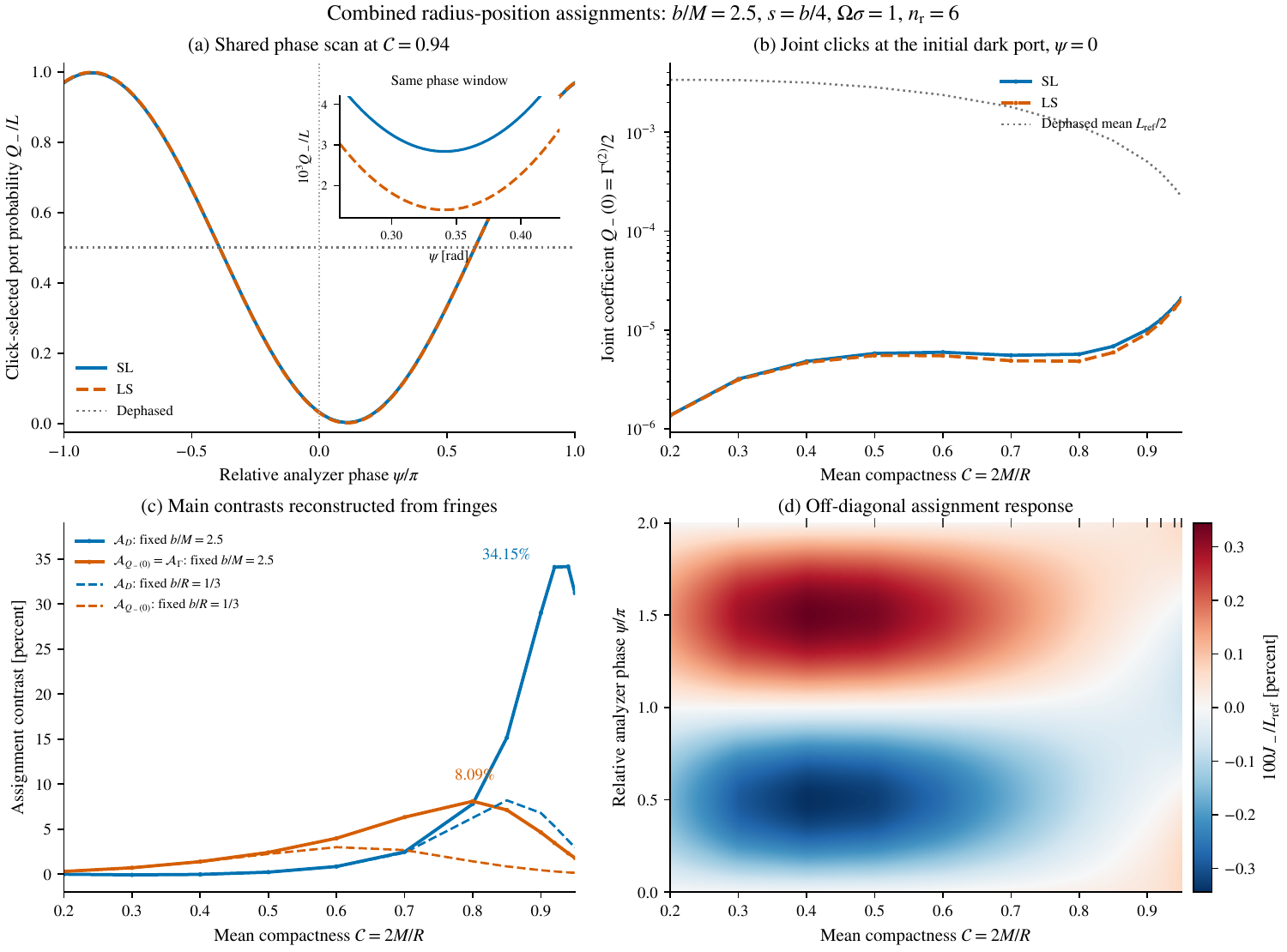}
 \caption{Interference readout of the combined radius-position result,
 with $M=0.20$, $\sigma=\Omega=1$, $b/M=2.5$, $s=b/4$, the radius pair of
 Eq.~\eqref{eq:adaptive-compactness-pair}, and physical $n_{\rm r}=6$ readout.
 (a) SL and LS minus-port fringes at $\mathcal C=0.94$. The inset uses one
 shared phase window to resolve their different minima, without aligning
 them independently. The vertical guide marks $\psi=0$.
 (b) Joint dark-port click coefficients at $\psi=0$; physical probabilities
 are $\lambda^2Q_{-,\mathfrak r}+\order_{\mathfrak r}(\lambda^4)$.
 The dotted curve is the mean dephased coefficient.
 (c) Reconstruction of the main contrasts from
 Eqs.~\eqref{eq:assignment-fringe-distance} and
 \eqref{eq:assignment-dark-port-influence}, including the fixed-$b/R=1/3$
 control. The $34.15\%$ distance contrast and $8.09\%$ joint-click contrast
 reproduce the existing results at $\mathcal C=0.94$ and $0.80$, respectively.
 (d) The off-diagonal assignment residual $100J_-/L_{\rm ref}$, normalized
 by a common positive yield rather than a phase-dependent difference.
 Only the sampled DEC-compatible core through $\mathcal C=0.95$ is shown.
 Lines connect computed samples; upper-edge ticks in (d) mark the 12
computed compactness values. The map fill linearly interpolates the real
and imaginary parts of $-(C_{\rm SL}-C_{\rm LS})/(2L_{\rm ref})$
between those samples solely for display, with analytic phase dependence.
The underlying resolution comes from the nonuniform compactness grid;
the pulse width $\sigma$ does not set display-cell widths. This interpolation
adds no physical samples or independent evidence between them.
All panels are evaluated algebraically from the physical Gram entries used for the combined radius-position results, with no additional scattering solves. The
 residual map is a diagnostic and is not assigned a new row-wise resolution
 certificate.}
 \label{fig:combined-configuration-interference}
\end{figure*}

\subsection{Separating geometry interference from the free detector clock}
\label{sec:geometry-interference-isolated}

The phase retained in the physical readout can be separated from the
geometry-dependent excitation response. For each assignment define
\begin{align}
 C_o^{\rm I}&:=e^{-\ii\Delta_D}C_o,
 \qquad \mu_o^{\rm I}:=
 \frac{C_o^{\rm I}}{\sqrt{N_{1,o}N_{2,o}}},
 \nonumber\\
 \beta&:=\psi+\Delta_D.
 \label{eq:geometry-calibrated-overlap}
\end{align}
The same $\Delta_D$ applies to SL and LS because their radii and clock
protocol agree. Thus one common analyzer calibration
$\psi=\beta-\Delta_D$ removes the deterministic post-switch detector phase;
the interaction phases of the two assignments are not separately aligned.
This also defines a reconstruction from the measured analyzer scan at the
original fixed readout. Normalizing by $\sqrt{N_{1,o}N_{2,o}}$ separately
removes the branch-yield balance, so
$|\mu_o^{\rm I}|=V_o/\mathcal B_o^{\rm y}$.
This separation retains geometry-dependent excitation filters, including
their branch-lapse dependence, as well as globally matched shell modes.
It isolates the free detector phase rather than every lapse effect.

For coherent balanced preparation, let
$\mathcal M_o(\psi):=(Q_{+,o}-Q_{-,o})/L_o$ be the click-selected port
asymmetry. A useful residual after subtracting the phase-aligned identical-
excitation control fringe is
\begin{align}
 H_o(\beta)&:=\cos\beta-
 \frac{\mathcal M_o(\beta-\Delta_D)}{\mathcal B_o^{\rm y}}
 \nonumber\\
 &=\operatorname{Re}\!\left[e^{\ii\beta}(1-\mu_o^{\rm I})\right].
 \label{eq:geometry-fringe-residual}
\end{align}
Here $H=0$ is the coherent reference $\mu_o^{\rm I}=1$, with identical
excitation states and aligned relative phase. Equality of rays alone fixes
$|\mu_o^{\rm I}|=1$ but can leave a relative phase that contributes to $H$.
The assignment
quadrature difference after the same clock and yield corrections is
\begin{equation}
 R(\beta):=-\frac12\operatorname{Re}\!\left[
 e^{\ii\beta}(\mu_{\rm SL}^{\rm I}-\mu_{\rm LS}^{\rm I})\right].
 \label{eq:geometry-assignment-quadrature}
\end{equation}
These quantities expose small changes of the complex overlap; they are
calibrated overlap residuals rather than raw joint click probabilities.
As for $J_-$, an arbitrary-phase assignment residual alone does not perform
the radius-only and position-only subtractions needed for a nonadditivity
test of the complex interference kernel.

\begin{figure*}[t]
 \centering
 \includegraphics[width=\textwidth]{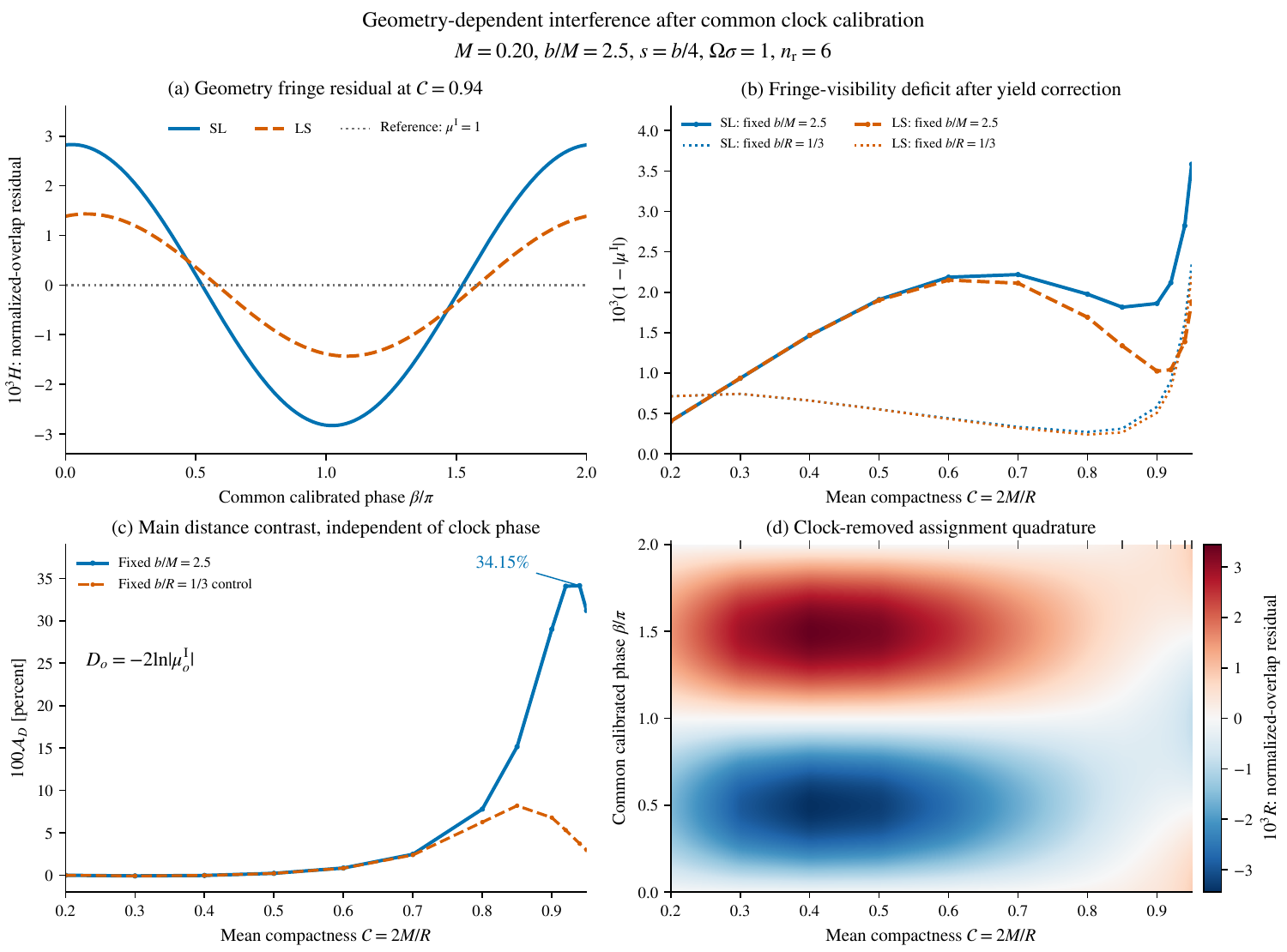}
 \caption{Geometry-dependent interference after one common detector-clock
 calibration and branch-yield correction, for the combined configurations
 with $M=0.20$, $\sigma=\Omega=1$, $s=b/4$, and $n_{\rm r}=6$.
 (a) Normalized-overlap residual $H$ at $\mathcal C=0.94$, $b/M=2.5$,
 from Eq.~\eqref{eq:geometry-fringe-residual}. The zero reference is the
 coherent phase-aligned identical-excitation fringe, $\mu^{\rm I}=1$.
 Neither SL nor LS has its individual
 interaction phase removed.
 (b) The phase-independent corrected visibility deficit $1-|\mu_o^{\rm I}|$,
 including the fixed-$b/R=1/3$ control. These are overlap deficits, not
 \% changes of total click probability.
 (c) The distance contrast reconstructed using $D_o=-2\ln|\mu_o^{\rm I}|$,
 reproducing the main $34.15\%$ result without the free detector phase.
 (d) Normalized-overlap assignment residual $10^3R$ from
 Eq.~\eqref{eq:geometry-assignment-quadrature}; this is distinct from the
 raw-yield residual $J_-/L_{\rm ref}$ in
 Fig.~\ref{fig:combined-configuration-interference}.
 Only the sampled DEC-compatible core through $\mathcal C=0.95$ is used.
 Upper-edge ticks in (d) mark the 12 computed compactness values. The fill
 is a piecewise-linear interpolation of the real and imaginary parts of
 $-(\mu_{\rm SL}^{\rm I}-\mu_{\rm LS}^{\rm I})/2$ between those samples,
 with analytically evaluated phase dependence. This display interpolation
 removes the steps of a constant-cell rendering; it adds no physical
 compactness samples. The coarse nonuniform compactness grid sets the
 underlying resolution. Constant-column boundaries on that grid are display
 discretization, rather than features of the switching convolution or
 physical discontinuities. Lines and markers in (b),(c) retain the computed values.
 No new mode solves or row-wise error certificates are introduced for $H$
 or $R$.}
 \label{fig:geometry-interference-isolated}
\end{figure*}

Figure~\ref{fig:geometry-interference-isolated} shows these residuals on
their natural $10^{-3}$ scale. At $\mathcal C=0.94$, $b/M=2.5$, the corrected
visibility deficits are
$1-|\mu_{\rm SL}^{\rm I}|=0.00282264$ and
$1-|\mu_{\rm LS}^{\rm I}|=0.00138658$.
Their exact conversion $D_o=-2\ln|\mu_o^{\rm I}|$ reproduces the
$34.15\%$ assignment contrast independently of the clock phase.
There is also a cancellation that does not require knowing $\Delta_D$:
\begin{align}
 r&:=\frac{\mu_{\rm SL}^{\rm r}}{\mu_{\rm LS}^{\rm r}}
   =\frac{\mu_{\rm SL}^{\rm I}}{\mu_{\rm LS}^{\rm I}},
 \nonumber\\
 D_{\rm SL}-D_{\rm LS}&=-2\ln|r|,
 \label{eq:geometry-clock-cancelling-ratio}
\end{align}
where $\mu_o^{\rm r}:=C_o/\sqrt{N_{1,o}N_{2,o}}$.
At the same point $\arg r=-1.51973\times10^{-4}$ rad is the surviving
relative assignment phase, while the common detector-clock phase is
$-0.341100$ rad. The original physical $8.09\%$ influence contrast remains
the result for the uncompensated readout in
Eq.~\eqref{eq:assignment-dark-port-influence}; the calibrated diagnostics
are not substituted for that physical attenuation.

\section{Scope and limitations}
\label{sec:limitations}

The common incoming-mode map defines the controlled-field model. A different
branch-space isometry can alter the cross response while leaving its diagonal
entries unchanged. The displaced construction additionally assumes one
asymptotic reference frame, and the numerical tests use collinear positions.
The result is a relational detector response rather than a
background-independent position operator or a device-independent witness of
quantum geometry. Unequal masses, shell motion, backreaction, and a dynamical
preparation of the shell control are outside the prescribed static model.

The scalar is massless, minimally coupled, and transparent to the shell, in
the stipulated static incoming vacuum. The compact pulse, fixed proper-time
gap, control analyzer, and common readout are operational inputs. A known
deterministic detector phase is physical clock marking and can change
port redistribution and probe-induced attenuation; its phase envelope is not
the prediction of an independently optimized pair of readout protocols.
The combined scan follows finite paths with $s=b/4$ and finite $b/M$.
The validated $b/M=2.5$ extension is not an optimization over offsets.

Throughout this study, the shells are always larger than their Schwarzschild radii, so
every sample lies outside $R=2M$. 
However, we have considered DEC-violating shell stresses, shown as the grey-shaded regions in the figures.
The  principal relational effects of the combined scan already occur in the DEC-compatible core, but tail magnitudes do not describe
conventional shell matter. This manuscript draws no terminal near-horizon
scaling conclusion from these finite-grid results.

The plotted quantities are leading response coefficients or their ratios
for one switching window. First-order vectors and Gram coefficients admit
the stated pointlike limits. Full detector outputs and finite-coupling
remainders are defined only at fixed common regulator
$\mathfrak r=(\kappa,\Lambda,J)$, as proved in
Appendix~\ref{app:uv-prescription}. Numerical quadrature convergence is not a
bound on a higher-order interacting-detector remainder, and a click yield
becomes a rate only after a repetition protocol has been specified.

\section{Conclusions}
\label{sec:conclusion}

We have shown that 
detector response in 
a locally flat shell interior depends on the
global superposed geometry through matched mode amplitude and scattering phase. In the
specified common incoming algebra, the positive excitation Gram matrix gives
a consistent description of coherent control-port probabilities, conditional
detector response, and the dephased mixture. In the uncompensated protocol, propagation to a common finite post-switch
event retains the detector-clock phase, whose fringe translation is distinct
from phase-scan visibility. A separate coherent phase-gate protocol
compensates this free-clock contribution in both excitation and decay,
revealing the interaction-level mode-phase structure over an extended gap
range. Unequal-lapse spectral weighting and finite switching remain, so
clock compensation alone does not define a detector-independent scattering
observable. In both protocols, tracing the control recovers the mixture
without a spurious factor-of-two enhancement.

The centered response is exactly $s$ wave. A common incoming direction basis
extends the construction to a lab-fixed detector and translated shell centers
while preserving positivity and common-translation covariance. The equal-radius
benchmark produces conditional changes of order $25\%$ without dark-port
amplification. The dependence is on branch-relative detector distances, not
on an absolute displacement of the complete configuration.

The finite mixed radius-position response remains nonzero in the fixed-$b/R$
control, with reflection enforcing the midpoint null and sign reversal.
The relative normalized-state contrast reaches $34.15\%$ among the computed samples;
the relative physical probe-influence contrast peaks at $8.09\%$ for the
declared $n_{\rm r}=6$ readout. The roughly $39\%$ value belongs to a separate
shared-clock-phase envelope. These are contrasts of small divergences and
leading influence coefficients (Sec.~\ref{sec:combined-results}); normalization
removes click yield from the first, while the second retains yield imbalance
and clock-resolved cross coherence. Both peaks occur within the DEC-compatible
shell domain. The derivations and numerical checks make the field identification, reference
frame, switching, readout, and regulator assumptions explicit. The results
establish coherent, relational response structure in this controlled static
shell model. They do not assign a detector state to a horizon, determine a
dynamical quantum spacetime, or replace the declared apparatus by a
background-independent measurement.

\section*{Related Work Declaration}

In finalizing this manuscript, we became aware of independent work by G.F. do Valle, S.S. Shariff, N. S. M{'o}ller, and N. Yokomizo, entitled "Unruh-DeWitt detectors in a delocalized massive shell". They investigate an Unruh-DeWitt detector inside a massive shell in a quantum superposition of spatial locations. Using a quantum reference frame transformation, they map the superposed geometries to a single, definite shell geometry with the detector in a superposition of positions. This construction allows the cross-branch field correlations contributing to the interference terms in the postselected detector response to be evaluated using a common vacuum state. Their work is being submitted to arXiv concurrently with the present work.

\section*{Acknowledgements}
This work was supported in part by the Natural Sciences and Engineering Research Council of Canada. Research is also supported by Veterans Affairs Canada.

Research at Perimeter Institute is supported in part by the Government of Canada through the Department of Innovation, Science and Economic Development and by the Province of Ontario through the Ministry of Colleges and Universities.

\bibliographystyle{apsrev4-2}
\bibliography{biblio}

\appendix

\section{Radial modes, junction data, and normalization}
\label{app:junction}

For completeness, let the unit-coefficient interior solution be
$\psi_{\rm in}=j_\ell(k_i r)$.  At the shell,
\begin{align}
  \psi_R&=j_\ell(k_iR_i),\\
  \psi'_{R,\rm in}&=k_i j_\ell'(k_iR_i),\\
  \psi'_{R,\rm out}&=\frac{\psi'_{R,\rm in}}{\sqrt{f_i}}
\end{align}
where $f_i$ is defined in \eqref{eq:lapses}.
The exterior Regge-Wheeler variables are initialized as
\begin{align}
  u(R_i)&=R_i\psi_R,\\
  \left.\frac{\dd u}{\dd r_*}\right|_{R_i}
  &=f_i\left(\psi_R+R_i\psi'_{R,\rm out}\right).
  \label{eq:rw-initial-data}
\end{align}
If the solution obtained from these data has raw fitted asymptotic amplitude
$C_{\omega\ell}^{(i)}$, the normalized interior coefficient is
\begin{equation}
  A_{\omega\ell}^{(i)}=\frac{2}{C_{\omega\ell}^{(i)}}.
\end{equation}
Substitution back into Eqs.~\eqref{eq:field-junction} and
\eqref{eq:derivative-junction} provides two independent matching residuals.

The factor of two in Eq.~\eqref{eq:standing-normalization}, together with the
prefactor in Eq.~\eqref{eq:global-mode}, gives
\begin{equation}
  \bigl(v^{(i)}_{\omega\ell m},
  v^{(i)}_{\omega'\ell'm'}\bigr)_{\rm KG}
  =\delta(\omega-\omega')\delta_{\ell\ell'}\delta_{mm'}.
  \label{eq:kg-normalization}
\end{equation}
In the Minkowski limit this convention yields
$A_{\omega\ell}=2\omega$ and
$\delta_{\omega\ell}=-\ell\pi/2\pmod{2\pi}$, fixing both the normalization
and the phase convention used by the cross-branch kernel.

For the centered detector, $f_i\to1$ and
$\mathcal A_{\omega0}\to2\omega$ reduce
Eq.~\eqref{eq:center-gram-integral} to
\begin{equation}
  \F_0(\Omega)=\int_0^\infty\frac{\dd\omega\,\omega}{4\pi^2}
  |\widehat\chi_c(\Omega+\omega)|^2.
  \label{eq:minkowski-integral}
\end{equation}
For the compact switching, set $a=\sigma/2$ and
$x=a(\Omega+\omega)$.  Equation~\eqref{eq:minkowski-integral} becomes
\begin{equation}
  \F_0(\Omega)=\frac{A_\chi^2\sigma^2}{2\pi a^2}
  \int_{a\Omega}^{\infty}\dd x\,
  (x-a\Omega)\sinc^{24}x .
  \label{eq:minkowski-compact}
\end{equation}
This convergent one-dimensional integral provides a normalization test
independent of shell matching.

\subsection{Proof of radial spectral completeness}
\label{app:radial-completeness}

We prove here that the continuum used in Eq.~\eqref{eq:field-expansion} is the
entire one-particle spectrum.  Fix a branch, suppress its index, and hold
$R>2M$ fixed.  Then $N\geq\sqrt{f_R}>0$, the static slice is complete, and
$t$ is a global static time.  For fixed $\ell$, separation of
Eq.~\eqref{eq:kg-equation} gives the Sturm-Liouville expression
\begin{align}
  L_\ell\psi
  &=\frac{1}{w}\left[-(p\psi')'+q_\ell\psi\right],
  \label{eq:radial-sl-operator}\\
  p&=\frac{Nr^2}{B},\qquad
  w=\frac{Br^2}{N},\qquad
  q_\ell=NB\,\ell(\ell+1),
  \label{eq:radial-sl-coefficients}
\end{align}
on
$\mathcal H_\ell=L^2((0,\infty),w\,\dd r)$.  Its physical quadratic form is
\begin{equation}
  \mathfrak q_\ell[\psi]
  =\int_0^\infty\!\dd r\,
  \left(p|\psi'|^2+q_\ell|\psi|^2\right)\geq0.
  \label{eq:radial-positive-form}
\end{equation}
Let $\mathcal D_0$ be the compactly supported piecewise-smooth functions that
are continuous at $R$ and regular at the center.  The nonnegative preform in
Eq.~\eqref{eq:radial-positive-form} is closable.  Indeed, under the unitary
Liouville map in Eq.~\eqref{eq:radial-unitary-map}, its restriction to
$\mathcal D_0$ is the restriction of Eq.~\eqref{eq:liouville-positive-form} to
a core in $H_0^1(\mathbb R_+)$: Hardy's inequality controls the inverse-square
term, the remaining regular potential is locally bounded and short range, and
the one-point trace is infinitesimally form bounded.  Its closure, equipped
with the form norm
$(\|\psi\|_{\mathcal H_\ell}^2+\mathfrak q_\ell[\psi])^{1/2}$, is therefore
densely defined, closed, and nonnegative.  The first representation theorem
supplies the associated self-adjoint Friedrichs operator.

Integration by parts shows explicitly which operator domain that form
selects.  On each side of the shell its elements have $\psi$ and $p\psi'$
locally absolutely continuous, satisfy $L_\ell\psi\in\mathcal H_\ell$, are
regular at $r=0$, and obey
\begin{equation}
  [\psi]_R=0,\qquad [p\psi']_R=0,
  \label{eq:sl-shell-domain}
\end{equation}
where $[g]_R=g(R^+)-g(R^-)$.  The Green boundary form
$p(\overline\psi\chi'-\overline{\psi'}\chi)$ then cancels between the two
sides of the shell.  Since
$p(R^-)=\sqrt{f_R}R^2$ and $p(R^+)=f_RR^2$,
Eq.~\eqref{eq:sl-shell-domain} is exactly
$\psi'_{\rm in}=\sqrt{f_R}\psi'_{\rm out}$, together with field continuity,
as in Eqs.~\eqref{eq:field-junction} and
\eqref{eq:derivative-junction}.  Infinity is the limit point.  At the center the
regular condition is $u(0)=0$ for $u=r\psi$; it is the Friedrichs boundary
condition for $\ell=0$, while the singular solution is already excluded by
square integrability for $\ell\geq1$.  Thus no unstated endpoint extension is
present.

The shell condition is most transparent in a global optical coordinate,
\begin{equation}
  \begin{aligned}
    \xi(r)&=
    \begin{cases}
      r/\sqrt{f_R},&0\leq r\leq R,\\
      R/\sqrt{f_R}+r_*(r)-r_*(R),&r\geq R,
    \end{cases}\\
    \frac{\dd\xi}{\dd r}&=\frac{B}{N}.
  \end{aligned}
  \label{eq:global-optical-radius}
\end{equation}
Writing $\xi_R=R/\sqrt{f_R}$, the map
\begin{equation}
  \begin{aligned}
    U:&\mathcal H_\ell\longrightarrow L^2(\mathbb R_+,\dd\xi),\\
    &(U\psi)(\xi)=u(\xi)=r(\xi)\psi(r(\xi)).
  \end{aligned}
  \label{eq:radial-unitary-map}
\end{equation}
is unitary because
$w|\psi|^2\dd r=|u|^2\dd\xi$.  Away from the shell,
$UL_\ell U^{-1}=-\partial_\xi^2+V^{\rm reg}_\ell$, where
\begin{equation}
  V^{\rm reg}_\ell(\xi)=
  \begin{cases}
    \ell(\ell+1)/\xi^2,&0<\xi<\xi_R,\\[2pt]
    f(r)\left[\ell(\ell+1)/r^2+2M/r^3\right],
      &\xi>\xi_R.
  \end{cases}
  \label{eq:global-radial-potential}
\end{equation}
Equation~\eqref{eq:sl-shell-domain} does not make
$\partial_\xi u$ continuous.  Since
$r_\xi(R^-)=\sqrt{f_R}$ and $r_\xi(R^+)=f_R$, it gives
\begin{equation}
  \begin{aligned}
    {}[u]_{\xi_R}&=0,\qquad
    [u_\xi]_{\xi_R}=\gamma_R u(\xi_R),\\
    \gamma_R&=\frac{f_R-\sqrt{f_R}}{R}<0.
  \end{aligned}
  \label{eq:geometric-point-interaction}
\end{equation}
Thus the Liouville normal form contains the geometrically induced point term
$\gamma_R\delta(\xi-\xi_R)$.  It is not an added scalar interaction, but it
must be included in the spectral problem.  In particular,
$\gamma_R\xi_R=\sqrt{f_R}-1\in(-1,0)$.

Although this point coefficient is negative, the operator is not unstable.
Transforming Eq.~\eqref{eq:radial-positive-form} gives the exact factorization
\begin{align}
  \mathfrak q_\ell[U^{-1}u]
  &=\int_0^\infty\!\dd\xi\,
    \left|u_\xi-\frac{r_\xi}{r}u\right|^2
  \notag\\
  &\quad+\ell(\ell+1)\int_0^\infty\!\dd\xi\,
    \frac{N^2}{r^2}|u|^2
  \notag\\
  &=\int_0^\infty\!\dd\xi\,|u_\xi|^2
    +\int_{\xi\ne\xi_R}\!\dd\xi\,
       V^{\rm reg}_\ell|u|^2
  \notag\\
  &\quad+\gamma_R|u(\xi_R)|^2\geq0.
  \label{eq:liouville-positive-form}
\end{align}
Distributional integration by parts gives the final equality: the jump
in $r_\xi$ produces the point term. The factorization shows that the
full form remains nonnegative despite the negative point coefficient.

We next determine its spectrum.  Let
$H_\ell^0=-\partial_\xi^2+\ell(\ell+1)/\xi^2$ with the regular condition at
the origin.  The difference from $H_\ell^0$ vanishes in the interior, apart
from the point term at $\xi_R$, and, because
$\xi=r_*+\text{constant}=r+\order(\ln r)$,
\begin{equation}
  V^{\rm reg}_\ell(\xi)-\frac{\ell(\ell+1)}{\xi^2}
  =\order\!\left(\frac{\ln\xi}{\xi^3}\right)
  \quad (\xi\longrightarrow\infty).
  \label{eq:bessel-short-range-remainder}
\end{equation}
The remainder and its first moment are integrable.  Cutting once to the right
of the shell leaves a finite interval with compact resolvent and an exterior
half-line whose potential is real and in $L^1$.  Restoring the cut and the
single real jump changes the resolvent by finite rank.  Weyl's theorem and
Weyl sequences supported arbitrarily far out therefore give
\begin{equation}
  \sigma_{\rm ess}(L_\ell)=[0,\infty).
  \label{eq:radial-essential-spectrum}
\end{equation}
Here $\sigma(L_\ell)$ denotes the spectrum of $L_\ell$. The subscript `ess' denotes the essential spectrum, while `ac', `pp', and `sc' denote its absolutely continuous, pure point, and singular continuous parts, respectively. 
For completeness, we also verify the spectral type rather than infer it from
the tail alone.  For every $k>0$, successive approximation in the exterior
Volterra equation
\begin{equation}
  F_+(k,\xi)=e^{\ii k\xi}
  +\int_\xi^\infty\!\dd s\,
  \frac{\sin k(s-\xi)}{k}
  V^{\rm reg}_\ell(s)F_+(k,s)
  \label{eq:jost-volterra}
\end{equation}
converges because the exterior potential is in $L^1$; it gives
$F_+=e^{\ii k\xi}+o(1)$ and
$\partial_\xi F_+=\ii k e^{\ii k\xi}+o(1)$.
The conjugate $F_-=\overline{F_+}$ is independent since
$W[F_+,F_-]=-2\ii k$.  Hence every exterior solution is a linear combination
of $F_+$ and $F_-$, and no nonzero such combination lies in $L^2$.
There are therefore no positive embedded eigenvalues.

Let $\phi_\ell(k,\xi)$ be the real solution with a fixed regular normalization
at the origin.  It propagates uniquely through
Eq.~\eqref{eq:geometric-point-interaction}.  That jump has transfer matrix
$\left(\begin{smallmatrix}1&0\\ \gamma_R&1\end{smallmatrix}\right)$ of
determinant one, so it preserves Wronskians.  Define the Jost function
$\mathcal J_\ell(k)=W[F_+(k),\phi_\ell(k)]$ on the exterior.  It cannot vanish
for real $k>0$: otherwise $\phi_\ell=cF_+$, and the reality of $\phi_\ell$
would give $W[\phi_\ell,\overline{\phi_\ell}]=0$, contradicting
$|c|^2W[F_+,F_-]=-2\ii k|c|^2$.  The Green-kernel jump across the real axis
and Stone's formula consequently give
\begin{equation}
  \frac{\dd\rho_\ell}{\dd\lambda}(k^2)
  =\frac{c_\ell k}{|\mathcal J_\ell(k)|^2}>0,
  \qquad k>0,
  \label{eq:jost-spectral-density}
\end{equation}
where the positive constant $c_\ell$ is fixed by the chosen origin
normalization.  On each compact interval in $k>0$, the Jost construction gives
locally uniform boundary values of the limiting resolvent kernel.  The regular
solution is cyclic for this scalar half-line Sturm-Liouville operator.
Stone's formula therefore exhausts its entire scalar spectral measure on that
interval by the density in Eq.~\eqref{eq:jost-spectral-density}; hence the
positive spectrum is absolutely continuous, has multiplicity one, and has no
singular-continuous component.
This is the usual half-line Jost/Weyl-Titchmarsh argument
\cite{Teschl2014,Weder2015}, applied only after the finite shell matching has
been carried explicitly.

Nonnegativity excludes negative spectrum.  If zero were an $L^2$ eigenvalue,
Eq.~\eqref{eq:liouville-positive-form} would vanish.  For $\ell\geq1$ its
second nonnegative term forces $u=0$.  For $\ell=0$ its first term gives
$u=Cr(\xi)$, or equivalently a spatially constant $\psi=C$; because
$r(\xi)\sim\xi$, this is not in $L^2(\dd\xi)$ unless $C=0$.  There is also no
half-bound state at the threshold.  For $\ell=0$ the unique regular zero-energy
solution is the same growing $u=Cr$.  For $\ell\geq1$, the first-moment
estimate in Eq.~\eqref{eq:bessel-short-range-remainder} and zero-energy
asymptotic integration give a basis with behavior
$\xi^{\ell+1}(1+o(1))$ and $\xi^{-\ell}(1+o(1))$.  A bounded regular threshold
solution would therefore be proportional to the latter and would lie in
$L^2$, contradicting the absence of a zero eigenfunction.  Since a continuous spectral measure cannot be supported at the single
point zero, these results give
\begin{equation}
  \sigma(L_\ell)=\sigma_{\rm ac}(L_\ell)=[0,\infty),
  \qquad
  \sigma_{\rm pp}(L_\ell)=\sigma_{\rm sc}(L_\ell)=\varnothing,
  \label{eq:complete-radial-spectrum}
\end{equation}
with radial multiplicity one.

It remains to fix the spectral measure.  Regularity at the origin determines
one real solution at each $\omega>0$ up to scale.  Its nonzero Jost
coefficients give a nonzero standing-wave amplitude, which we set to two as in
Eq.~\eqref{eq:standing-normalization}; the additive offset between $\xi$ and
$r_*$ only shifts the phase.  Green's identity for frequencies $\omega$ and
$\omega'$ has no origin term, and its two shell terms cancel by
Eq.~\eqref{eq:geometric-point-interaction}.  Explicitly, for
$X>\xi_R$ it gives
\begin{align}
  &(\omega^2-\omega'^2)
  \int_0^X\!\dd\xi\,
  \widetilde u_{\omega\ell}^*\widetilde u_{\omega'\ell}
  \notag\\
  &\quad=\left[
  \widetilde u_{\omega\ell}^*\partial_\xi\widetilde u_{\omega'\ell}
  -(\partial_\xi\widetilde u_{\omega\ell})^*
   \widetilde u_{\omega'\ell}\right]_X.
  \label{eq:finite-lagrange-normalization}
\end{align}
Taking $X\to\infty$ after smearing in positive frequencies and using the
amplitude-two standing asymptotics gives
\begin{equation}
  \int_0^\infty\!\dd r\,w(r)
  \left[\frac{\widetilde u_{\omega\ell}(r)}{r}\right]^*
  \frac{\widetilde u_{\omega'\ell}(r)}{r}
  =2\pi\delta(\omega-\omega').
  \label{eq:radial-delta-normalization}
\end{equation}
The coefficient $2\pi$ is the singular diagonal contribution obtained from
the outer Wronskian in Green's identity: after division by
$\omega^2-\omega'^2$, the difference-frequency term converges to
$2\pi\delta(\omega-\omega')$, while the sum-frequency term vanishes
distributionally on $\mathbb R_+^2$.  One must not replace the exact modes by
their sine asymptotes over the entire half-line: when $\delta_\omega$ depends
on frequency, that raw sine integral also contains nonsingular principal-value
terms, which are cancelled by the finite-radius part of the exact Green
identity.  Expressing the resolution in $\omega$, rather than
$\lambda=\omega^2$, incorporates $\dd\lambda=2\omega\,\dd\omega$.

Because Eq.~\eqref{eq:complete-radial-spectrum} leaves no orthogonal spectral
sector, the generalized Fourier map
\begin{equation}
  (\mathcal U_\ell h)(\omega)
  =\frac{1}{\sqrt{2\pi}}
  \int_0^\infty\!\dd r\,w(r)
  \left[\frac{\widetilde u_{\omega\ell}(r)}{r}\right]^*h(r)
  \label{eq:radial-fourier-map}
\end{equation}
is unitary from $\mathcal H_\ell$ onto
$L^2(\mathbb R_+,\dd\omega)$ and diagonalizes $L_\ell$ as multiplication by
$\omega^2$.  Its inverse is the corresponding mode integral.  Equivalently,
\begin{equation}
  \int_0^\infty\frac{\dd\omega}{2\pi}
  \frac{\widetilde u_{\omega\ell}(r)}{r}
  \left[\frac{\widetilde u_{\omega\ell}(r')}{r'}\right]^*
  =\frac{\delta(r-r')}{w(r')}
  \label{eq:radial-resolution-identity}
\end{equation}
in the weighted-distribution sense.  No bound-state sum is present.

Finally, spherical-harmonic completeness gives
\begin{equation}
  L^2\!\left(\Sigma,\frac{\sqrt h}{N}\dd^3x\right)
  =\bigoplus_{\ell=0}^\infty\bigoplus_{m=-\ell}^{\ell}
   \mathcal H_\ell,
  \label{eq:angular-radial-direct-sum}
\end{equation}
and summing Eq.~\eqref{eq:radial-resolution-identity} against
$Y_{\ell m}(\Omega)Y_{\ell m}^*(\Omega')$ gives the explicit full identity
\begin{align}
  &\sum_{\ell,m}\int_0^\infty\frac{\dd\omega}{2\pi}
  Y_{\ell m}(\Omega)Y_{\ell m}^*(\Omega')
  \notag\\[-2pt]
  &\quad\times\frac{\widetilde u_{\omega\ell}(r)}r
  \left[\frac{\widetilde u_{\omega\ell}(r')}{r'}\right]^*
  =\frac{\delta(r-r')}{w(r')}\delta_{S^2}(\Omega,\Omega').
  \label{eq:full-spatial-resolution}
\end{align}
Using Eq.~\eqref{eq:radial-delta-normalization} in the Klein-Gordon product
and writing $\langle a,b\rangle_w=\int_0^\infty w a^*b\,\dd r$ yields
\begin{align}
  (v_{\omega\ell m},v_{\omega'\ell'm'})_{\rm KG}
  &=\frac{\omega+\omega'}{4\pi\sqrt{\omega\omega'}}
    \delta_{\ell\ell'}\delta_{mm'}
  \notag\\
  &\quad\times
    \left\langle\frac{\widetilde u_{\omega\ell}}r,
    \frac{\widetilde u_{\omega'\ell}}r\right\rangle_w
  \notag\\
  &=\delta(\omega-\omega')\delta_{\ell\ell'}\delta_{mm'},
  \label{eq:kg-normalization-derived}
\end{align}
while the positive/negative-frequency product vanishes.  Thus, for either
branch, the modes map the complete coefficient space
$\mathfrak h=\bigoplus_{\ell,m}L^2(\mathbb R_+,\dd\omega)$ unitarily onto the
positive-frequency solution space.  Quantization on the symmetric Fock space
$\mathcal F_s(\mathfrak h)$ therefore gives all field oscillators and the
commutator in Eq.~\eqref{eq:common-commutator}, with no missing discrete
creation operator.

The branch modes are therefore complete, while their cross-branch identification remains a separate physical convention. Concentric
branches use the common $(\omega,\ell,m)$ basis; translated centers use the incoming-direction basis, with the unitary angular transforms and translation phases established in Appendix~C. The shared asymptotic time, tortoise-coordinate convention, and unit incoming amplitude specify the preparation without making it unique. These statements hold at fixed $R>2M$; no uniform $R\to2M^+$ limit is claimed.

\section{Compact switching and spectral normalizability}
\label{app:uv-normalizability}

This appendix proves that the finite-readout one-particle vectors used in the
paper belong to the field Hilbert space.  It also isolates the defect in a
Gaussian stopped sharply at a finite nonzero tail.  The proof is independent
of detector-clock phase and therefore introduces no phase locking.

\subsection{Exact pulse, endpoints, and transform}

Write $x_+:=\max(x,0)$.  The twelvefold box convolution in
Eq.~\eqref{eq:compact-switching} has the exact truncated-power representation
\begin{equation}
 \begin{aligned}
 b_a^{*12}(\tau)
 &=\frac{1}{\sigma^{12}11!}
 \sum_{k=0}^{12}(-1)^k\binom{12}{k}\\[-2pt]
 &\quad\times(\tau+6\sigma-k\sigma)_+^{11},
 \qquad a=\frac{\sigma}{2}
 \end{aligned}
 \label{eq:compact-switching-truncated-power}
\end{equation}
with $x_+ := \max(x,0)$, thus $(y)_+^{11} = \max (y,0)^{11}$.
This follows by taking the twelfth power of the Laplace transform of a
normalized interval density, or equivalently by induction under convolution.
It proves directly that the support is $[-6\sigma,6\sigma]$, that the pulse is
piecewise degree eleven and globally $C^{10}$, and that it is not $C^{11}$.
On the two intervals adjacent to the support endpoints, \([-6\sigma,-5\sigma]\) and \([5\sigma,6\sigma]\), the pulse reduces to intervals,
\begin{equation}
 b_a^{*12}(\tau)=
 \frac{(6\sigma-|\tau|)^{11}}{11!\sigma^{12}},
 \qquad 5\sigma\leq|\tau|\leq6\sigma.
 \label{eq:compact-switching-endpoint}
\end{equation}
Thus the pulse and its first ten derivatives vanish at both endpoints; the
eleventh derivative has the expected spline jump.

Regard each factor $b_a$ as the probability density of an independent random
variable uniform on $[-a,a]$.  Convolution preserves unit area, and variances
add, so the sum of twelve such variables has variance
$12a^2/3=\sigma^2$.  Multiplication by
$A_\chi\sqrt{2\pi}\sigma$ proves
Eq.~\eqref{eq:compact-switching-moments}.  Fourier transformation of each box
gives
\begin{equation}
 \widehat b_a(\nu)=\sinc(a\nu),\qquad
 \widehat\chi_c(\nu)=A_\chi\sqrt{2\pi}\sigma
 \sinc^{12}(a\nu),
 \label{eq:compact-transform-proof}
\end{equation}
which proves Eq.~\eqref{eq:compact-switching-transform}.  The nonzero zeros
$\nu_k=2\pi k/\sigma$ have order twelve, and the intensity zeros have order
twenty-four.  In particular the transform does not change sign.  For every
fixed integer $s\geq0$,
\begin{equation}
 \left|\widehat\chi_c^{(s)}(\nu)\right|
 \leq C_s(1+|\nu|)^{-12}.
 \label{eq:compact-transform-derivative-bound}
\end{equation}
The bound follows by differentiating
$\sin^{12}(a\nu)/(a\nu)^{12}$: derivatives of the bounded numerator do not
worsen the twelfth-power denominator, while derivatives of the denominator
only improve it.  The alternating formula
Eq.~\eqref{eq:compact-switching-truncated-power} is used for the proof, not for
floating-point evaluation near a knot; the production calculation evaluates
the stable analytic transform in Eq.~\eqref{eq:compact-transform-proof}.

\subsection{Flat-space norm and its ultraviolet tail}

In the Minkowski limit $\mathcal A_{\omega0}=2\omega$, and hence
\begin{equation}
 \|\eta_0\|^2
 =\int_0^\infty\frac{\dd\omega\,\omega}{4\pi^2}
 |\widehat\chi_c(\Omega+\omega)|^2.
 \label{eq:compact-flat-norm}
\end{equation}
Because $|\sinc x|\leq\min(1,|x|^{-1})$, the integrand has an
$O(\omega^{-23})$ envelope.  More explicitly, for
$Y=\Lambda+\Omega$ and $\Omega\geq0$,
\begin{align}
 \int_\Lambda^\infty\frac{\dd\omega\,\omega}{4\pi^2}
 |\widehat\chi_c(\Omega+\omega)|^2
 &\leq\frac{A_\chi^2 2^{23}}{\pi\sigma^{22}}
 \left(\frac{Y^{-22}}{22}-\frac{\Omega Y^{-23}}{23}\right).
 \label{eq:compact-flat-tail-bound}
\end{align}
This proves absolute ultraviolet convergence without averaging oscillations.
For completeness, expanding $\sin^{24}x$ into its constant Fourier component
$\binom{24}{12}/2^{24}$ and finitely many cosines, then integrating the cosine
terms once by parts, gives the sharper asymptotic
\begin{align}
 \int_\Lambda^\infty\frac{\dd\omega\,\omega}{4\pi^2}
 |\widehat\chi_c(\Omega+\omega)|^2
 &=\frac{A_\chi^2\binom{24}{12}}{44\pi}
 [\sigma(\Lambda+\Omega)]^{-22}
 \nonumber\\
 &\quad+O_{\sigma\Omega}\!\left(
 A_\chi^2[\sigma(\Lambda+\Omega)]^{-23}\right).
 \label{eq:compact-flat-tail-asymptotic}
\end{align}

\subsection{Matched-shell modes and radius derivatives}

We next control the matched spectral amplitude and its radius derivatives
under the spectral integral.  Fix a multipole $\ell$ and a compact radius interval
$K\Subset(2M,\infty)$.  The matched coefficients obey, uniformly for
$R\in K$ and large $\omega$,
\begin{equation}
 \sup_{R\in K}
 \left|\partial_R^j\mathcal A_{\omega\ell}(R)\right|
 \leq C_{j,K,\ell}(1+\omega)^{j+1},
 \qquad j=0,1,\ldots .
 \label{eq:matched-mode-radius-bound}
\end{equation}
To prove this bound without differentiating a wrapped phase, let
$F_\pm(\omega,r_*)$ be the exterior Jost pair, normalized by
$F_\pm\sim e^{\pm\ii\omega r_*}$ and
$W[F_+,F_-]=-2\ii\omega$.  Expand the real solution generated by unit
interior data as
\begin{align}
 u_{\omega\ell}^{\rm raw}&=c_+F_++c_-F_-,
 \nonumber\\
 c_-&=-\frac{W[F_+,u_{\omega\ell}^{\rm raw}]}{2\ii\omega}
 \nonumber\\
 &=\frac{\ii C_{\omega\ell}^{\rm raw}}{2}
 e^{-\ii\delta_{\omega\ell}}.
 \label{eq:incoming-jost-coefficient}
\end{align}
The conventions in Eqs.~\eqref{eq:standing-normalization} and
\eqref{eq:incoming-mode} then give the exact relation
\begin{equation}
 \mathcal A_{\omega\ell}=\frac{\ii}{c_-}.
 \label{eq:matched-amplitude-jost-quotient}
\end{equation}

Put
$\theta_R=\omega R/\sqrt{f_R}-\ell\pi/2$ and
$L(R)=R/\sqrt{f_R}-r_*(R)$.  The fixed-$\ell$ spherical-Bessel expansions and
the junction data in Eqs.~\eqref{eq:field-junction} and
\eqref{eq:derivative-junction} give, uniformly for $R\in K$,
\begin{align}
 u_{\omega\ell}^{\rm raw}(R)
 &=\frac{\sqrt{f_R}}{\omega}\sin\theta_R
   +O_{K,\ell}(\omega^{-2}),
 \nonumber\\
 \partial_{r_*}u_{\omega\ell}^{\rm raw}(R)
 &=\sqrt{f_R}\cos\theta_R+O_{K,\ell}(\omega^{-1}).
 \label{eq:raw-shell-high-frequency-data}
\end{align}
The Volterra equation for $F_+$ and its $r_*$ derivatives yields
\begin{align}
 F_+(\omega,r_*(R))
 &=e^{\ii\omega r_*(R)}[1+O_{K,\ell}(\omega^{-1})],
 \nonumber\\
 \partial_{r_*}F_+(\omega,r_*(R))
 &=\ii\omega e^{\ii\omega r_*(R)}+O_{K,\ell}(1),
 \label{eq:jost-high-frequency-data}
\end{align}
with the corresponding differentiated symbol estimates.  The exterior
potential itself is independent of $R$ in the common tortoise convention;
radius differentiation acts on the moving evaluation point and the shell
data.  Substitution in Eq.~\eqref{eq:incoming-jost-coefficient} gives
\begin{align}
 c_-(\omega,R)
 &=\frac{\ii\sqrt{f_R}}{2\omega}
 e^{-\ii[\omega L(R)-\ell\pi/2]}
 [1+O_{K,\ell}(\omega^{-1})],
 \nonumber\\
 C_{\omega\ell}^{\rm raw}(R)
 &=\frac{\sqrt{f_R}}{\omega}+O_{K,\ell}(\omega^{-2}),
 \label{eq:incoming-jost-asymptotic}
\end{align}
and hence $|c_-|\geq c_{K,\ell}/\omega$ at sufficiently large $\omega$.
The differentiated estimates give
$\partial_R^j c_-=O_{j,K,\ell}(\omega^{j-1})$.  Repeated differentiation of
the quotient in Eq.~\eqref{eq:matched-amplitude-jost-quotient}, using this
lower bound, proves Eq.~\eqref{eq:matched-mode-radius-bound}.  In particular,
\begin{equation}
 \mathcal A_{\omega\ell}(R)
 =\frac{2\omega}{\sqrt{f_R}}
 e^{\ii[\omega L(R)-\ell\pi/2]}
 [1+O_{K,\ell}(\omega^{-1})],
 \label{eq:matched-amplitude-high-frequency}
\end{equation}
which retains the scattering/optical phase.

Put $\nu_R(\omega)=\Omega+\omega/\sqrt{f_R}$.  Combining
Eqs.~\eqref{eq:compact-transform-derivative-bound} and
\eqref{eq:matched-mode-radius-bound}, and applying the product and chain rules
to Eq.~\eqref{eq:finite-readout-amplitude}, yields
\begin{equation}
 \begin{aligned}
 \partial_R^r q_R^I(\omega)
 &=O_K\!\left(\omega^{r-23/2}\right),\\
 \left|\partial_R^r q_R^I(\omega)\right|^2
 &=O_K\!\left(\omega^{2r-23}\right).
 \end{aligned}
 \label{eq:readout-state-radius-decay}
\end{equation}
The exponent is integrable for $r=0,1,\ldots,10$.  In particular, the state
norm has an $O(\omega^{-23})$ envelope and the first-radius-derivative norm
needed by centered radius differentiation has an $O(\omega^{-21})$ envelope.  The bounds are
uniform on $K$, so dominated convergence permits the radius derivative to pass
through the centered spectral integral.  Cross entries follow by
Cauchy-Schwarz.

Every fixed finite multipole sum in the translated calculation obeys the
preceding termwise estimate.  To treat the exact untruncated translated
response, let $P_R=L_R^{1/2}$ and let $E_R^P(I;x,x)$ denote the diagonal
kernel of its spectral projector on a frequency interval $I$.  Uniformly for
$R\in K$ and detector points $x$ in a compact interior set a positive optical
distance from the shell,
\begin{equation}
 E_R^P([n,n+1);x,x)\leq C_{K,x}(1+n)^2,
 \qquad n=1,2,\ldots .
 \label{eq:translated-local-spectral-cluster}
\end{equation}
This is a local frequency-cluster estimate, not a derivative of a cumulative
Weyl asymptotic.  To prove it without a nontrapping assumption, choose an even,
nonnegative Schwartz function $\varphi$ with
$\varphi\geq\mathbf 1_{[-1,1]}$ and with $\widehat\varphi$ supported in a time
interval shorter than the uniform optical distance from the detector set to
the shell.  For $n\geq1$ set
\begin{equation}
 m_n(\lambda)=\varphi(\lambda-n)+\varphi(\lambda+n).
\end{equation}
Then $m_n$ is even and nonnegative and majorizes
$\mathbf 1_{[n,n+1)}$ on the nonnegative spectrum.  Hence the spectral theorem
gives $E_R^P([n,n+1);x,x)\leq m_n(P_R)(x,x)$.  Because $m_n$ is even, Fourier
inversion writes its operator kernel entirely in terms of the finite-speed
cosine propagator,
\begin{equation}
 m_n(P_R)=\frac{1}{\pi}\int
 \widehat\varphi(t)\cos(nt)\cos(tP_R)\,\dd t
 \label{eq:even-cluster-majorant}
\end{equation}
up to the fixed Fourier convention.  Finite propagation therefore keeps the
kernel in the smooth uniformly elliptic interior region.  The local
three-dimensional wave parametrix bounds its diagonal by
$C_{K,x}(1+n)^2$, uniformly in $R\in K$.  Thus trapping outside that
short-time neighborhood does not enter the estimate.

With Klein-Gordon normalization, the one-particle measure is
$(2\omega)^{-1}\dd E_R^P(\omega;x,x)$, up to the fixed convention factor
already included in the mode amplitudes.  Since
$|\widehat\chi_c(\Omega+\omega/\sqrt{f_R})|^2
\leq C_K(1+\omega)^{-24}$, Eq.~\eqref{eq:translated-local-spectral-cluster}
gives, for $n\geq1$,
\begin{align}
 &\int_{[n,n+1)}
 \frac{|\widehat\chi_c(\Omega+\omega/\sqrt{f_R})|^2}{2\omega}
 \,\dd E_R^P(\omega;x,x)
 \nonumber\\
 &\qquad\leq C_{K,x}\,n^{-1}(1+n)^{-24}(1+n)^2
 =O_{K,x}(n^{-23}).
 \label{eq:translated-readout-cluster-tail}
\end{align}
The sum over $n\geq1$ converges absolutely.  The bounded interval
$0<\omega\leq1$ can be controlled without interchanging a channelwise
threshold expansion and the infinite multipole sum.  On every compact
$K\Subset(2M,\infty)$ the full three-dimensional quadratic forms are uniformly
comparable with the Euclidean form and their measures are uniformly comparable
with Euclidean volume: the only coefficient jump is the transparent shell
interface and $f_R$ is bounded away from zero.  The Euclidean Sobolev
inequality consequently gives the uniform Nash inequality
\begin{equation}
 \|u\|_2^{10/3}\leq C_K\,\mathfrak q_R[u]\,\|u\|_1^{4/3}.
 \label{eq:translated-nash}
\end{equation}
For $v(t)=e^{-tL_R}g$, form differentiation gives
$-\partial_t\|v(t)\|_2^2=2\mathfrak q_R[v(t)]$, while the Markov property gives
$\|v(t)\|_1\leq\|g\|_1$.  Equation~\eqref{eq:translated-nash} therefore yields
$\|e^{-tL_R}\|_{1\to2}\leq C_Kt^{-3/4}$; self-adjointness and the semigroup
property give $\|e^{-tL_R}\|_{1\to\infty}\leq C_Kt^{-3/2}$.  In particular,
the diagonal heat kernel satisfies this bound on the compact detector set.

Let
$F_R(\varepsilon;x)=E_R^P([0,\varepsilon];x,x)$.  Positivity and the spectral
theorem, with $t=\varepsilon^{-2}$, now give the full-angular-momentum estimate
\begin{equation}
 F_R(\varepsilon;x)
 \leq e\,e^{-L_R/\varepsilon^2}(x,x)
 \leq C_{K,x}\varepsilon^3,
 \qquad 0<\varepsilon\leq1.
 \label{eq:translated-low-frequency-projector}
\end{equation}
Writing $W_R(\omega)=
|\widehat\chi_c(\Omega+\omega/\sqrt{f_R})|^2$, both $W_R$ and $W_R'$ are
uniformly bounded on this interval.  Stieltjes integration by parts and
Eq.~\eqref{eq:translated-low-frequency-projector} give
\begin{align}
 \mathcal N_{R,x}^{(0,1]}
 &:=\int_{(0,1]}\frac{W_R(\omega)}{2\omega}\,
 \dd F_R(\omega;x)
 \nonumber\\
 &=\left[\frac{W_RF_R}{2\omega}\right]_{0}^{1}
 \nonumber\\
 &\quad-\int_0^1F_R(\omega;x)\,
 \dd\!\left(\frac{W_R(\omega)}{2\omega}\right)<\infty,
 \label{eq:translated-low-frequency-kg}
\end{align}
because the boundary term at zero is $O(\omega^2)$ and the remaining
integrand is $O(\omega)$.  This proves the infrared bound for the complete
multipole sum.  Mixed translated Gram entries follow by Cauchy-Schwarz, so
the full translated response is normalizable.  The radius-derivative estimates stated here apply only to the centered
fixed-$\ell$ sector established above.

It remains to close the infrared endpoint.  For the centered $s$ wave, the
absence of a threshold eigenstate or half-bound state proved in
Appendix~\ref{app:radial-completeness} means that the zero-energy Jost function
does not vanish.  In the unit-interior normalization this gives
$C_{\omega0}^{\rm raw}=c_R/\omega+O_K(1)$ with $c_R$ bounded away from zero on
$K$.  Equations~\eqref{eq:incoming-jost-coefficient} and
\eqref{eq:matched-amplitude-jost-quotient} therefore give
$\mathcal A_{\omega0}=O_K(\omega)$ as $\omega\downarrow0$.  The zero-energy
Jost data depend smoothly on $R$ and remain uniformly nonzero on $K$;
differentiating their threshold expansion consequently gives
$\partial_R^j\mathcal A_{\omega0}=O_{j,K}(\omega)$ for every fixed $j$ used
below.  Moreover,
$\partial_R^j(\Omega+\omega/\sqrt{f_R})=O_{j,K}(\omega)$ for $j\geq1$.
Since $\widehat\chi_c$ and all its derivatives are bounded near $\Omega$, the
product and chain rules yield
\begin{equation}
 |\partial_R^j q_R^I(\omega)|^2=O_{j,K}(\omega),
 \qquad j=0,1,\ldots,10,
 \qquad \omega\downarrow0.
 \label{eq:readout-state-infrared-bound}
\end{equation}
For a translated detector, the full-multipole heat-kernel estimate
\eqref{eq:translated-low-frequency-kg} supplies the corresponding infrared
statement without requiring a uniform-in-$\ell$ Jost expansion.  Thus both
endpoints are controlled and the finite post-switch vectors used in the paper
lie in the one-particle Hilbert space.  None of these estimates is asserted
uniformly in the unsampled limit $R\to2M^+$.

At an isolated sinc zero, logarithmic densities and a separately unwrapped
phase are poor coordinates.  The response can instead be formed directly from the
complex $L^2$ vectors. Isolated zeros therefore do not create an artificial
phase jump in the Gram integrand.

\subsection{Exact switch-off and sharply truncated tails}

Let $g$ be a Gaussian and suppose it is accumulated only to a finite time
$T$ with $g(T)\neq0$:
\begin{equation}
 G_T(\nu):=\int_{-\infty}^{T}\dd\tau\,
 g(\tau)e^{\ii\nu\tau}.
\end{equation}
Two integrations by parts give
\begin{equation}
 G_T(\nu)=\frac{g(T)e^{\ii\nu T}}{\ii\nu}+O(\nu^{-2}).
 \label{eq:truncated-gaussian-asymptotic}
\end{equation}
Consequently its flat one-particle norm has the ultraviolet behavior. This expression gives the ultraviolet divergence at fixed \(\Omega\) and \(\sigma\). For \(\Omega>0\), the logarithm may equivalently be written as \(\log(1+\Lambda/\Omega)\), as they differ only by a bounded term at large $\Lambda$ with the difference absorbed into \(O(1)\).
\begin{equation}
 \int_0^\Lambda\frac{\dd\omega\,\omega}{4\pi^2}
 |G_T(\Omega+\omega)|^2
 =\frac{|g(T)|^2}{4\pi^2}\log(\sigma\Lambda)+O(1).
 \label{eq:truncated-gaussian-divergence}
\end{equation}
For $g(T)=A_\chi e^{-T^2/(2\sigma^2)}$, the logarithmic coefficient is
$A_\chi^2e^{-T^2/\sigma^2}/(4\pi^2)$.  The complete Gaussian is Schwartz and
has no such defect; the divergence is caused by stopping a nonzero tail
sharply and then interpreting that truncated integral as a state.

For the compact pulse, by contrast, $\chi_c$ and its first ten derivatives
vanish at $T_\chi=6\sigma$.  Therefore every readout satisfying
$\tau_i^{\rm r}\geq T_\chi$ has exactly the same interaction-picture vector,
and
\begin{align}
 q_i^{S}(\omega;t_{\rm r})
 &=e^{-\ii\omega t_{\rm r}}e^{-\ii\Omega\tau_i^{\rm r}}
   q_i^I(\omega),
 \nonumber\\
 q_i^{\rm r}(\omega;t_{\rm r})
 &=e^{+\ii\omega t_{\rm r}}q_i^S(\omega;t_{\rm r})
  =e^{-\ii\Omega\tau_i^{\rm r}}q_i^I(\omega)
  \in L^2(\mathbb R_+,\dd\omega).
 \label{eq:post-switch-l2-state}
\end{align}
At equality, the single support endpoint has zero measure; no half-weight
prescription is required.  By contrast, a genuine readout at
$T_\chi-\varepsilon$ has
$\chi_c(T_\chi-\varepsilon)=O(\varepsilon^{11})\neq0$ and a sharply truncated
transform with a $1/\nu$ tail.  Its flat norm still diverges logarithmically,
with coefficient $O(\varepsilon^{22})$; smallness does not restore Hilbert-space
membership.  The implementation's spacing-relative sixteen-ulp window is
therefore only roundoff identification: such an input is clamped to the exact
common $T_\chi$ event before any detector phase is formed, while an input
earlier by more than $16|\operatorname{spacing}(T_\chi)|$ is rejected.  Later
readouts change the full state by common free-field evolution and by the
physical detector-clock scalar.  The common field unitary cancels from a
fixed-event Gram matrix. The detector scalar cancels from a normalized overlap
magnitude, but coherent quantities such as $\Gamma_{12}^{(2)}(t_{\rm r})$
generally change with the readout event. This is exact switch-off, not phase
locking.

\section{Direction-basis reduction and collinear quadrature}
\label{app:direction-basis}

Fix one asymptotic Cartesian frame, shell centers $\bm c_i$, and a detector at
$\bm x$.  The branch-relative displacement and distance are
\begin{equation}
  \bm\rho_i:=\bm x-\bm c_i,
  \qquad \rho_i:=|\bm\rho_i|<R_i.
  \label{eq:branch-distance}
\end{equation}
The constant interior lapse gives $t=\tau/\sqrt{f_i}$ throughout branch $i$.
On the control space define
\begin{equation}
  \widehat{\bm C}:=\sum_i\bm c_i|i\rangle\langle i|,
  \qquad
  \widehat D_{\bm x}:=|\bm x-\widehat{\bm C}|
  =\sum_i\rho_i|i\rangle\langle i|.
  \label{eq:center-distance-operator}
\end{equation}
Its distribution in the prepared control state is
\begin{equation}
  p_{\bm x}(d)=|\alpha_C|^2\delta(d-\rho_1)
  +|\beta_C|^2\delta(d-\rho_2).
  \label{eq:distance-distribution}
\end{equation}
This phase-independent observable is relational to the stipulated control
basis; coherence instead enters the complementary readout through
$\F_{12}^{\rm r}(t_{\rm r})$
\cite{GiacominiCastroRuizBrukner2019,
deLaHametteKabelCastroRuizBrukner2023}.  For intrinsically identical branches,
the dephased response is
\begin{equation}
  \begin{aligned}
    \F_{\rm mix}(\bm x)&=\sum_i w_i
      \F_{\rm loc}(|\bm x-\bm c_i|),\\
    (w_1,w_2)&=(|\alpha_C|^2,|\beta_C|^2),
  \end{aligned}
  \label{eq:center-mixture-convolution}
\end{equation}
a distribution of branch-relative distances rather than a response about one
  effective center.  The common-center, lab-fixed, and center-following protocols
respectively use $\bm x=\bm c_1=\bm c_2$, one $\bm x$ with distinct
$\bm c_i$, and controlled trajectories $\bm x_i=\bm c_i$
\cite{FooOnoeZych2020}.  For symmetric scans we write
\begin{equation}
  \bm c_{1,2}=\bm c_0\mp\frac{\bm b}{2},
  \qquad \bm x=\bm c_0+\bm s,
  \label{eq:symmetric-center-parameterization}
\end{equation}
so only $\bm b$ and $\bm s$ enter.

Let $\hat{\bm n}$ label a delta-normalized incoming asymptotic position
direction, so the incident rays propagate along $-\hat{\bm n}$.  The unitary
spherical-harmonic transform obeys
\begin{align}
  a_{\omega\hat{\bm n}}
  &:=\sum_{\ell,m}Y_{\ell m}(\hat{\bm n})a_{\omega\ell m},
  \nonumber\\
  a_{\omega\ell m}
  &=\int\dd\Omega_{\hat{\bm n}}\,
    Y_{\ell m}^*(\hat{\bm n})a_{\omega\hat{\bm n}},
  \nonumber\\
  [a_{\omega\hat{\bm n}},a^\dagger_{\omega'\hat{\bm n}'}]
  &=\delta(\omega-\omega')
    \delta^{(2)}(\hat{\bm n},\hat{\bm n}').
  \label{eq:direction-basis-commutator}
\end{align}
The translation phase follows directly from the common-mass asymptotics:
\[
 r_*(|r\hat{\bm n}-\bm c_i|)
 =r_*(r)-\hat{\bm n}\cdot\bm c_i+o(1),\qquad r\to\infty.
\]
The shared Schwarzschild logarithm contributes only $o(1)$ to this difference.
A branch-centered incoming annihilation mode therefore carries
$e^{+\ii\omega\hat{\bm n}\cdot\bm c_i}$ relative to the common incoming
wave $e^{-\ii\omega(t+r_*)}/r$.  Multiplication by the inverse phase fixes
its common incident amplitude; conjugation gives the positive center phase
in the excitation coefficient.  With $k_i=\omega/\sqrt{f_i}$, the scalar
spherical-wave translation and addition identities
\cite{Wittmann1988SphericalTranslation,NISTDLMF} give
\begin{multline}
  q_i(\omega,\hat{\bm n};\bm x)=
  \frac{\widehat\chi_c(\Omega+k_i)}{\sqrt{4\pi\omega}}
  e^{+\ii\omega\hat{\bm n}\cdot\bm c_i}\\
  {}\times\sum_{\ell=0}^{\infty}\frac{2\ell+1}{4\pi}
  \mathcal A^{(i)*}_{\omega\ell}
  j_\ell(k_i\rho_i)
  P_\ell(\hat{\bm n}\cdot\hat{\bm\rho}_i).
  \label{eq:directional-amplitude}
\end{multline}
The conjugated common-incoming coefficient and positive translation phase are
fixed by Eq.~\eqref{eq:incoming-mode}.  At $\rho_i=0$ the expression is defined
continuously by its $\ell=0$ term.  Substitution into
Eq.~\eqref{eq:relational-gram} gives a Hermitian positive response without
pairing angular labels about distinct origins.

The scalar translation theorem and Rayleigh/addition identities
\cite{Wittmann1988SphericalTranslation,NISTDLMF} fix the conjugation and phase
in Eq.~\eqref{eq:directional-amplitude}.  In the flat limit the sum reduces to
$(2\omega/4\pi)e^{\ii\omega\hat{\bm n}\cdot(\bm x-\bm c_i)}$; the explicit
center phase leaves the common plane wave
$e^{\ii\omega\hat{\bm n}\cdot\bm x}$, proving common-translation covariance.
Coincident centers recover the standard partial-wave sum by angular
orthogonality.  An extra $(-1)^\ell$, omitted conjugation, or reversed
translation phase would fail these checks.

For the near-center expansion, first impose finite frequency and angular
cutoffs, set $\widehat\chi_i=\widehat\chi_c(\Omega+k_i)$, and expand the first
three regular spherical Bessel functions.  The cutoff removal and the displayed
remainder are justified below; no condition $k_i\rho_i\ll1$ is imposed
uniformly on the unbounded frequency integral.
Equation~\eqref{eq:directional-amplitude} becomes
\begin{multline}
  q_i=\frac{\widehat\chi_i
  e^{\ii\omega\hat{\bm n}\cdot\bm c_i}}
  {4\pi\sqrt{4\pi\omega}}
  \biggl[\mathcal A^{(i)*}_{\omega0}
  +k_i\mathcal A^{(i)*}_{\omega1}
  (\hat{\bm n}\cdot\bm\rho_i)\\
  {}-\frac{k_i^2\rho_i^2}{6}\mathcal A^{(i)*}_{\omega0}
  +\frac{k_i^2\rho_i^2}{3}\mathcal A^{(i)*}_{\omega2}
  P_2(\hat{\bm n}\cdot\hat{\bm\rho}_i)
  +\order((k_i\rho_i)^3)\biggr].
  \label{eq:near-center-amplitude}
\end{multline}
Substituting Eq. \(\eqref{eq:near-center-amplitude}\) into the diagonal \(i=j\) case of Eq. \(\eqref{eq:relational-gram}\), the translation phase cancels. Angular integration then eliminates the linear and quadrupole cross terms and gives
\begin{multline}
  \F_{ii}(\rho_i;\Omega)=\F_{ii}(0;\Omega)\\
  {}+\rho_i^2\int_0^\infty
  \frac{\dd\omega}{48\pi^2\omega}
  |\widehat\chi_i|^2k_i^2
  \left(
  |\mathcal A^{(i)}_{\omega1}|^2
  -|\mathcal A^{(i)}_{\omega0}|^2
  \right)+\order(\rho_i^4).
  \label{eq:near-center-local-response}
\end{multline}
We justify the cutoff removal and the remainder at finite order.  For a
Cartesian multi-index $\alpha$, let
\begin{equation}
 E_{R_i,\alpha}^{P}(I;\bm x)
 :=
 \left.
 \partial_{\bm x}^{\alpha}\partial_{\bm x'}^{\alpha}
 E_{R_i}^{P}(I;\bm x,\bm x')
 \right|_{\bm x'=\bm x}.
 \label{eq:translated-differentiated-spectral-projector}
\end{equation}
On a fixed compact interior neighborhood $U_i$ of the branch center, differentiating the
short-time local parametrix used in
Eq.~\eqref{eq:translated-local-spectral-cluster} gives
\begin{equation}
 E_{R_i,\alpha}^{P}([n,n+1);\bm x)
 \leq C_{\alpha}(1+n)^{2+2|\alpha|},
 \qquad n\geq1 .
 \label{eq:translated-differentiated-spectral-cluster}
\end{equation}
At low frequency a weaker bound suffices, and follows from interior elliptic
regularity.  Choose bounded neighborhoods
$U_i\Subset V_i\Subset W_i\Subset\{\rho_i<R_i\}$.
Realize the projector on $[0,\zeta]$ in spectral coefficient space, with
evaluation vector $\mathcal K_\zeta(\bm x)$.  Its squared norm is
$E_{R_i}^{P}([0,\zeta];\bm x,\bm x)$, so
Eq.~\eqref{eq:translated-low-frequency-projector} first constructs
$\mathcal K_\zeta\in L^2(W_i;\mathfrak h)$ with norm at most
$C\zeta^{3/2}$.  In the flat interior the spectral theorem gives the
distributional equation
$-f_i\Delta_{\bm x}\mathcal K_\zeta
=\mathsf M_\omega^2\mathcal K_\zeta$, where $\mathsf M_\omega$ is
frequency multiplication on the restricted coefficient space and has norm
at most $\zeta$.  Hilbert-valued interior elliptic estimates, obtained
componentwise and integrated over spectral coefficients, therefore give
\[
 \|\mathcal K_\zeta\|_{H^6(V_i;\mathfrak h)}
 \leq C\sum_{j=0}^{3}
 \|\mathsf M_\omega^{2j}\mathcal K_\zeta\|_{L^2(W_i;\mathfrak h)}
 \leq C\zeta^{3/2}.
\]
The constants depend on the fixed geometry and neighborhoods, not on
$0<\zeta\leq1$.  Sobolev embedding $H^6(V_i;\mathfrak h)
\hookrightarrow C^4(U_i;\mathfrak h)$ supplies the pointwise derivatives,
without assuming a differentiated heat-kernel bound.  Squaring their norms gives
\begin{equation}
 E_{R_i,\alpha}^{P}([0,\zeta];\bm x)
 \leq C_{\alpha}\zeta^3,
 \qquad 0<\zeta\leq1,\quad |\alpha|\leq4 .
 \label{eq:translated-differentiated-low-frequency}
\end{equation}
Since
$|\widehat\chi_c(\Omega+\omega/\sqrt{f_i})|^2
 \leq C(1+\omega)^{-24}$,
the contribution of the $n$th unit-frequency block to the squared
one-particle norm of $\partial_{\bm x}^{\alpha}q_i$ is
\begin{equation}
 O\!\left(n^{-1}n^{-24}n^{2+2|\alpha|}\right)
 =O\!\left(n^{2|\alpha|-23}\right).
 \label{eq:translated-spatial-derivative-tail}
\end{equation}
This is summable for $|\alpha|\leq4$.  At the infrared endpoint, Stieltjes
integration by parts with
Eq.~\eqref{eq:translated-differentiated-low-frequency} is finite as well:
the boundary term is $O(\omega^2)$ and the remaining integrand is $O(\omega)$,
as in Eq.~\eqref{eq:translated-low-frequency-kg}.  The estimates are uniform
on the chosen interior neighborhood, so dominated
differentiation gives
\begin{equation}
 \begin{aligned}
  q_i&\in C^4\!\left(
    U_i;\,
    L^2(\mathbb R_+\times S^2,
        \dd\omega\,\dd\Omega_{\hat{\bm n}})
  \right),\\
  \F_{ii}&=\|q_i\|^2\in C^4(U_i) .
 \end{aligned}
 \label{eq:translated-center-c4}
\end{equation}
Rotational symmetry makes the diagonal response even about its branch center.
Taylor's theorem therefore yields
Eq.~\eqref{eq:near-center-local-response}, including its
$\order(\rho_i^4)$ remainder.  This is a finite-order regularity result: the
polynomial ultraviolet envelope does not establish a convergent all-orders
Taylor series.  A mixed response need not be even about an externally selected
midpoint unless a branch-exchange symmetry enforces it.  Away from the
near-center regime, the converged multipole sum is used directly.  The response
remains finite through either center, and the width in
Eq.~\eqref{eq:center-mixture-convolution} is relational, not a regularization
of a classical singularity or detector smearing.

For a diagonal element the center phase has unit modulus and angular
orthogonality removes the double multipole sum.  Equation~\eqref{eq:relational-gram}
therefore gives the useful independent check
\begin{multline}
  \F_{ii}(\rho_i;\Omega)=\int_0^\infty
  \frac{\dd\omega}{16\pi^2\omega}
  |\widehat\chi_c(\Omega+k_i)|^2\\
  {}\times\sum_{\ell=0}^{\infty}(2\ell+1)
  |\mathcal A^{(i)}_{\omega\ell}|^2j_\ell(k_i\rho_i)^2.
  \label{eq:relational-local-sum}
\end{multline}
At $\rho_i=0$, $j_\ell(0)=\delta_{\ell0}$ and this is precisely
Eq.~\eqref{eq:center-gram-integral} with $i=j$.  In particular, the
direction $\hat{\bm\rho}_i$ never needs to be assigned at a branch center.

For the numerical profile, rotate the common line to the $z$ axis, denote the
scalar center coordinates by $c_i$, the detector coordinate by $s$, and set
$u:=\hat{\bm n}\cdot\hat{\bm z}$.  With
$\epsilon_i:=\operatorname{sgn}(s-c_i)$ away from the center, define the
truncated transfer function
\begin{equation}
  \mathcal T_i^{(L)}(\omega,u;s):=
  e^{\ii\omega u c_i}
  \sum_{\ell=0}^{L}\frac{2\ell+1}{4\pi}
  \mathcal A^{(i)*}_{\omega\ell}
  j_\ell(k_i|s-c_i|)P_\ell(\epsilon_i u).
  \label{eq:collinear-transfer}
\end{equation}
At $s=c_i$ the same continuous $\ell=0$ prescription is used.  The response
then becomes
\begin{multline}
  \F_{ij}^{(L)}(s;\Omega)=\int_0^\infty
  \frac{\dd\omega}{4\pi\omega}
  \widehat\chi_c(\Omega+k_i)^*
  \widehat\chi_c(\Omega+k_j)\\
  {}\times2\pi\int_{-1}^{1}\dd u\,
  \mathcal T_i^{(L)}(\omega,u;s)^*
  \mathcal T_j^{(L)}(\omega,u;s).
  \label{eq:collinear-gram-quadrature}
\end{multline}
This is the form evaluated by Gauss-Legendre quadrature.  A common shift
$c_i\mapsto c_i+a$, $s\mapsto s+a$ multiplies every transfer function by
$e^{\ii\omega ua}$ and hence leaves Eq.~\eqref{eq:collinear-gram-quadrature}
unchanged.  Exchanging the two branches sends
$\F_{12}\mapsto\F_{21}=\F_{12}^*$.  The multipole cutoff $L$, angular rule,
frequency grid, and frequency cutoffs are logically independent convergence
parameters; agreement under only one of these refinements is not sufficient.

\subsection{Relational-center benchmark}
\label{sec:relational-numerics}

To isolate translation from a radius change, both branches use $M=1$, $R=3$
and centers $\bm c_{1,2}=(0,0,\mp0.5)$.  A detector with
$\sigma=\Omega=1$ is scanned along $\bm x=(0,0,s)$ for
$-1.5\leq s\leq1.5$, with equal control weights and
$\phi-\theta=\pi/4$.  The dephased profile is then symmetric while both
readout ports remain away from a dark denominator.  The full-scan mixture
varies by about $0.87\%$ of its maximum; at the midpoint the higher-multipole fraction is
$3.085\%$ and $|\mu_{12}|=0.97923$, while higher multipoles vanish at either
branch's own center.  Appendix~\ref{app:numerics} gives the angular, multipole,
frequency, and covariance tests.

\section{Positivity and phase-scan bound}
\label{app:positivity}

The phase-scan visibility of the joint probability is
\begin{equation}
  \mathcal V_Q=
  \frac{Q_{\max}-Q_{\min}}{Q_{\max}+Q_{\min}}
  =\frac{2|c_1c_2\F_{12}^{\rm r}|}{L_{\rm r}}\leq1.
  \label{eq:joint-contrast}
\end{equation}
For arbitrary complex numbers $z_1,z_2$, the Gram definition gives
\begin{equation}
  \sum_{i,j}z_i^*\F_{ij}^{\rm r}z_j
  =\left\|\sum_i z_i|\zeta_i(t_{\rm r})\rangle\right\|^2\geq0.
\end{equation}
Hence $|\F_{12}^{\rm r}|\leq
\sqrt{\F_{11}^{\rm r}\F_{22}^{\rm r}}$; together with the
arithmetic-geometric mean inequality this bounds the numerator of
Eq.~\eqref{eq:joint-contrast} by $L_{\rm r}$.  Positivity gives
$Q_\pm(t_{\rm r})\geq0$
because each joint coefficient is a squared field-vector norm.

Completeness of the complementary readout projectors gives
Eq.~\eqref{eq:control-trace}.  The joint identity remains valid at a dark port;
the weighted conditional identity is written only where the conditionals are
defined.  An unweighted mean of conditional responses is not generally the
dephased mixture.  If one branch weight vanishes, the weighted cross term
vanishes even though $\F_{12}$ itself need not.  If the branches are identical,
\begin{equation}
  \begin{aligned}
    \F_{11}^{\rm r}&=\F_{22}^{\rm r}=\F_{12}^{\rm r}=\F_0,\\
    Q_\pm(t_{\rm r})&=P_\pm^{(0)}(t_{\rm r})\F_0,\\
    \F_\pm^{\mathrm{cond}}(t_{\rm r})&=\F_0
      \quad(P_\pm^{(0)}(t_{\rm r})>0).
  \end{aligned}
  \label{eq:identical-branch}
\end{equation}
Thus postselection doesn't double the conditional response of two identical
branches. The geometry-sensitive excess
\begin{equation}
  \Delta\F_\pm^{\mathrm{geom}}
  :=\F_\pm^{\mathrm{cond}}(t_{\rm r})-\F_{\mathrm{mix}}(t_{\rm r})
  \label{eq:geometry-excess}
\end{equation}
vanishes in this limit.  Neither an isolated zero of
Eq.~\eqref{eq:geometry-excess} nor a zero cross contribution at one control
phase implies that the control state is incoherent.

\section{Regulated detector outputs and coefficient limits}
\label{app:uv-prescription}

The completed first-order vector is square integrable, but this doesn't by itself define products of point fields at coincident detector times.
So we specify the following regulator for every full detector output
and every higher-order expansion in this paper.  In the common incoming
one-particle space $\mathfrak h$, choose
$\mathfrak r=(\kappa,\Lambda,J)$ with
$0<\kappa<\Lambda<\infty$ and a finite nonnegative integer $J$.  Define
\begin{equation}
 \begin{aligned}
 Q_{\mathfrak r}&=\mathbf1_{[\kappa,\Lambda]}(\omega)P_J,\\
 P_J&=\sum_{\ell=0}^J\sum_{m=-\ell}^{\ell}|\ell m\rangle\langle\ell m|.
 \end{aligned}
 \label{eq:detector-mode-regulator}
\end{equation}
The angular harmonics are referred to one fixed common incoming frame.
For displaced shells the branch modes are first expressed in that frame;
we do not truncate each branch's centered harmonics separately.  The same
$Q_{\mathfrak r}$ is used on both branches and at every radius in a derivative
stencil.  It is a contraction commuting with the common free-field evolution.
This is a prescribed mode regulator, not a claim of local or covariant
finite-cutoff dynamics.  Write $\mathfrak r\to\mathrm{pt}$ for
$\kappa\downarrow0$, $\Lambda\uparrow\infty$, $J\uparrow\infty$; then
$Q_{\mathfrak r}\to1$ strongly.  The auxiliary infrared cutoff avoids any
endpoint-domain assumption in constructing the regulated propagator.

Replace the field at the detector by
$\Phi_{i,\mathfrak r}(t)=a(h_{i,\mathfrak r}(t))+
a^\dagger(h_{i,\mathfrak r}(t))$, where $h_{i,\mathfrak r}$ is the
$Q_{\mathfrak r}$-projected creation-mode coefficient of
Eq.~\eqref{eq:field-expansion} at $x_i(t)$.  Finitely many angular modes on
a compact positive-frequency interval give
$h_{i,\mathfrak r}(t)\in\mathfrak h$ continuously in $t$.  With
$w_i(t)=\sqrt{f_i}\chi_c(\sqrt{f_i}t)$ and
$V_{i,\mathfrak r}(t)=w_i(t)\mu_i^{(t)}(t)\Phi_{i,\mathfrak r}(t)$, set
\begin{equation}
 B_{i,\mathfrak r}:=\int\dd t\,|w_i(t)|
       \|h_{i,\mathfrak r}(t)\|<\infty .
 \label{eq:regulated-interaction-bound}
\end{equation}
The free detector
gap $\Omega$, switching, geometry, and readout are held fixed.  At finite
$\mathfrak r$ we use this Hamiltonian without additional counterterms.

Here is a constructive control of the Dyson expansion, rather than an
operator-norm expansion of an unbounded interaction.  Let $\mathsf N_\Phi$
be field particle number and $P_{\leq K}$ its spectral projection.  Replace
$V$ temporarily by $P_{\leq K}VP_{\leq K}$; it is bounded, Hermitian, and
integrable in time, hence has a unitary propagator for real $\lambda$.
On a vector $\psi$ with at most $m$ particles, the creation/annihilation
bounds and the ordered time simplex give, uniformly in $K$,
\begin{equation}
 \|U_{i,\mathfrak r}^{(n)}\psi\|
 \leq\frac{(2B_{i,\mathfrak r})^n}{n!}
       \sqrt{\frac{(m+n)!}{m!}}\,\|\psi\| .
 \label{eq:regulated-dyson-bound}
\end{equation}
Indeed each field factor on the $q$-particle sector has norm at most
$2\sqrt{q+1}\|h\|$, and the scalar ordered integral is $B^n/n!$.
The coefficients with and without the number cutoff agree once $K\geq m+n$.
The bound is summable with any factor $|\lambda|^n(m+n+1)^s$, for fixed
$s\geq0$, so the series converges in all number graph norms on finite-particle
vectors, locally uniformly in $\lambda$ and time.  The bound
$\|V(t)\psi\|\leq2|w(t)|\|h(t)\|
\|(\mathsf N_\Phi+1)^{1/2}\psi\|$ then permits passage of the integral
Schr\"odinger equation to the limit.  Dominated convergence gives strong
limits of both the forward and reverse number-cutoff
propagators.  Their norm-one bounds allow extension from the dense
finite-particle domain and passage of both inverse identities to the limit.
Thus $U_{i,\mathfrak r}$ is unitary and its vacuum-state Dyson series is
entire in $\lambda$.  In particular,
\[
 \|U_{i,\mathfrak r}^{(n)}G\|
 \leq(2B_{i,\mathfrak r})^n/\sqrt{n!}.
\]
Known free propagation to the common readout changes neither these bounds
nor detector parity.  Every interaction flips $g\leftrightarrow e$;
therefore $P_gE_{i,\mathfrak r}$ is even and $P_eE_{i,\mathfrak r}$ is odd
in $\lambda$.  The state remainders are norm bounds, and density-operator
remainders are trace-norm bounds, at fixed $\mathfrak r$.  In particular,
the scalar overlaps and click probabilities have no odd powers.  This
justifies the $\order_{\mathfrak r}(\lambda^4)$ remainders used in the main
text.  Their constants may grow without bound as the regulator is removed.

That qualification is necessary even for an inertial detector in the
four-dimensional massless Minkowski vacuum, at rest in the frequency-cutoff
frame.  For the same real compact pulse define
$C_\chi(s)=\int\dd u\,\chi_c(u)\chi_c(u-s)$.  With only a frequency cutoff
(the harmless infrared cutoff has been removed), the vacuum component of
the second-order no-click vector is
\begin{equation}
 c_\Lambda^{(2)}=-\frac1{4\pi^2}\int_0^\Lambda\omega\,\dd\omega
       \int_0^\infty\dd s\,C_\chi(s)e^{-\ii(\Omega+\omega)s}.
 \label{eq:flat-second-order-vacuum-cutoff}
\end{equation}
The autocorrelation is real, even, compactly supported, and sufficiently
smooth for three integrations by parts; $C_\chi(0)>0$ and $C_\chi'(0)=0$.
Consequently its one-sided transform is
$-\ii C_\chi(0)/k+O(k^{-3})$ for $k\geq\Omega>0$.  Its real part is exactly
$|\widehat\chi_c(k)|^2/2$.  Hence
\begin{align}
 \operatorname{Re}c_\Lambda^{(2)}&=-\frac12N_\Lambda,\qquad
 N_\Lambda\longrightarrow N<\infty,
 \nonumber\\
 \operatorname{Im}c_\Lambda^{(2)}
 &=\frac{C_\chi(0)}{4\pi^2}
   \left[\Lambda-\Omega\ln(1+\Lambda/\Omega)\right]+O(1).
 \label{eq:flat-second-order-uv-phase}
\end{align}
The last remainder stays bounded because
$\int_0^\infty\omega(\Omega+\omega)^{-3}\dd\omega<\infty$.
Thus finite first-order norm does not supply a pointlike second-order
state.  Coincident-time propagator and vertex divergences are also discussed
in Ref.~\cite{ColosiOecklZampeli2024}, Sec.~VI.C; no renormalization scheme
from that work is imported here.  We retain the vacuum contractions required
by Eq.~\eqref{eq:dyson-unitarity-identity}.  Subtracting a common phase is
not a proof of a renormalized propagator, and branch-dependent phases may
not be silently absorbed into the stipulated control preparation.

The coefficient limits needed here do exist.  Because the regulator is a
common projection independent of time and geometry,
\begin{equation}
 \eta_{i,\mathfrak r}=Q_{\mathfrak r}\eta_i\longrightarrow\eta_i,
 \qquad
 \zeta_{i,\mathfrak r}=Q_{\mathfrak r}\zeta_i\longrightarrow\zeta_i
 \quad\hbox{in }\mathfrak h.
 \label{eq:first-order-regulator-limit}
\end{equation}
Cauchy-Schwarz gives convergence of every first-order Gram entry. Normalized overlaps converge wherever response norms remain
positive. The attenuation coefficient is consequently defined by the ordered
limits
\begin{equation}
 \Gamma_{12}^{(2)}
 =\lim_{\mathfrak r\to\mathrm{pt}}\lim_{\lambda\to0^+}
 \frac{1-|\mathcal I_{12,\mathfrak r}(\lambda)|}{\lambda^2}.
 \label{eq:coefficient-first-point-limit}
\end{equation}
The leading probability coefficients are obtained in the same order from
the corresponding regulated probabilities. No full point-detector
propagator, cutoff-independent finite-coupling output, or cutoff-uniform
remainder is asserted. The numerical frequency cutoffs test convergence of
these finite first-order integrals; they are not a renormalization
prescription or a higher-order detector-error estimate.

\section{Numerical protocols and validation}
\label{app:numerics}

Tests below distinguish numerical convergence of leading Gram coefficients from finite-coupling control. All displayed plots use $A_\chi=1$.
Far-domain matching checks, frequency refinements, symmetry limits, and exact
post-switch phase propagation are separate diagnostics.

\subsection{Radial solver and centered benchmark}

This subsection describes the an uncompensated excitation benchmark of Fig.~\ref{fig:publication-response}; the extended compensated calculation is specified separately below.

The baseline takes $\sigma=1$, the unit pulse normalization $A_\chi=1$, and
\begin{equation}
  M=0.20,\qquad R_1=5.0,\qquad R_2=6.0,
  \label{eq:benchmark-shells}
\end{equation}
with $\alpha_C=\cos(\pi/8)$,
$\beta_C=\sin(\pi/8)$ and fixed $\phi-\theta=0$ at the pulse center.
Figure~\ref{fig:publication-response}(b) varies the detector-clock phase while
holding those preparation and analyzer phases fixed; no subsequent phase
alignment is applied.  The
primary readout uses Eq.~\eqref{eq:common-readout-schedule}.  The response uses
$0.1\leq\Omega\sigma\leq3.5$ and
\begin{equation}
  1.25\times10^{-4}\leq\omega\sigma\leq10.
  \label{eq:frequency-grid}
\end{equation}
The centered calculation is exactly $\ell=0$.

For each $(R_i,\omega,\ell)$ the unit-coefficient regular interior mode fixes
$u(R_i)$ and $\dd u/\dd r_*(R_i)$ through Appendix~\ref{app:junction}.  The
exterior equation is integrated with an eighth-order adaptive Dormand-Prince
method at relative and absolute tolerances $10^{-9}$ and $10^{-11}$.  The
field and its tortoise derivative are fitted simultaneously to sine and cosine
components over the final two wavelengths.  The first-order asymptotic phase
used in that fit is
\begin{equation}
  \Theta_\ell(r)=\omega r_*+
  \frac{\ell(\ell+1)}{2\omega r}+
  \frac{M}{2\omega r^2},
  \label{eq:asymptotic-fit-phase}
\end{equation}
where the last term is generated by the $2M/r^3$ part of the scalar
potential.  The domain extends at least 24 wavelengths beyond the shell and
not less than $4R_i$; 256 fit samples are used.  Representative modes are
recomputed on 24-, 48-, and 96-wavelength domains.  The acceptance gates are
relative changes below $10^{-4}$ in $A$, $10^{-4}$ rad in phase,
$1.5\times10^{-4}$ in $\mathcal A=Ae^{\ii\delta}$, and a fit residual below
$10^{-4}$.

The centered response uses 801 quadratically clustered frequencies on the
interval in Eq.~\eqref{eq:frequency-grid}.  Nested 201-, 401-, and 801-point
grids, a halved infrared cutoff, and a raised ultraviolet cutoff are evaluated
independently for $\F_{11}^{\rm r}$, $\F_{22}^{\rm r}$, complex
$\F_{12}^{\rm r}(t_{\rm r})$, $\F_{\rm mix}$, $Q_\pm(t_{\rm r})$, and both
conditional responses.  $\Delta_D$ also reconstructs the phase-demodulated interaction cross entry and the algebraic attenuation
envelope.  The separate post-switch diagnostic computes $n_{\rm r}=6,7,8$
and compares precisely the three phase-demodulated response/Gram quantities in
Eq.~\eqref{eq:post-switch-readout-diagnostics}.  Their maximum relative
envelope is measured and compared with the declared $10^{-3}$ validation
tolerance; this is a propagation comparison for the response coefficients. 

The clock-resolved conditional curves remain away from dark denominators and
obey the control-trace identity at every readout phase.  Across the gap scan the
phase-invariant ray overlap remains $|\mu_{12}|=0.99937$-$0.99990$; its
scattering contribution $\arg\mu_{12}$ is reported separately from the total
readout phase $\Delta_D(t_{\rm r})+\arg\mu_{12}$.

Across the centered response observables and gap scan at the declared readout,
the largest archived finest-grid or cutoff change is $5.50\times10^{-7}$.
The worst far-domain change in $Ae^{\ii\delta}$ is $3.42\times10^{-8}$, and
the largest fit residual over the complete 801-row base mode bank is
$5.61\times10^{-10}$. The centered audit also checks the direct/Gram identity,
the phase-envelope bound, and the $n_{\rm r}=6,7,8$ phase-demodulated
response/Gram invariance envelope in
Eq.~\eqref{eq:post-switch-readout-diagnostics} against a separate $10^{-3}$
validation tolerance. Agreement of the physical $\Gamma_{12}^{(2)}$ across
different readout schedules is not a validation criterion.

\subsection{Extended compensated centered response}
\label{app:compensated-centered-numerics}

Figures~\ref{fig:deexcitation-response}
and~\ref{fig:excitation-decay-contrast} use a separate extended calculation
with $0\leq E\sigma\leq200$ in steps of $0.05$ and field frequencies
$0\leq\omega\sigma\leq600$. The field grid has 1025 quadratically
clustered points on $[0,3]$ and spacing $0.025$ above $3$, for 24905 points
in total; the zero-frequency integrand is its limiting value zero.
The centered response is exactly $\ell=0$. Below $\omega\sigma=3$, the
outgoing Schwarzschild Jost solution is propagated inward with DOP853 at
relative and absolute tolerances $2\times10^{-11}$ and
$2\times10^{-13}$. At higher frequencies its inverse-radius expansion is
evaluated at the shells through order 24, with representative comparisons
to orders 18 and 30 and direct integration. The shell Wronskian fixes the
complex common-incoming normalization. Simpson quadrature evaluates the
interaction Gram entries and the compensated observables.

At $E\sigma=0,1,3.5,35.225$ and $10,20,\ldots,200$, coarsening the field
grid by a factor of two changes the compensated conditional responses by
at most $7.1\times10^{-7}$ relatively and the contrasts by at most
$2.1\times10^{-5}$ percentage points. Comparing field cutoffs 450 and 600
changes the conditional responses by at most $5.7\times10^{-11}$ relatively.
Independent forward real amplitude--phase integration checks representative
high-frequency complex modes. Gram positivity, direct positive-vector
versus Gram evaluation, identical-branch limits, and control tracing pass.
The actual pre/post gate sequence is also checked in both transition and
no-transition sectors. Recalibrating the gates for endpoint schedules
$T=n_{\rm r}\sigma/\min_i\sqrt{f_i}$ with $n_{\rm r}=6,7,8$
($n_{\rm r}=6$ being the ideal boundary-gate limit) leaves the
compensated conditional coefficients unchanged within
$1.4\times10^{-15}$ relatively. These are numerical checks of leading
coefficients, not bounds on finite-coupling remainders.

\subsection{Relational-center convergence}

The relational benchmark uses 401 frequencies on
$2.5\times10^{-4}\leq\omega\leq7$, $\ell_{\max}=8$, and a 128-point
Gauss-Legendre direction rule.  Independent checks use 201 and 401
frequencies, $\ell_{\max}=4,6,8$, 32, 64, and 128 angular points, a raised
ultraviolet cutoff, and a halved infrared cutoff.  Nine modes with
$\ell=0,4,8$ and $\omega\sigma=0.2,1,3.5$ are recomputed on the three far
domains.  Publication gates require agreement with the analytic local
partial-wave sum, Gram positivity, continuity and local evenness at both
centers, common-translation and branch-exchange covariance, the concentric
and Minkowski limits, and Eq.~\eqref{eq:branch-distance}.  The finest response
refinements pass $2\times10^{-4}$ and the complex radial coefficient passes
$1.5\times10^{-4}$.

All accepted relational refinements are below $2.26\times10^{-7}$ across the eight
response observables.  The worst complex radial-coefficient envelope is
$6.42\times10^{-5}$; even-center fits have linear slope below
$1.5\times10^{-15}$ and relative residual below $1.2\times10^{-9}$.

\subsection{Joint radius-translation convergence}

The combined benchmark uses 20 compactness rows, the same adaptive radius pair
as Eq.~\eqref{eq:adaptive-compactness-pair}, the fixed-$b/M$ set through the
validated $b/M=2.5$ extension, and the fixed-$b/R$ control.  Production integration
uses 257 frequencies on $2.5\times10^{-4}\leq\omega\leq7$,
$\ell_{\max}=2$, and a 16-point Gauss-Legendre rule.  Independent refinements
compare 129 and 257 frequencies, 8 and 16 angular points, halved infrared and
raised ultraviolet cutoffs, and every assignment Gram entry.  The
$\ell=1\to2$ change is retained as a multipole-necessity ablation rather than a
  convergence estimate; $\ell=2\to3$ changes at the new $b/M=2.5$ peaks are
  $7.57\times10^{-9}$ for the phase-invariant $\mathcal A_D$.  The
clock-resolved $\mathcal A_\Gamma(t_{\rm r})$ and both exact shared-phase
extrema are instead recomputed under the nested frequency-density and angular
rules and the $\ell=2\to3$ spot check. The independent cutoff trials were
computed for interaction-ray Gram entries; the restricted propagation to the
physical readout and phase extrema is given below.
All $\ell=0,1,2$ radial coefficients at representative compactnesses are
also recomputed on 24-, 48-, and 96-wavelength domains.

The acceptance tests cover Gram positivity, common-translation and
branch-exchange covariance, midpoint nulls, the full profile identity
$D_{\rm SL}(s)=D_{\rm LS}(-s)$, rowwise assignment contrasts, and the
fixed-$b/R$ control.  Equation~\eqref{eq:assignment-influence-direct} is
integrated directly at the common readout and checked against clock-dressed Gram
subtraction as a total positive coefficient.  The post-switch diagnostic evaluates the two diagonal assignment entries and phase-demodulated cross entry at $n_{\rm r}=6,7,8$; their envelope is tested as a propagation diagnostic for the demodulated response/Gram
entries under the dedicated response tolerance.
This audit is distinct from the physical
schedule dependence of $\mathcal A_\Gamma(t_{\rm r})$.  Separately, a
closed-form common-phase extremization is verified by reconstructing both
extrema, and the phase-averaged attenuation identity follows algebraically from
Eq.~\eqref{eq:attenuation-phase-envelope}.  A readout-dependent assignment contrast receives a
  resolved marker only when its ratio denominator is nonvanishing and the two
  exact shared-phase ratio extrema remain stable under the frequency-density,
  angular, and $\ell=2\to3$ refinements.  The numerator may cross zero within
  the displayed band; that physical crossing is not a failure of resolution.
  All 80 fixed-$b/M$ rows and all 20 fixed-$b/R$ rows pass these gates for
  $\mathcal A_D$, the physical $n_{\rm r}=6$ $\mathcal A_\Gamma$, and both
  shared-phase endpoints.  The largest fixed-$b/M$ refinement bounds are
  $5.13\times10^{-6}$ for $\mathcal A_D$, $1.77\times10^{-6}$ for the physical
  $\mathcal A_\Gamma$, and $(1.60,1.49)\times10^{-5}$ for the lower and upper
  phase endpoints; the corresponding fixed-$b/R$ bounds are
  $3.64\times10^{-6}$, $5.47\times10^{-6}$, and
  $(6.56,6.20)\times10^{-5}$.
  The direct/Gram, post-switch, and phase-envelope checks are applied row by row.
The production three-readout invariance audit has maximum relative residual
$4.18\times10^{-16}$ among the audited response entries and maximum absolute
propagated residual $7.67\times10^{-12}$ in $\mathcal A_\Gamma$.  For this test, each post-switch cross entry is
  demodulated and then transported to the common physical $n_{\rm r}=6$ clock
  phase before comparison with the production value; it does not compare the
  physical $\mathcal A_\Gamma(t_{\rm r})$ values at different readout times.

The cutoff audit covers only $\mathcal C=0.20$, $0.80$, $0.85$, $0.90$,
$0.92$, $0.95$, and $0.999$: 28 fixed-$b/M$ and seven fixed-$b/R$
points per trial. Physical propagation below is restricted to the
fixed-$b/M$ calculations, for which the base interaction and readout
representations are compared to bound their numerical discrepancy.
No corresponding comparison is reported for the fixed-$b/R$ control.
The infrared endpoint is lowered to $1.25\times10^{-4}$ or the
ultraviolet endpoint raised to $8$, each retaining the production grid
and adding 16 nodes. The diagonal, cross-magnitude, and circular-phase
errors $e_d,e_m,e_\phi$ are maxima over the trial set, with magnitude
errors normalized by componentwise reference maxima. A common bound $B=0.006763$ exceeds every reference diagonal and
cross magnitude. Clock dressing preserves complex difference bounds. Writing
$\Gamma_o=\Gamma_o^{(2)}$ for each assignment coefficient, every common phase obeys
\begin{equation}
 |\delta\Gamma_o(\phi)|\leq e_o
 :=B\bigl[e_d+e_m+2\sin(e_\phi/2)\bigr]+\beta_o.
 \label{eq:cutoff-influence-bound}
\end{equation}
Here $\beta_o$ bounds the absolute base/readout difference of diagonal means
plus that of demodulated cross entries. Writing
$S=\Gamma_{\rm SL}+\Gamma_{\rm LS}$, $E=e_{\rm SL}+e_{\rm LS}$,
and $\mathcal A=\mathcal A_\Gamma$, the ratio identity gives
\begin{equation}
 |\delta\mathcal A(\phi)|\leq
 \frac{(1+|\mathcal A(\phi)|)E}{S(\phi)-E},\qquad S(\phi)>E.
 \label{eq:cutoff-contrast-bound}
\end{equation}
For the extrema in Eq.~\eqref{eq:assignment-common-phase-extrema}, let
$A_*=\max(|r_-|,|r_+|)$ and $S_{\min}=d_0-\|\bm d\|$. Uniformly in phase,
\begin{equation}
 |\delta r_\pm|\leq\sup_\phi|\delta\mathcal A(\phi)|
 \leq\frac{(1+A_*)E}{S_{\min}-E},\qquad S_{\min}>E.
 \label{eq:cutoff-extrema-bound}
\end{equation}
This includes shifts of the extremizing phase.

At the physical $8.09\%$ peak, the infrared and ultraviolet bounds are
$2.807\times10^{-4}$ and $2.220\times10^{-11}$ in contrast, including the
$2.49\times10^{-15}$ direct/Gram reference discrepancy; the infrared bound
is $0.0281$ percentage points. The corresponding bounds on each phase
extremum are $6.857\times10^{-4}$ and $5.421\times10^{-11}$. At the
$39.468\%$ phase maximum ($\mathcal C=0.92$, $b/M=2.5$), the infrared
extremum bound is $1.793\times10^{-3}$: it meets the maximum's error budget
but is too loose for the small minimum. The cutoff trials cover only the compactness values listed above. These bounds quantify observed cutoff-trial
differences, not the entire omitted tails or unseen cavity layers, and do
not enlarge the original figure-marker scope.

\end{document}